\documentclass[10pt,aps,prx,twocolumn,amsmath,amssymb,longbibliography,superscriptaddress]{revtex4-2}

\pdfpageattr {/Group << /S /Transparency /I true /CS /DeviceRGB>>}
\usepackage{graphicx}
\usepackage{color}
\usepackage[hidelinks]{hyperref}
\usepackage{orcidlink}
\usepackage{amsmath}
\usepackage{amsthm}
\usepackage{array}
\usepackage{physics}
\usepackage{amsfonts}
\usepackage{times,txfonts}
\usepackage{mathtools}
\usepackage{soul}
\usepackage{cancel}
\hypersetup{colorlinks=true, linkcolor=black, citecolor=magenta, urlcolor=magenta}

\begin{document}
\author{George Mihailescu\,\orcidlink{0000-0002-0048-9622}}
\email[]{george.mihailescu96@gmail.com}
\affiliation{School of Physics, University College Dublin, Belfield, Dublin 4, Ireland}
\affiliation{Centre for Quantum Engineering, Science, and Technology, University College Dublin, Dublin 4, Ireland}

\author{Uesli Alushi\,\orcidlink{0009-0000-6888-0532}}

\author{Roberto Di Candia\,\orcidlink{0000-0001-9087-2125}}
\affiliation{Department of Information and Communications Engineering, Aalto University, Espoo 02150, Finland}

\author{Simone Felicetti\,\orcidlink{0000-0002-1098-9006}}
\email[]{simone.felicetti@cnr.it}
\affiliation{Institute for Complex Systems, National Research Council (ISC-CNR), Via dei Taurini 19, 00185 Rome, Italy}
\affiliation{Physics Department, Sapienza University, P.le A. Moro 2, 00185 Rome, Italy}

\author{Karol Gietka\,\orcidlink{0000-0001-7700-3208}}
\email[]{karol.gietka@uibk.ac.at}
\affiliation{Institut f\"ur Theoretische Physik, Universit\"at Innsbruck, Technikerstra{\ss}e\,21a, A-6020 Innsbruck, Austria} 

\title{Critical Quantum Sensing: a tutorial on parameter estimation near quantum phase transitions}


\begin{abstract}
Quantum phenomena offer the possibility of measuring physical quantities with precision beyond classical limits. However, current progress is constrained by scalability, environmental noise, and challenges in practical integration. This highlights the necessity for novel approaches. An emerging paradigm in this direction is \emph{critical quantum metrology}---which harnesses the enhanced susceptibility and nonclassical correlations naturally occurring near quantum phase transitions as resources for quantum-enhanced precision. This tutorial provides a pedagogical introduction to key concepts and a detailed overview of prominent quantum sensing strategies that exploit critical phenomena in metrology. Through examples of increasing complexity, the reader is guided through various critical quantum sensing protocols applied to different critical systems. Special emphasis is placed on the optimal scaling of estimation precision with respect to fundamental resources. Finally, we discuss how critical quantum metrology extends from idealized models to realistic open-system, dissipative regimes, and strongly correlated fermionic systems, outlining both the challenges and opportunities for future quantum technologies.
\end{abstract}
\date{\today}
\maketitle

\tableofcontents

\graphicspath{{./introductin/}}
\section{Introduction}
Quantum metrology~\cite{Maccone2006quantummetrology, maccone2011advancesQM, toth2014quantum, TAYLOR20161QMinBIOLOGY, Degen2017Quantumsensing, Pirandola2018photonicQS, rmpTreutlein2018QMwithatoms, sciarrino2020photonicQM, 2022BarberPRX_OQMtut} is an exciting and rapidly evolving field that leverages the unique properties of quantum mechanics---such as superposition, coherence, and entanglement---to measure physical quantities with unprecedented precision. By surpassing the classical limits of measurement, quantum metrology has emerged as a cornerstone of modern quantum technologies, offering both deep theoretical insights and transformative practical applications~\cite{quantumtechnologies2003Milburn, QT2015Schmiedmayer}. One of the most striking successes of quantum metrology is its role in the detection of gravitational waves by the LIGO interferometer~\cite{ligo1992,LIGO_Aasi_2015}. In this groundbreaking experiment, squeezed vacuum states are used to reduce quantum noise, enabling measurements beyond the standard quantum limit---the fundamental precision bound for classical systems~\cite{Giovannetti2004SQL}. This achievement not only represents a triumph of quantum metrology but also exemplifies its potential to revolutionize precision measurements in fields ranging from fundamental physics to advanced sensing technologies.

Over the past few decades, researchers have made remarkable progress in pushing the boundaries of precision beyond the standard quantum limit. Platforms such as trapped ions~\cite{ionsreview2019, ionsRMP2021, Rey2024ioncrusytalmetrology}, cold atoms~\cite{rmpTreutlein2018QMwithatoms, PhysRevResearch.4.043074, dai2021cold, MitchisonPRL, brattegard2025correlateddecoherencethermometrymobile, glatthard2022optimal,agarwal2025quantumsensingultracoldsimulators,PhysRevLett.132.240803}, and photonic systems~\cite{DemkowiczDobrzaski2015limitsint, sciarrino2020photonicQM, 2022BarberPRX_OQMtut} have demonstrated quantum-enhanced measurement techniques, showcasing how quantum mechanics can improve precision. However, these advancements are often restricted to small-scale or mesoscopic quantum systems. The leap to macroscopic systems---where quantum effects could be harnessed on a much larger scale---remains one of the most ambitious challenges in the field. Macroscopic systems face significant hurdles, such as decoherence, technical noise, and the difficulty of generating and maintaining large-scale quantum correlations. Even state-of-the-art technologies like atomic clocks \cite{dai2021cold}, which achieve extraordinary precision, primarily rely on unentangled ensembles of atoms and classical statistical averaging to improve accuracy. Overcoming these barriers requires innovative approaches with superior scalability and greater robustness to noise.

One promising strategy is \emph{critical metrology}~\cite{zanardi2007criticalscaling, zanardi2008criticalityresource,
zanardi2008informationgeometry}, an approach that exploits the extreme sensitivity of quantum systems near critical points of phase transitions. At these critical points, the system statical and dynamical properties becomes highly responsive to external perturbations, providing a natural mechanism for amplifying small changes in parameters of interest. The divergent susceptibility of critical systems is already exploited in scientific and technological sensing applications, such as in bubble chambers~\cite{Perkins2000} and transition edge photodetectors~\cite{IrwinHilton2005}. In particular, \emph{critical quantum sensing} is an approach that exploits the nonclassical properties developed in proximity of critical points, in order to achieve a metrological quantum advantage~\cite{sun2010fisher,mihailescu2023multiparameter,mihailescu2025metrologicalsymmetriessingularquantum,tsang2013quantumTED,lorenzo2017quantum,zakrz2018criticalb,bin2019mass,garbe2020criticalmetrology,mihailescu2024quantumsensingnanoelectronicsfisher,2020saulo,witko2021CQMferro,abol2021criticalityglobalsensing,liu2021experimental,e23101353,Gietka2022understanding,Pavlov_2023,plenio2022PRX,Garbe2022critical,garbe2022exponential,ding2022enhanced,gietka2023overcoming,salvia2023CQMrealtimefeedback,ilias2023criticalityenhanced,Mihailescu_Thermometry,paris2023jointloos,paraoanu2023parametriccritical,stefan,salvia2023CQMrealtimefeedback,Gietka2024combining,gietka2024tempcqm,tang2023enhancement,li2024quantummetrologicalcapabilityprobe,alushi2024optimality,relation2024zhang,zicari2024criticalityamplifiedquantumprobingspontaneous,alushi2024collectivequantumenhancementcritical,paris2024spinchain,salvia2023CQMrealtimefeedback,niu2025rolelongrangeinteractioncritical,beaulieu2025Criticality,2020saulo,MONTENEGRO20251,sahoo2023localization,PhysRevApplied.23.014019,Sarkar_2024,cavazzoni2025frequencyestimationfrequencyjumps,alderete2025nonlinearquantumsensingfrustrated,free2022sarkar,he22023stark,Manshouri2025quantumenhanced, parlato2025quantum}.

\begin{table*}[htb!]
 \caption{A readers guide on how to best use this tutorial.}
  \label{tab:image-table}
  \centering
  \includegraphics[width=\linewidth]{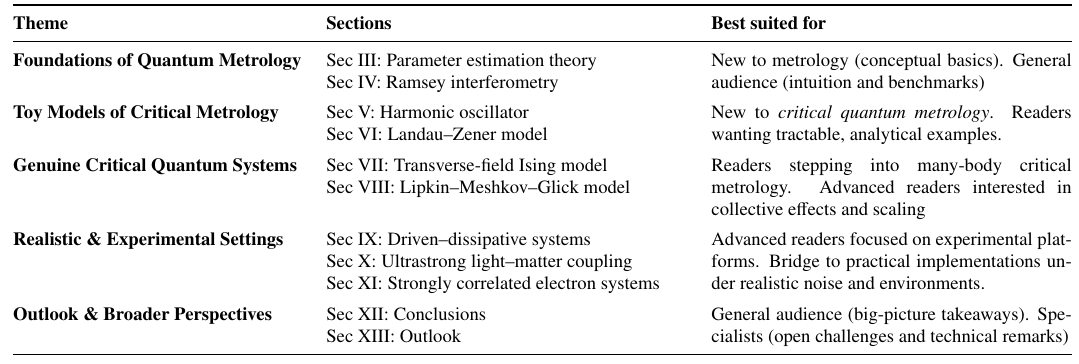}
\end{table*}

Critical quantum metrology occupies a unique position at the intersection of quantum metrology, quantum phase transitions, and many-body physics. By leveraging the universal properties of systems near criticality, it holds the potential to unlock scalable precision measurement technologies. Unlike traditional quantum metrology, which often relies on creating and preserving complex entangled states, critical metrology takes advantage of collective phenomena that emerge naturally near critical points. These effects---which may also involve many-body entanglement---can amplify signals in ways that might be more robust and accessible for macroscopic systems. At the same time, operating close to criticality with engineered critical states presents challenges: critical slowing down limits dynamical response, and the inherent difficulty of controlling a system near a phase transition can complicate practical implementations~\cite{PhysRevLett.29.1080,PhysRevB.32.516}.

This tutorial aims to provide a clear and comprehensive introduction to critical metrology, including its foundational principles, practical challenges, illustrative examples and future opportunities. We will explore how quantum Fisher information connects metrology to critical phenomena, the trade-offs involved in implementing critical metrology and strategies for addressing key challenges like optimal measurements and preparation times. Furthermore, we will discuss how critical metrology extends the reach of quantum-enhanced precision to larger systems, bridging the gap between fundamental physics and practical technologies.  Whether you are new to quantum metrology or an experienced researcher, this tutorial aims to inspire fresh ideas and collaborations. By combining the power of quantum criticality with the demands of precision measurements, critical metrology promises to redefine the frontiers of what can be achieved in quantum technologies.

\section{How to use this document}
Before delving into the details, let us briefly outline the structure of this tutorial and how it may best be read. The material is organized to gradually build up from the fundamentals of parameter estimation theory to the forefront of experimental realizations of critical metrology, see Tab.~\ref{tab:image-table}. Readers primarily interested in the core ideas may choose to focus on the sections introducing critical models, while those seeking a broader perspective will find the introductory and concluding parts equally valuable. While every effort has been made to minimize redundancy, some degree of overlap between sections is unavoidable. This is deliberate, as the tutorial is structured to accommodate selective reading: certain concepts are reintroduced in different contexts to ensure coherence and accessibility, regardless of the reader’s chosen path through the material.

Section~\ref{sec:parameter_estimation_theory} offers a concise introduction to parameter estimation theory, covering key notions such as the signal-to-noise ratio, Fisher information, quantum Fisher information, and their multiparameter extensions. In Section~\ref{sec:ramsey_example}, we explore these concepts in depth using the paradigmatic case of Ramsey interferometry. This example illustrates the standard quantum limit, Heisenberg scaling, the Heisenberg limit, and the influence of decoherence and open-system effects. Readers already familiar with parameter estimation may skip these sections, though we suggest consulting Section~\ref{sec:ramsey_example}, as it offers a valuable contrast for the discussion of critical metrology that follows.

In Section~\ref{sec:QPT_HO}, we introduce the minimal critical model—a tunable quantum harmonic oscillator—as a pedagogical platform to illustrate how criticality can enhance metrological performance, while emphasizing the role of preparation times and finite-size effects. Section~\ref{sec:QPT} extends the discussion to the Landau–Zener model, bridging single-mode physics and genuine quantum phase transitions, while also enabling exploration of geometric interpretations, thermal effects and multiparameter scenarios. Section~\ref{sec:Ising_Model} focuses on the transverse-field Ising model as a paradigmatic many-body system, incorporating both local and collective measurement strategies. Finally, Section~\ref{sec:LMG_Model} examines the all-to-all interacting Lipkin–Meshkov–Glick model, highlighting its connection to harmonic oscillator physics and exploring dynamical protocols as well as excited-state phase transitions. Together, these four sections form the core of the tutorial: they introduce critical metrology through analytically or semi-analytically tractable, well-established toy models that nevertheless can be implemented in a variety of experimental platforms.

Section~\ref{sec:HO} explores driven–dissipative systems, examining both their steady-state and dynamical regimes, and discussing how environmental coupling shapes the prospects for critical metrology under realistic conditions. Section~\ref{sec:ultrastrong} focuses on ultrastrongly coupled light–matter systems, particularly the quantum Rabi model, as a cutting-edge experimental platform where critical metrological ideas can be directly tested. Section~\ref{sec_mrlm} introduces strongly correlated fermionic systems as a contrasting platform for critical quantum sensing, focusing on an exactly solvable point of the 2 channel Kondo model. Together, these sections collocate critical metrology within state-of-the-art experimental realizations, highlighting approaches that capture the essence of critical phenomena while accounting for the inevitable influence of the environment—providing the most realistic pathways toward experimental critical metrology. 

Finally, Section~\ref{sec:conc} summarizes the main insights of the tutorial, and Section~\ref{sec:outlook} outlines remaining open challenges, and points to promising directions for future research in critical metrology. 

\graphicspath{{./Parameter_Estimation_Figs/}}

\section{A brief primer on parameter estimation theory}\label{sec:parameter_estimation_theory}
Many quantities of interest are not directly accessible, either in principle, or due to experimental constraints. This is particularly true for Hamiltonian parameters or quantities such as temperature which have no associated observable, and therefore cannot, even in principle, be directly measured~\cite{paris2009review,maccone2011advancesQM,toth2014quantum,Degen2017Quantumsensing,Maccone2006quantummetrology}. Instead, we must rely on some indirect measurement strategy to infer their values. In practice, this involves constructing a dataset from repeated measurements on identically prepared systems, followed by statistical analysis to extract the parameter of interest. This procedure naturally falls within the framework of quantum parameter estimation theory, which provides a rigorous approach to quantify precision limits and optimal strategies for estimating quantities encoded in quantum states~\cite{helstrom1969quantum,Holevo,caves1994statisticaldistance,paris2009review,Maccone2006quantummetrology}.

There are two main paradigms for parameter estimation: local and global estimation theory~\cite{caves1994statisticaldistance,HELSTROM1967101,CRB_Book,vanTrees2013,mukhopadhyay2024saturableglobalquantumsensing,DemkowiczDobrzaski2015limitsint,Mukhopadhyay_2025}. In local estimation theory, it is assumed that some prior procedure has narrowed the possible values of the parameter to a small interval. The goal is then to determine the parameter value within this local neighbourhood with maximum precision. A widely used tool in this context is the (quantum) Fisher information, which quantifies parameter sensitivity—that is, how well the parameter can be estimated—and the (quantum) Cramér-Rao bound, which sets the fundamental precision limit for any given estimation strategy. In contrast, global estimation theory makes minimal assumptions about prior information, aiming instead to identify or localize the parameter across the entire domain of known possibilities. In this tutorial, our primary focus will be on local estimation theory, and we will use the quantum Fisher information as the main benchmark for precision. For a more comprehensive review of local and global estimation theory, we refer the reader to the seminal works~\cite{Holevo,paris2009review} and~\cite{Holevo,vt,rubio2020bayesian,PhysRevA.110.L030401,Rubio_2019,10.1088/2058-9565/ae08e1,qbn1-p6bq,zhang2025robustefficientestimationglobal,FAU2021107917,PhysRevA.95.012136,Mukhopadhyay_2025}, respectively. In the following section, we provide a brief overview of the key concepts and tools of quantum parameter estimation theory.

\begin{figure*}[t!]
	\centering
	\includegraphics[width=1\linewidth]{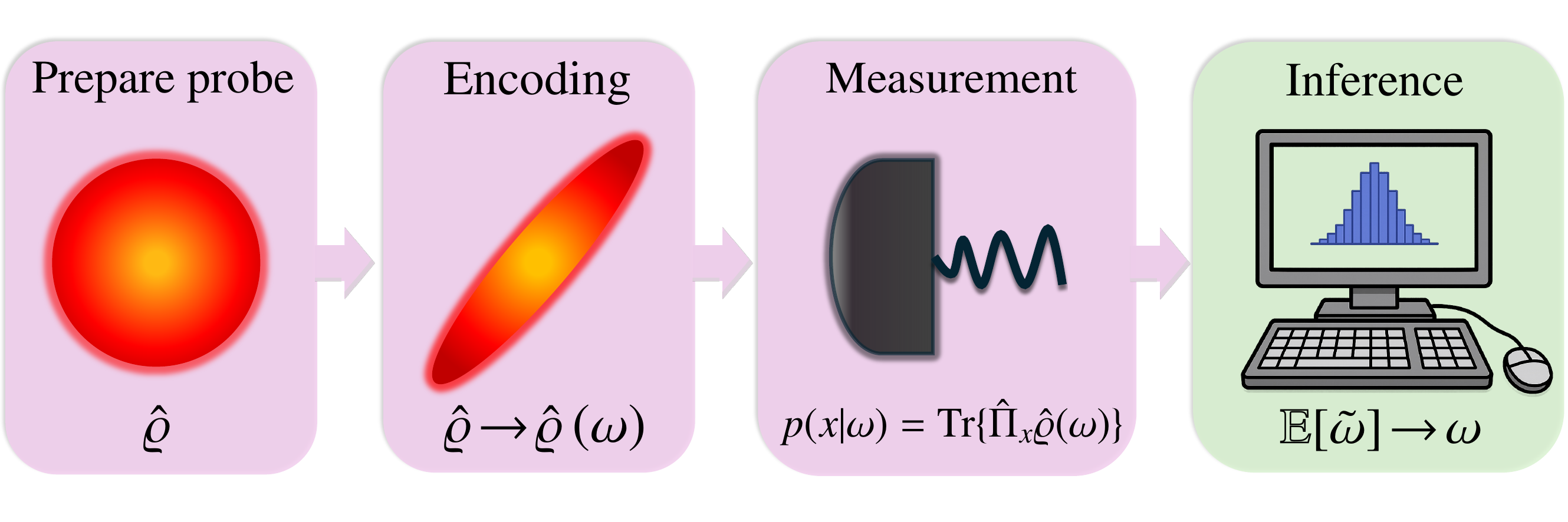}
	\caption[Illustration of a general sensing protocol.]{\textbf{Schematic illustration of a general parameter estimation protocol.} A probe state $\hat{\varrho}$ is initially prepared. The state is then allowed to evolve in time, and the information regarding the target parameter is encoded on the quantum state, $\hat{\varrho} \!\to\! \hat{\varrho}(\omega)$. After the system evolves in time, a measurement—whose outcome probabilities depend on the unknown parameter—is performed. By repeating this process $m$ times, the resulting outcomes can be used to construct a statistical model, from which an estimator $\tilde{\omega}$ can be derived to infer the parameter of interest.}
	\label{fig:param_esti_general_sensing_depict}
\end{figure*}

\subsection{General sensing protocol}\label{sub_sec:general_sensing_protocol}
Sensors are deeply embedded in our everyday lives: from navigational tools and environmental monitoring to medical diagnostics and astronomical imaging. To build intuition, let us begin with perhaps the most familiar example of a sensor: the thermometer. A traditional thermometer consists of a liquid, such as mercury, enclosed in a glass tube. The liquid’s response—expanding or contracting with changes in temperature—allows us to infer the surrounding temperature simply by observation. In this example, the liquid serves as a \emph{probe}: its physical response to an external perturbation encodes information about an unknown parameter (temperature). Ideally, the probe operates noninvasively, faithfully reflecting the properties of the system without disturbing them.

Quantum metrology generalizes this same idea to the quantum domain. Instead of a macroscopic liquid, we use well-controlled quantum systems whose states are sensitive to the parameter of interest. In this tutorial, such a \emph{probe state} is described by a density matrix $\hat{\varrho}$, which may be pure or mixed~\footnote{Throughout this tutorial, both operators and matrices are denoted with a hat; the meaning should be clear from the context.} . The parameters to be estimated are denoted collectively by $\omega$. With these definitions in place, we can now outline a general quantum sensing protocol. Based on these basic principles, a general sensing protocol can then be defined as follows (see Fig.~\ref{fig:param_esti_general_sensing_depict} for a schematic representation): 
\begin{itemize}
    \item Preparation of the probe state \(\hat\varrho\), independent of the parameters of interest.
    \item Evolution, during which the probe interacts with the system, and the parameters of interest are dynamically encoded,  \(\hat{\varrho}\to\hat{\varrho}(\omega)\).
    \item Measurement of the parameter-dependent probe state \(\hat{\varrho}(\omega)\).
  \item Repetition of the previous steps a large number of times $\mathcal{M}$, to construct a statistical dataset. From this dataset, the target parameter $\omega$ can be inferred using a suitable estimator, denoted by $\tilde{\omega}$ throughout this manuscript.

\end{itemize}
To achieve increasingly precise estimates of the target quantities, each step of the sensing protocol can, in principle, be optimized. This tutorial focuses on the attainability of such optimal strategies under practical resource constraints—such as limited particle number, finite interaction time, or restricted access to non-classical states—and on how closely they can approach the fundamental bounds. Equally important, we examine their robustness against realistic sources of noise, including dephasing, particle loss, and fluctuations in control fields.

\subsection{Signal-to-noise ratio}\label{subsec_error_prop_formula}
A central goal of quantum parameter estimation theory is to devise measurement strategies that extract the maximum possible information from a given quantum probe. This usually means enhancing the measurement signal, reducing the noise, or optimally balancing both. In many cases, the resulting theoretical optimal strategies require the implementation of measurements that are not practically accessible. A more fundamental task in quantum metrology, however, is to quantify the uncertainty in estimating an unknown parameter $\omega$ utilizing a general observable $\hat{\mathcal{O}}$ that is readily available and straightforward to implement. This can be captured by the signal-to-noise ratio (SNR)
\begin{equation}
    \label{eq:error_prop_formula_def}
    S_\omega \equiv \frac{\mathcal{M}\,|\partial_\omega \langle \hat{\mathcal{O}} \rangle|^2}{\text{Var}(\hat{\mathcal{O}})} = \frac{1}{\text{Var}(\tilde{\omega})}\;,
\end{equation}
which follows from the \emph{error propagation formula} and where $\langle \hat{\mathcal{O}} \rangle = \text{Tr}\{ \hat{\mathcal{O}} \hat{\varrho}(\omega) \}$ is the expectation value of the observable $\hat{\mathcal{O}}$, $\text{Var}(\hat{\mathcal{O}}) = \langle \hat{\mathcal{O}}^2 \rangle - \langle \hat{\mathcal{O}} \rangle^2$ is the measurement uncertainty~\footnote{Throughout this manuscript, the notations $\langle \hat{\mathcal{O}} \rangle$ and $\mathrm{Tr}\{\hat{\mathcal{O}} \hat{\varrho}(\omega)\}$ will be used interchangeably.}, and $\mathcal{M}$ is the number of measurement taken to construct the observable $\langle \hat{\mathcal{O}} \rangle$. Here, precision is quantified by the sensitivity of the observable $\hat{\mathcal{O}}$ to small variations in the target parameter $\omega$, together with the magnitude of the variance $\hat{\mathcal{O}}$. The larger the value of \(S_\omega\), the smaller the variance $\text{Var}(\tilde{\omega})$, and consequently the higher the measurement precision. 

\subsection{Fisher information}\label{sub_sec:Fisher_Info}
Parameter estimation theory provides a rigorous framework for quantifying the performance of estimating an unknown parameter~\cite{paris2009review,caves1994statisticaldistance,Holevo,HELSTROM1967101,helstrom1969quantum,1055103,Liu_2020_QFIM_MPE_Review,PhysRevLett.108.233602,Giovannetti2004SQL,Maccone2006quantummetrology,maccone2011advancesQM}. In this framework, information about the target parameter $\omega$ is extracted by performing measurements on the parameter-dependent probe state $\hat{\varrho}(\omega)$. We consider a set of measurements $\{ \hat{\Pi}_x \}$, where each $\hat{\Pi}_x$ is a positive operator-valued measure (POVM) element associated with outcome $x$. These elements satisfy the completeness relation $\int_x \hat{\Pi}_x dx = \mathbb{I}$ for continuous outcomes, and $\sum_x \hat{\Pi}_x = \mathbb{I}$ for discrete outcomes, where $\mathbb{I}$ denotes the identity operator. The conditional probability of obtaining outcome $x$, given that the parameter takes the value $\omega$, is then determined by the Born rule: 
\begin{equation}
    p(x | \omega ) = \text{Tr}[ \hat{\Pi}_x \hat{\varrho}(\omega) ]\;,
\end{equation}
whereby repeating the measurement procedure \(\mathcal{M}\) times yields a dataset $\left\{\vec{x}\,\right\} = (x_1, x_2 ,\dots,x_\mathcal{M})$.
This dataset can then be used to construct a frequency distribution of outcomes which approximates the true probability distribution when $\mathcal{M}\gg1$, ${p}(\{ \vec{x}\, \} | \omega) \approx p(x | \omega)$, and from which an estimator $\tilde{\omega}$ can be defined. An estimator is a mapping $\tilde{\omega} = \tilde{\omega}(x_1,x_2,\dots,x_\mathcal{M})$ from the set $\{\vec{x}\,\}$ of measurements outcomes into the space of parameters. In the limit $\mathcal{M}\gg1$, the sensitivity of this distribution to infinitesimal variations of $\omega$ is quantified by the Fisher information~\cite{Holevo,paris2009review}, which, for a continuous set of outcomes, is
\begin{align}
     \mathcal{F}_\omega &= \int  \frac{[\partial_\omega p(x|\omega)]^2}{p(x|\omega)} dx\;,
\end{align}
while for a discrete set becomes
\begin{align}
    \label{eq:def_fi_intro}
    \mathcal{F}_\omega &= \sum_i  \frac{[\partial_\omega p(x_i|\omega)]^2}{p(x_i|\omega)}\;,
\end{align}
where $p(x_i|\omega)$ is the probability of obtaining a discrete outcome labelled $x_i$ given $\omega$. The Fisher information quantifies the sensitivity of the conditional probabilities with respect to the parameter to be estimated. A large value of the Fisher information indicates that the measurement procedure is highly sensitive in the proximity of $\omega$. It is important to note that the Fisher information depends on the \emph{particular} choice of the measurement being employed. Statistical errors arising from the uncertainty of the measurement outcomes are unavoidable, and the precision with which the parameter of interest can be estimated is bounded by the Cram\'{e}r-Rao bound (CRB)~\cite{Holevo,paris2009review}:
\begin{equation}
    \label{eq:crb_normal_def}
    \text{Var}(\tilde{\omega}) \geq \frac{1}{\mathcal{M}\mathcal{F}_\omega}\;.
\end{equation}
For simplicity, we set $\mathcal{M}=1$ in all the following expressions, unless explicitly noted otherwise. The CRB provides an asymptotically valid lower bound on the statistical error expressed through the variance of the parameter, and holds for unbiased estimators. 


\subsection{Quantum Fisher information}\label{sub_sec:QFI}
Naturally, one may ask: given a probe $\hat{\varrho}$, what is the best possible precision with which the target parameter $\omega$ can be estimated? To address this, the starting point is the classical precision bound. For a given measurement basis \(\{ \hat{\Pi}_x \}\), the measurement outcomes define a probability distribution $p(x|\omega)$. The associated Fisher information quantifies the sensitivity of these outcomes to changes in \(\omega\), and, through the CRB, sets a lower limit on the variance of any unbiased estimator. Since the Fisher information depends explicitly on the particular measurement choice, we can better formulate the originally posed question: Given a probe \(\hat{\varrho}\), which measurement basis produces the probability distribution that is most sensitive to changes in \(\omega\)? 

Addressing this question requires the optimization of the Fisher information over all possible measurement strategies, i.e., over the full set of POVMs. Under regularity conditions (allowing derivatives and traces to be interchanged), this can be easily accomplished~\cite{paris2009review}. The result of this optimization is the quantum Fisher information (QFI), which represents the ultimate sensitivity achievable by the quantum state itself, independent of the choice of measurement. The QFI is thus defined as~\cite{paris2009review,helstrom1969quantum,caves1994statisticaldistance,Holevo}
\begin{equation}
    \mathcal{I}_\omega = \text{Tr}\left[\hat{\varrho}(\omega) \hat{L}_\omega^2 \right]\;,
\end{equation}
where the operator \(\hat{L}_\omega\) is the \emph{symmetric logarithmic derivative} (SLD), defined implicitly via the derivative of the probe state:
\begin{equation}
    \partial_\omega\hat{\varrho}(\omega)=\frac{1}{2} \left( {\hat{L}_\omega\hat{\varrho}(\omega)+\hat{\varrho}(\omega)\hat{L}_\omega}\right)\;.
\end{equation}
From the QFI one obtains the \emph{quantum} Cram\'{e}r-Rao bound (QCRB)~\cite{Holevo,paris2009review}
\begin{equation}
    \label{eq:ch_intro_CRB}
    \text{Var}(\tilde{\omega}) \geq \frac{1}{\mathcal{I}_\omega}\;,
\end{equation}
which represents the ultimate precision limit for estimating $\omega$ using quantum probes. The QCRB is saturated in the asymptotic dataset limit for procedures that use both optimal measurements and optimal estimator post processing~\cite{Holevo,paris2009review,helstrom1969quantum,HELSTROM1967101}.

The SLD operator $\hat{L}_\omega$ plays a crucial role in the parameter estimation task. In particular, it defines the optimal measurement basis for estimating the parameter $\omega$; that is, any operator $\hat{\mathcal{O}}$ satisfying $[\hat{L}_\omega,\hat{\mathcal{O}} ] = 0$ corresponds to an optimal measurement. The SLD itself is a complicated object, and may not always be practical from an experimental perspective. It can be obtained as the solution to the Lyapunov equation
\begin{equation}
    \label{eq:sld_lyap}
    \hat{L}_\omega = 2 \int_0^{\infty} e^{-t \hat{\varrho}(\omega)} \partial_\omega \hat{\varrho}(\omega)e^{-t \hat{\varrho}(\omega)}dt\;,
\end{equation}
which, in terms of the spectral decomposition of the probe \(\hat{\varrho}(\omega)=\sum_ip_i|\psi_i\rangle\langle\psi_i|\), reads explicitly~\cite{Holevo,paris2009review,helstrom1969quantum,HELSTROM1967101}
\begin{equation}
    \hat{L}_\omega = 2 \sum_{i,j} \frac{\langle \psi_j|\partial_\omega \hat{\varrho}(\omega) | \psi_i\rangle}{p_i+p_j}|\psi_j\rangle\langle\psi_i|\;.
\end{equation}
Notice that this expression is valid only for $p_i + p_j \neq 0$, and that, in principle, both the eigenvalues and the eigenvectors of the probe state depend on the target parameter. Using this decomposition, the QFI can then be expressed directly in terms of the eigenvalues and eigenvectors of the probe state as~\cite{paris2014cqmLMG}
\begin{equation}
    \label{eq:def_qfi}
    \mathcal{I}_\omega = \underbrace{\sum_i \frac{\left( \partial_\omega p_i \right)^2}{p_i}}_{\text{Population-dependent}} + \underbrace{2\sum_{m \neq n} \sigma_{n,m} | \langle \psi_m| \partial_{\omega} \psi_n \rangle |^2}_{\text{Basis-dependent}}\;,
\end{equation}
with $\sigma_{nm} = (p_n - p_m)^2/(p_n+p_m)$. In Eq.~\eqref{eq:def_qfi} we have separated the QFI into population-dependent and basis-dependent contributions. The first term depends on how the eigenvalues $p_i$ of the state $\hat{\varrho}(\omega)$ vary with $\omega$ and quantifies the information carried by population shifts, independent of coherence. The second term captures coherent, basis-dependent contribution when the eigenvectors $|\psi_i\rangle$ change with small variations in $\omega$, and is deeply tied to the geometry of quantum states. We note here that the first term in Eq.~\eqref{eq:def_qfi} bears a similarity to the Fisher information defined in Eq.~\eqref{eq:def_fi_intro} and, for this reason, is often referred to as the ``classical'' contribution to the QFI. Indeed, these two quantities are structurally identical if one performs a fixed projective measurement in the $\omega$-dependent eigenbasis of $\hat{\varrho}(\omega)$ treating the populations $p_i$ as classical probabilities. However, we remark that the term ``classical'' contribution can be somewhat misleading, as the measurement required to saturate the bound may itself be quantum and parameter-dependent. In the case of pure states $\hat{\varrho}(\omega) = |\psi\rangle\langle\psi|$, for which $\text{Tr} \{ \hat{\varrho}(\omega) \} = 1$, the SLD becomes
\begin{equation}
    \hat{L}_\omega = 2 \left(|\psi\rangle \langle\partial_\omega \psi | + | \partial_\omega \psi \rangle \langle \psi| \right)\;,
\end{equation}
and, consequently, the QFI simplifies to~\cite{Holevo,paris2009review,helstrom1969quantum,HELSTROM1967101}
\begin{equation}
    \label{eq:intro_pure_state_qfi_def}
    \mathcal{I}_\omega = 4 \left[ \langle \partial_\omega \psi | \partial_\omega \psi\rangle - \langle \partial_\omega \psi | \psi \rangle \langle \psi | \partial_\omega \psi\rangle\right]\;.
\end{equation}

Finally, it is important to emphasize the connection between the QFI and the SNR. A particularly elegant and physically intuitive interpretation of the QFI is that it corresponds to the \textit{maximal SNR} achievable when estimating the parameter~$\omega$, optimized over all possible observables:
\begin{equation}
    \mathcal{I}_\omega = \max_{\hat{\mathcal{O}}} S_\omega, \quad \text{where} \quad S_\omega = \frac{|\partial_\omega \langle \hat{\mathcal{O}} \rangle|^2}{\text{Var}(\hat{\mathcal{O}})}\;.
\end{equation} 
This provides a clear operational interpretation: The optimal measurement is one whose expectation value is highly sensitive to small changes in~$\omega$, while simultaneously minimizing statistical fluctuations. In other words, the optimal measurement basis maximizes the SNR. Rigorous proofs for this intuitive identity can be found in~\cite{escher2012SNR,Gorecki2025}.

\subsection{Multiparameter estimation}\label{sub_sec:MP_Est}
The most general quantum sensing scenario arises when one seeks to estimate a set of $d$ unknown parameters, $\vec{\omega} =(
    \omega_1,\omega_2,\dots, \omega_d)$, through suitable measurements of a quantum probe~\cite{paris2009review,Liu_2020_QFIM_MPE_Review,albarelli2020perspective}. As in the single-parameter case, information about these parameters is imprinted on the probe state $\hat{\varrho}(\vec{\omega})$. For a given set of POVMs $\{ \hat{\Pi}_x  \}$, the uncertainty in estimating \(\vec \omega\) from $\mathcal{M}$ independent measurements is bounded by the multiparameter Cram\'{e}r-Rao inequality~\cite{Holevo,paris2009review,helstrom1969quantum,HELSTROM1967101}:
\begin{equation}
    \label{eq:def_mp_crb}
    \text{Cov}[\tilde{\vec{\omega }}] \geq  \hat{\mathcal{F}}^{-1}\;,
\end{equation}
which is now promoted to a matrix inequality. Both quantities in the above matrix inequality are positive semidefinite matrices. In Eq.~\eqref{eq:def_mp_crb}, the covariance matrix has entries $\text{Cov}(\omega_i,\omega_j) = \langle(\omega_i - \langle \omega_i \rangle )(\omega_j - \langle \omega_j \rangle) \rangle$, with diagonal elements corresponding to the variances $\text{Var}(\omega_i) \equiv \text{Cov}(\omega_i,\omega_i)$. The  Fisher information matrix (CFIM), associated with a specific choice of measurement basis, has entries given by
\begin{equation}
    \mathcal{F}_{ij} = \sum_k\frac{\partial_{\omega_i}p(x_k|\vec{\omega}) \partial_{\omega_j}p(x_k|\vec{\omega})}{p(x_k|\vec{\omega})}\;.
\end{equation}

As in the single parameter estimation case, the QFI matrix (QFIM) is obtained by maximizing the Fisher information over all possible POVMs. Its elements can be expressed in terms of the SLD operators as
\begin{equation}
    \label{eq:intro_SLD_all_defs}
    \mathcal{I}_{ij} = \frac{1}{2} \text{Tr} \left[ \hat{\varrho}(\vec{\omega}) \{ \hat{L}_i,\hat{L}_j \}  \right] \equiv \text{Tr} \left[ \hat{L}_j \partial_{\omega_i} \hat{\varrho}(\vec{\omega})  \right]\;,
\end{equation}
where $\hat{L}_i$ is the SLD operator associated with parameter $\omega_i$, defined implicitly by the self-adjoint Lyapunov equation Eq.~\eqref{eq:sld_lyap}. By construction, the QFIM is a real, symmetric, and positive semidefinite matrix. Notice that the diagonal elements of the QFIM reduce to the single-parameter QFIs, $\mathcal{I}_{ii} \equiv \mathcal{I}_{\omega_i}$,  while the off-diagonal elements capture correlations between parameters. Using the spectral decomposition $\hat{\varrho}(\vec{\omega}) = \sum_i p_i |\psi_i\rangle\langle\psi_i|$, the elements of the QFIM can be written in the convenient form~\cite{Liu_2020_QFIM_MPE_Review},
\begin{equation}
    \label{eq:def_qfim}
    \mathcal{I}_{ij} = \sum_{k,l} \frac{2~\text{Re}\left[ \langle \psi_k| \partial_{\omega_i} \hat{\varrho} | \psi_l\rangle \langle \psi_l | \partial_{\omega_j} \hat{\varrho} |\psi_k\rangle \right]}{p_k + p_l}\;,
\end{equation}
valid whenever $p_k + p_l \neq 0$. For pure states $\hat{\varrho}(\vec{\omega}) = |\psi\rangle\langle\psi|$, this simplifies to
\begin{equation}
    \mathcal{I}_{ij} = 4~\text{Re}\left[ \langle \partial_{\omega_i} \psi| \partial_{\omega_j} \psi \rangle - \langle \partial_{\omega_i} \psi | \psi \rangle \langle \psi|\partial_{\omega_j} \psi \rangle  \right]\;.
\end{equation}
For compatible SLDs (see below), the ultimate precision in multiparameter estimation is determined by the QCRB~\cite{paris2009review,caves1994statisticaldistance,Holevo,HELSTROM1967101,helstrom1969quantum},
\begin{equation}
    \label{eq:def_qcrb_mp}
    \text{Cov}[\tilde{\vec{\omega}}] \geq  \hat{\mathcal{F}}^{-1} \geq  \hat{\mathcal{I}}^{-1}\;,
\end{equation}
again understood as a matrix inequality. Similarly to the single-parameter case, saturation of this bound requires asymptotically large datasets, prior knowledge of the parameter neighbourhood, and the use of optimal unbiased estimators. 

As in single-parameter estimation, the SLD operators $\hat{L}_i$ define the optimal measurements for estimating the individual parameters $\omega_i$. Their matrix elements can be computed as 
\begin{equation}
    \label{eq:intro_sld_elements}
    \langle p_i| \hat{L}_i| p_j \rangle = \delta_{ij} \frac{\partial_{\omega_i} p_i}{p_i} - \frac{2(p_i - p_j)}{p_i + p_j} \langle \psi_i | \partial_{\omega_i} \psi_j\rangle\;,
\end{equation}
well-defined whenever $p_i + p_j \neq 0$. For pure states this reduces to $\hat{L}_i = 2(|\psi\rangle\langle\partial_{\omega_i} \psi| + |\partial_{\omega_i}\psi\rangle\langle\psi|)$. 

We remark that, while multiparameter estimation may appear as a straightforward extension of its single-parameter counterpart, important subtleties arise. For instance, the optimal measurements for different parameters may be incompatible, preventing the simultaneous saturation of precision bounds; there can be intrinsic trade-offs in precision due to statistical correlations between parameters; and, in some cases, the QFIM might be non-invertible. These additional complexities can play a significant and subtle role in the context of critical metrology, and we outline their mathematical underpinning in the following.

\subsection{Incompatibility problem in multiparameter estimation}\label{sub_sec:Inc_MP_Est}

In principle, the CRB is always attainable in both single and multiparameter scenarios. By contrast, while the QCRB is always attainable in the single parameter setting, this condition is not generally guaranteed in the multiparameter case. The reason originates from the possible \emph{incompatibility} of the SLD operators \(\hat{L}_i\) associated with different parameters \(\omega_i\)~\cite{Holevo,helstrom1969quantum,paris2009review,Liu_2020_QFIM_MPE_Review,demkowicz2020multi,1055103,Matsumoto_2002,pezze2017optimal,Albarelli2019Holevo,PhysRevA.100.032104,PhysRevX.11.011028,e22111197,PhysRevLett.126.120503,conlon,Belliardo_2021,carollo2020erratum,tsang2020quantum}. If the SLDs are incompatible, no common measurement basis can simultaneously extract the maximal information about all parameters. This phenomenon, known as measurement incompatibility, sets a fundamental limitation, preventing the saturation of the ultimate precision bound in Eq.~\eqref{eq:def_qcrb_mp}. In such cases, unavoidable trade-offs between the estimation precisions of different parameters emerge. The necessary and sufficient condition for the attainability of the multiparameter QCRB is 
\begin{equation}
    \text{Tr}\left( \hat{\varrho}(\vec{\omega}) \left[\hat{L}_i, \hat{L}_j \right] \right) = 0\;.
\end{equation}
When this condition is not satisfied, the problem of measurement incompatibility arises, and identifying optimal estimation strategies becomes, in general, difficult~\cite{ragy2016compatibility,Belliardo_2021,Carollo_2019}. Notably, the degree of measurement incompatibility in multiparameter estimation can be quantified through the quantumness parameter~\cite{Carollo_2019,carollo2020erratum}
\begin{equation}
R=\left|\left|\frac{1}{2}\text{Tr}\left(\hat{\varrho}(\omega)[\hat{L}_i,\hat{L}_j]\right) \hat{\mathcal{I}}^{-1}\right|\right|_\infty\;,
\end{equation}
where \(||\cdot||_\infty\) denotes the spectral norm, i.e. the largest eigenvalue in magnitude. By construction, \(0\leq R\leq 1\). The case \(R=0\) corresponds to fully compatible SLD operators, meaning that a common measurement basis exists and the QCRB is saturable. Conversely, \(R=1\) signals maximal incompatibility, where no single measurement setup can jointly saturate the QCRB for all parameters. In this sense, \(R\) acts as a figure of merit for the severity of incompatibility in the estimation problem. More broadly, measurement incompatibility reflects the intrinsically quantum nature of the estimation problem: it is a direct manifestation of quantum indeterminacy, arising from the non-commutativity of the optimal observables associated with different parameters.

For practical purposes, the problem of measurement incompatibility in the multiparameter setting means that saturation of the QCRB is in general not guaranteed. This has important implications, as the QCRB may then underestimate the true estimation error. To properly account for such incompatibilities and define more faithful precision bounds, it is often more appropriate to use the Holevo–Cramér–Rao bound (HCRB) as a benchmark for the ultimate precision limit in multiparameter estimation. The HCRB is defined as \cite{Holevo,Holevo1976}
\begin{equation}
\text{Tr}\left(W\text{Cov}[\tilde{\vec \omega}]\right)\geq C_H\left(W\right)\;,
\end{equation}
where \(W\) is a positive weight matrix---specifying the cost assigned to estimation errors of different parameters. The Holevo bound, \(C_H\left(W\right)\), is obtained from an optimization problem and is explicitly constructed to incorporate the effects of measurement incompatibility. Importantly, the HCRB is always guaranteed to be asymptotically attainable, even in situations where the QCRB is not. Naturally, when the SLDs are compatible, the issue of measurement incompatibility disappears, and the HCRB reduces to the QCRB. Lastly, it is important to emphasize that the evaluation of the HCRB is in general quite challenging. While analytical results are possible for the simplest of models, in general one typically relies on numerical approaches, such as semidefinite programming~\cite{albarelli2020perspective,kurdzialek2022using,PhysRevX.12.011039,Albarelli2019Holevo,gorecki2020optimal,MullerRigat2023certifyingquantum}. Only recently an efficient numerical method, valid for finite dimensional systems, has been proposed~\cite{Albarelli2019Holevo}. 

A further subtlety concerns the different assumptions underlying single versus multiparameter estimation. In the single-parameter setting, one seeks to infer a single target parameter \(\omega_i\), with the implicit assumption that all other relevant parameters are known with absolute certainty. By contrast, the multiparameter paradigm relaxes this assumption: several parameters $\vec{\omega}$ are to be estimated simultaneously (locally). Consequently, even when a common optimal measurement basis exists and the multiparameter bound is saturable, the precision for any \emph{individual} parameter is generally \emph{lower} than that in the corresponding single-parameter scenario. This follows from the matrix inversion in Eq.~\eqref{eq:def_mp_crb}. The effect can be seen straightforwardly in the case of two-parameter estimation, where $\vec{\omega} = ( \omega_1, \omega_2)$. From Eq.~\eqref{eq:def_qcrb_mp}, the attainable precision with respect to the parameter $\omega_i$ is given by
\begin{equation}
    \label{eq:qfim_degrade_eg}
     \text{Var}(\tilde{\omega}_i) \geq \frac{\mathcal{I}_{jj}}{\mathcal{I}_{ii}\mathcal{I}_{jj} - \mathcal{I}_{ij}^2}=\frac{\mathcal{I}_{jj}}{\text{det}[\hat{\mathcal{I}}]}\;,
\end{equation}
with \(i,j\in\{1,2\}\). In general, this represents a reduction in sensitivity compared to the precision attainable when all other parameters are assumed known, in which case the bound simplifies to $\text{Var}(\tilde{\omega}_i) \geq 1/\mathcal{I}_{ii}$. The loss of precision is governed by correlations between the multiple unknown parameters, as encoded in the off-diagonal elements of the QFIM. Consequently, strategies designed to optimize the single-parameter QFI are not necessarily optimal in the multiparameter setting.  Instead, effective approaches must explicitly account for these correlations in order to balance precision across all parameters of interest. In the case of critical quantum probes, this effect can become significantly pronounced as the enhanced sensitivity to parameter variations, arising from diverging susceptibilities near criticality, also amplifies the trade-off between parameters~\cite{di2022multiparameter,mihailescu2023multiparameter}. As a result, while criticality can boost the overall metrological potential, it may exacerbate the precision loss for individual parameters.

\subsection{Invertibility of the quantum Fisher information matrix}\label{sub_sec:Inv_MP_Est}
Beyond the issues discussed above, the invertibility of the Fisher information matrices plays a crucial role in multiparameter estimation. In the case of an invertible QFIM but singular (non-invertible) CFIM, the issue is associated to the incompleteness of the chosen measurement setup. Therefore, the solution is to enlarge the measurement space, for instance, by replacing projective measurements with generalized POVMs, or by employing sequential projective measurements~\cite{candeloro2024dimension,yang2024sequential,PhysRevResearch.7.023060}. On the other hand, a singular QFIM is more problematic as it directly implies that the CFIM is necessarily singular for \emph{all} possible choices of the measurement setup. Since the QCRB is only well defined for an invertible QFIM, we strictly require 
\begin{equation}
    \text{det}[\hat{\mathcal{I}}] \neq 0\;.
\end{equation}
When $\text{det}[\hat{\mathcal{I}}] = 0$, the QCRB becomes undefined and nothing meaningful can be said about the achievable precision for the parameter set $\vec{\omega}$~\cite{taming_singular,yang2025overcoming,mihailescu2025metrologicalsymmetriessingularquantum,frigerio2024overcomingsloppinessenhancedmetrology,candeloro2024dimension,mihailescu2023multiparameter,2001parameterestimationsingular}. A singular QFIM implies diverging variances and thus a vanishing SNR for parameter estimation. Such singularities typically arise from underlying dependencies between the original parameter set, and are such that signals corresponding to different parameters are indistinguishable in the measurement data~\cite{mihailescu2025metrologicalsymmetriessingularquantum}. Furthermore, recent works have demonstrated that under appropriate conditions, the CRB in the multiparameter case can be surpassed provided a probe has proximity to a singular point of the QFIM~\cite{mukhopadhyay2025beatingjointquantumestimation}.

\subsection{From abstract to applied}
To provide a practical intuition, in Section \ref{sec:ramsey_example} we revisit some of the core concepts of parameter estimation within the context of a concrete physical example: Ramsey interferometry. This paradigmatic quantum sensing protocol provides a natural setting to illustrate how quantities like the SNR, QFI, and the CRB arise in practice. We further investigate how measurement strategies and quantum resources, including entanglement and squeezing, can be exploited to enhance the sensitivity to the target parameters $\omega$. This allows us to introduce the fundamental precision limits of classical and quantum sensing: the \emph{standard quantum limit}, and the \emph{Heisenberg limit}. By working through this example step by step, we reinforce the abstract ideas introduced here,  and lay the groundwork for more advanced schemes discussed later in this tutorial.

Building on the paradigmatic sensing strategy introduced in Section~\ref{sec:ramsey_example}, we turn to the harmonic oscillator and Landau–Zener models as minimal settings to explore quantum criticality in metrology in Sections~\ref{sec:QPT_HO} and \ref{sec:QPT}, respectively. These models allow us to highlight the features that make critical systems powerful platforms for sensing, such as their geometric structure and diverging susceptibilities. We present a detailed analysis of both single- and multiparameter estimation, focusing on the fundamental precision trade-offs, measurement incompatibility, and the emergence of singularities in the QFIM and how to deal with them. Particular focus is devoted to the role of resources, such as interrogation time, and to the impact of thermal fluctuations on critical quantum sensing.

\graphicspath{{./Ramsey_Inter_Figs/}}

\section{Parameter estimation in quantum metrology on the example of Ramsey interferometry}\label{sec:ramsey_example}

To perform precision measurements, one must first measure a signal that depends on the unknown parameter of interest. Every measurement, however, is inevitably associated with some degree of noise. {This noise acts as a detrimental source of imprecision, and a central goal of metrology is to minimize its impact. Furthermore, certain measurement strategies may be inherently more informative than others, or less susceptible to the degrading effects of noise.} {From a practical perspective, noise can be introduced by various technical and environmental factors}, such as thermal fluctuations, vibrations, and electronic readout noise. At the most fundamental level, quantum fluctuations introduce a noise floor that cannot be avoided, setting the ultimate limit on measurement precision.

{In practice, noise arising from calibration errors, experimental imperfections, and unwanted coupling to the environment often forms the dominant bottleneck to measurement precision. Unlike fundamental quantum fluctuations, however, these sources are not intrinsic, and with careful isolation and experimental design they can be mitigated---allowing one to approach the ultimate quantum limits of precision. The QFI formalizes these limits, providing a bound on how accurately a parameter can be estimated with a given probe.} 

Importantly, the QFI can be thought of as the SNR optimized over all possible choices of measurement. {However, this optimal choice may not be directly implementable. Nonetheless, it provides a crucial benchmark: by establishing the ultimate precision limit, it allows us to assess how well feasible measurement strategies perform in comparison, and to understand how close we are to the fundamental quantum bound.} Consequently, a central concept in metrology and parameter estimation is the SNR---quantifying the quality of a {specific} measurement by comparing the strength of the signal to the noise that obscures it. With the QFI playing the role of an optimal SNR, this establishes the ultimate benchmark for precision and the link between abstract theoretical limits to real-world measurement challenges. 

In the following, we will analyze the signal and noise contributions in Ramsey interferometry for different measurement schemes, examining how the underlying physical processes ultimately determine the achievable metrological precision. We focus on Ramsey interferometry as it provides a paradigmatic setting to separate signal and noise contributions in quantum metrology. By comparing different measurement schemes within this framework, we can directly link the underlying physical processes to the achievable precision, thereby clarifying the fundamental limits of parameter estimation.

\subsection{Ramsey interferometry}\label{sub_sec:ramsey_inter}
To illustrate the principles of quantum metrology, we consider the \emph{Ramsey interferometer}—a cornerstone of precision measurement and the operating principle behind atomic clocks, which rank among the most accurate instruments ever built~\cite{RevModPhys.87.637}. Their extraordinary precision stems from the inherent stability of atomic transition frequencies: each atom functions as an almost perfect clock, and its ticks are dictated by fundamental physical laws, much like those of an ideal pendulum. By counting the oscillations of an atom’s quantum state, time can be tracked with remarkable accuracy. In practice, however, imperfections such as fluctuations in the probing laser’s intensity and frequency introduce unavoidable sources of imprecision.

To simplify the discussion, atoms are modelled as two-level quantum systems, where each atom has a ground state $|0\rangle$ and an excited state $|1\rangle$. The dynamics of these two states can be described using Pauli matrices, with the Hamiltonian of $N$ identical two-level atoms given by (throughout this manuscript we set $\hbar = 1$ for simplicity, except in a few formulas where it is explicitly retained for clarity)
\begin{align}\label{HRamsey}
    \hat H = \sum_{i=1}^N \frac{\omega}{2} \hat \sigma_z^{(i)}\;,
\end{align}
where $\omega = \Delta E / \hbar$ is the frequency corresponding to the energy difference $\Delta E$ between the two states, and $\hat \sigma_z^{(i)}$ is the Pauli $z$-operator for the $i$-th atom. 

\begin{figure*}[htb!]
    \centering
    \includegraphics[width=0.8\linewidth]{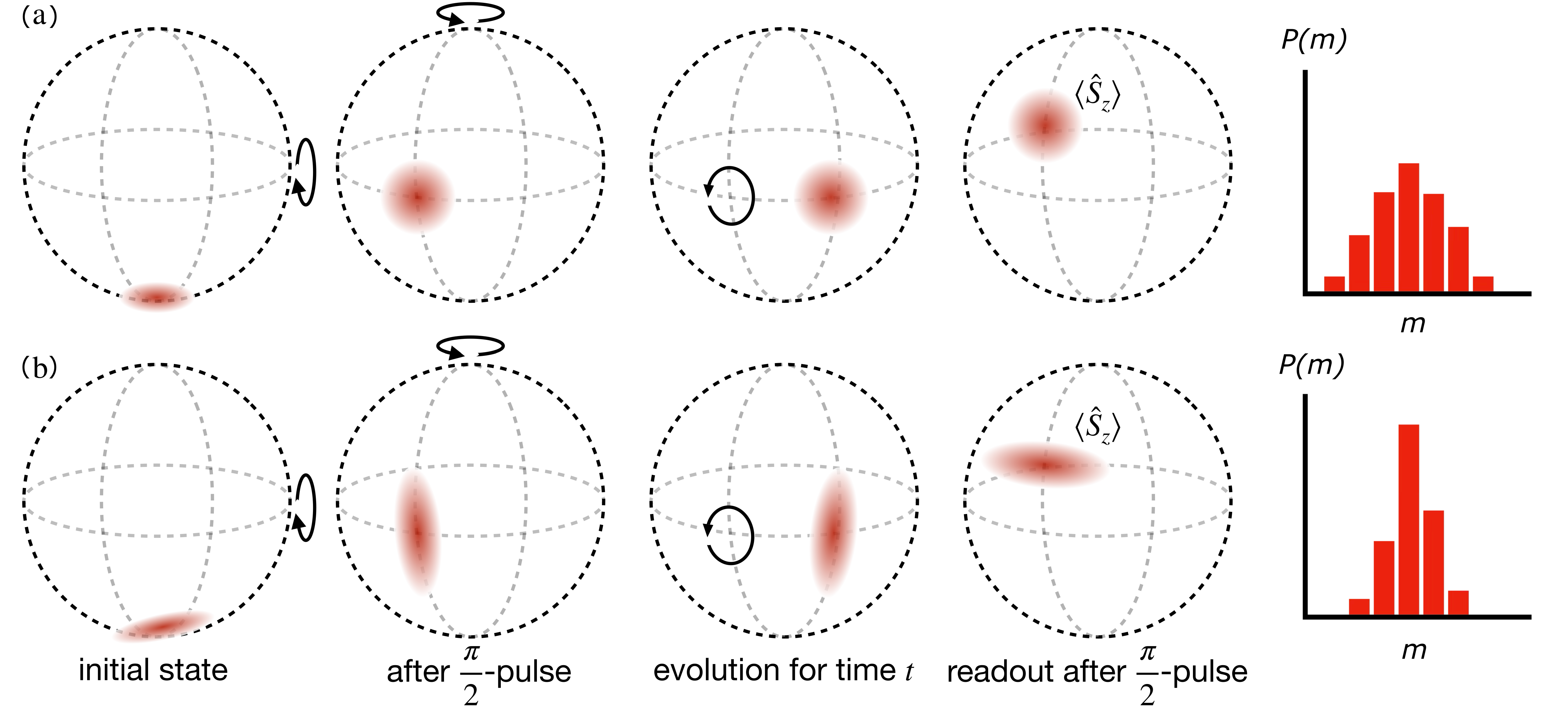}
    \caption{\textbf{Schematic representation of the Ramsey interferometry protocol}. The protocol is illustrated for (a) a classical (coherent) atomic ensemble and (b) a spin-squeezed atomic state. The red cloud-like shape indicates the quantum state's uncertainty distribution, and the curved arrows denote the axes and directions of rotation. The protocol begins with an initial state polarized along the $z$-axis. A first $\pi/2$ pulse rotates the state onto the equator of the Bloch sphere, aligning it along the $y$-axis. During the free evolution period of duration $t$, the state precesses around the $z$-axis by an angle $\omega t$, where $\omega$ is the frequency to be estimated. A second $\pi/2$ pulse, applied around the $x$-axis, converts the acquired phase information into a population imbalance along $z$, which is read out via a measurement of $\langle \hat{S}_z \rangle$. In the spin-squeezed case (b), the reduced quantum noise along the measurement axis suppresses uncertainty in the population imbalance, enhancing the SNR and thereby improving the precision of frequency estimation. The histograms on the right side represent the probability $P(m) = |\langle m|\psi\rangle|^2$ of measuring population imbalance $m$. The probability distribution is narrower for a squeezed state reflecting reduced fluctuations (squeezing).}
    \label{fig:Ramsey}
\end{figure*}
b

In a Ramsey interferometer [see Fig.~\ref{fig:Ramsey} for a schematic Bloch sphere representation], parameter estimation begins by preparing a quantum system in a well-defined initial state, typically the collective ground state. The first step is to apply a pulse that creates a coherent superposition of the two energy levels, represented as a quantum state vector on the Bloch sphere, halfway between the two energy eigenstates. 

Next, the system evolves freely under the influence of the unknown parameter, such as a frequency, magnetic or gravitational field. During this evolution, a relative phase accumulates between the components of the superposition state, encoding information about the parameter. The evolution time is carefully chosen to maximize sensitivity to small variations in the parameter. Because directly measuring the phase is challenging, a second pulse is applied after the evolution period. This pulse converts the accumulated phase information into a measurable quantity, typically the population difference between the two energy levels. The population difference depends on the unknown parameter, and serves as the interferometer's output signal. 

By repeating this process—preparing the system, allowing it to evolve, and measuring the population difference---a statistical dataset is collected. Techniques such as least squares method, {maximum likelihood estimation}, or {Bayesian inference} can then be used to analyze the data and estimate the parameter with high precision. In the following, we will consider the estimation qualities of three differently prepared initial states: coherent spin state, squeezed spin state, and maximally entangled spin state.

\subsection{Coherent spin state}
In Ramsey interferometry, states prepared in a classical superposition are called \emph{coherent spin states} (CSS)~\cite{Radcliffe_1971} [see Fig.~\ref{fig:Ramsey_states}(a)], and are often measured through a population difference $\sum_{i=1}^N \hat \sigma_z^{(i)}$. As atoms behave collectively, we define {collective spin operators} as
\begin{align}
    \hat S_\alpha = \frac{1}{2} \sum_{i=1}^N \hat \sigma_\alpha^{(i)}, \quad \text{for } \alpha = x, y, z \;.
\end{align}
For a CSS state 
\begin{align}
   |\psi\rangle  = \bigotimes_{i=1}^N \frac{1}{\sqrt{2}}\left[|0\rangle + |1\rangle \right]\;,
\end{align}
which rotates around the Bloch sphere with frequency $\omega$, the expectation value of the population difference after the second $\pi/2$ pulse is
\begin{align}
    \langle \hat S_z \rangle = \frac{N}{2} \sin (\omega t)\;.
\end{align}
The corresponding measurement noise arising from quantum fluctuations is given by
\begin{align}
    (\Delta \hat S_z)^2 = [\Delta \hat S_y(t=0)]^2 = \frac{N}{4}\;.
\end{align}
The precision in the estimation procedure when utilizing a CSS can then be quantified using the SNR as
\begin{align}
    S_\omega = \frac{[\partial_\omega \langle \hat S_z \rangle]^2}{(\Delta \hat S_z)^2} = N t^2 \cos^2 (\omega t) \leq N t^2\;.
\end{align}
The maximum SNR, ${S_\omega = N t^2}$, defines the \emph{standard quantum limit} for coherent dynamics. {This is a manifestation of the central limit theorem: each of the $N$ spins behaves independently---their measurement outcomes are averaged to reduce uncertainty. In this case the intrinsic uncertainty due to quantum measurement averages out like classical noise from repeated trials. The CSS, being an unentangled product state, is relatively easy to prepare and robust to noise such as decoherence effects, making it highly attractive from a practical point of view.} Atomic clocks use this principle to measure the detuning ${\delta = \omega_d - \omega}$, where $\omega_d$ is the laser frequency driving the system. By tuning the laser close to resonance [${\cos (\delta t) \approx 1}$], the SNR approaches the standard quantum limit. This enables atomic clocks to achieve extraordinary precision, detecting effects like gravitational shifts---energy level changes caused by the curvature of space-time. Over a timescale comparable to the age of the Universe, approximately 13.8 billion years, these clocks would accumulate an error of only about one second~\cite{Ye2022gravitationalredshiftclock}.

\subsection{Redistribution of noise with spin-squeezing}

\begin{figure}[t]
     \centering
    \includegraphics[width=1\linewidth]{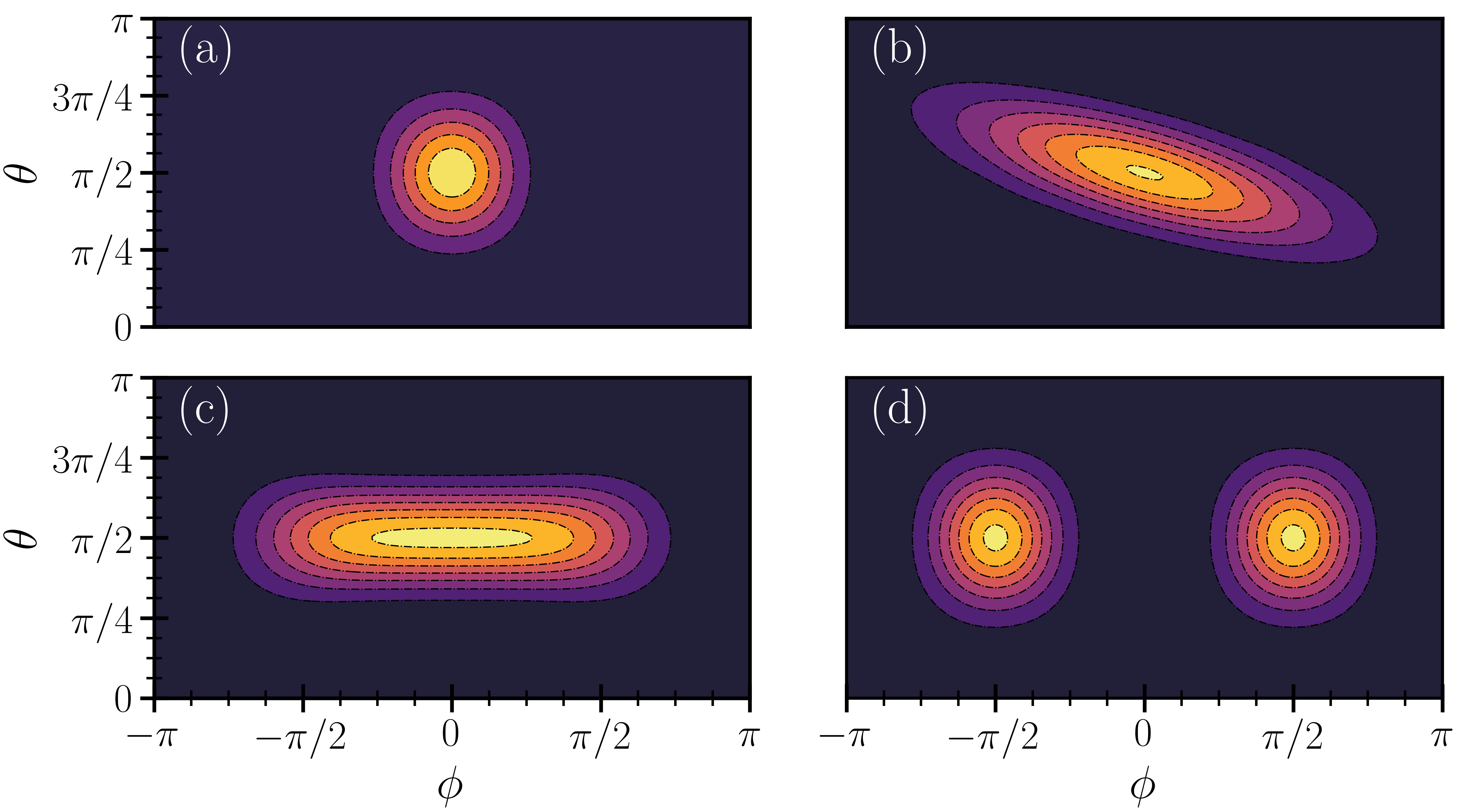}
    \caption{ \textbf{Mercator-projected phase-space (Husimi Q) distributions of quantum states used in Ramsey interferometry.} Shown are: (a) CSS achieving the standard quantum limit; (b) optimally squeezed state generated via a one-axis twisting Hamiltonian, capable of surpassing the standard quantum limit; (c) optimally squeezed state generated via a two-axis counter-twisting Hamiltonian, reaching the quadratic (Heisenberg) scaling; (d) maximally entangled state achieving the ultimate Heisenberg limit. To obtain the plots, we set $N=10$. 
    }
    \label{fig:Ramsey_states}
\end{figure}

While reaching the standard quantum limit is remarkable, further precision can be achieved by reducing quantum noise. This is done using {quantum squeezing} [see Fig.~\ref{fig:Ramsey_states}(b) and (c)], a form of entanglement where the uncertainty in one observable (for instance $\hat S_z$) is reduced at the expense of increased uncertainty in a conjugate observable (for instance $\hat S_y$)~\cite{MA201189SPINSQueezing}. The ultimate limit of squeezing is set by the Heisenberg uncertainty relation
\begin{align}
    (\Delta \hat S_1)^2 (\Delta \hat S_2)^2 \geq \frac{1}{4} \langle \hat S_3 \rangle^2\;,
\end{align}
where $\hat S_1, \hat S_2, \hat S_3$ are spin components along orthogonal axes. 

A natural question is how such squeezed states can be generated. The most widely studied mechanisms are one-axis twisting and two-axis counter-twisting, originally introduced by Kitagawa and Ueda~\cite{UedaKitagawa1993_SSS}. In one-axis twisting, the collective spin evolves under an interaction Hamiltonian of the form $\hat H \propto \hat S_z^2$. Geometrically, this causes the initially symmetric uncertainty distribution on the Bloch sphere to \emph{twist} around the $z$-axis. As a result, the quantum noise ellipse becomes squeezed along one direction and stretched along the orthogonal direction [see Fig.~\ref{fig:Ramsey_states}(b)]. In contrast, two-axis counter-twisting arises from a Hamiltonian of the form $\hat H \propto \hat S_x^2 - \hat S_y^2$. On the Bloch sphere, this corresponds to simultaneous twisting around two orthogonal axes, which redistributes quantum fluctuations more efficiently and can generate stronger squeezing than one-axis twisting [see Fig.~\ref{fig:Ramsey_states}(c)]. 

\begin{figure}[t]
    \centering     \includegraphics[width=\linewidth]{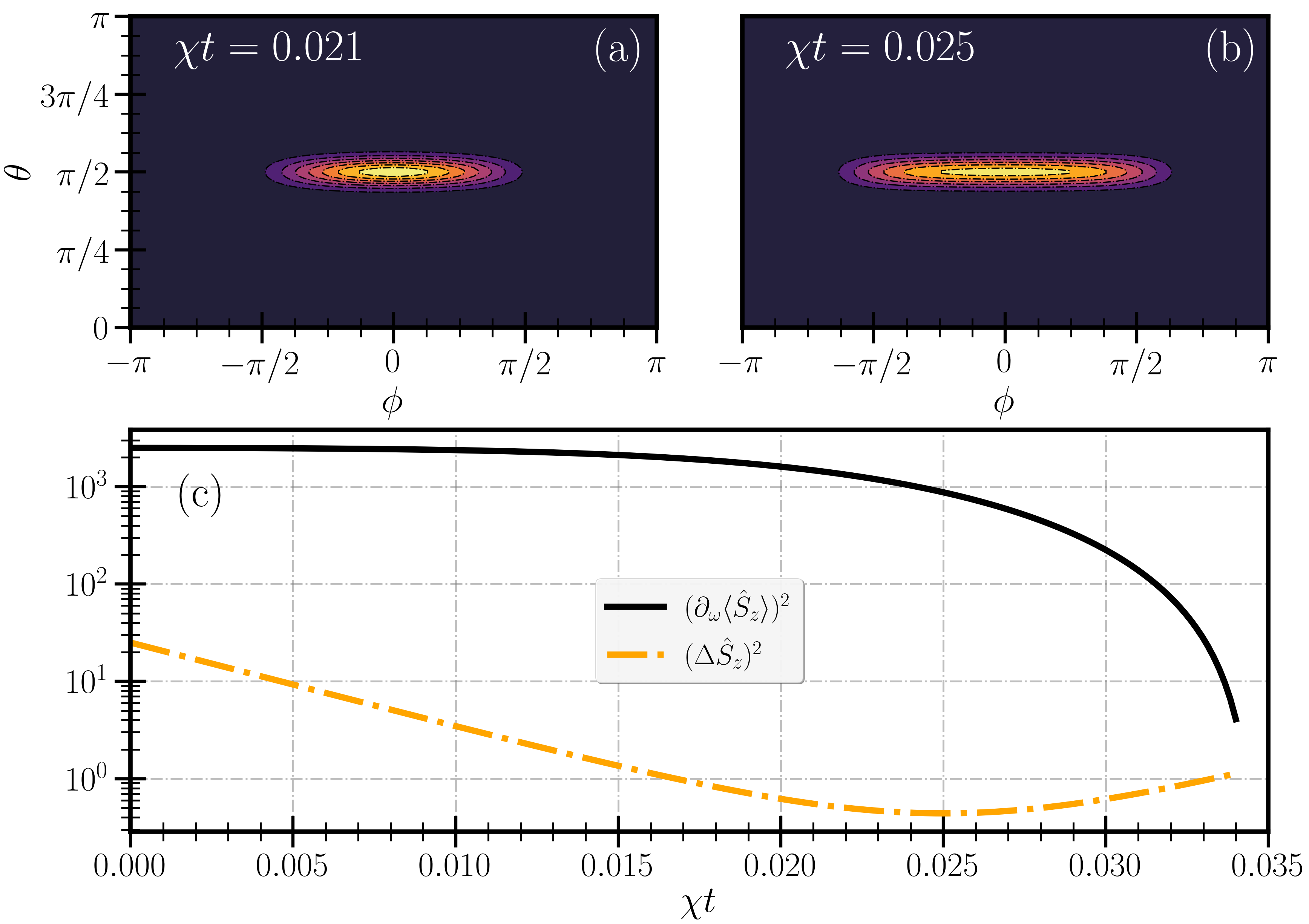}
    \caption{\textbf{Squeezing with two-axis counter-twisting.} Mercator-projected phase-space (Husimi Q) distributions of: (a) the state that maximizes SNR, and (b) the state with minimal quantum fluctuations (maximally squeezed). 
    We show that the state maximizing the SNR is not maximally squeezed, because excessive squeezing reduces the signal. Panel (c) illustrates the resulting trade-off between signal strength and noise reduction. To obtain the plots, we set $N=100$ and evolved an initial coherent state polarized along $x$ axis with a two-axis counter-twisting unitary $\hat U(t) =\exp[-i\chi t(\hat S_z^2-\hat S_y^2)]$. 
    }
    \label{fig:Ramsey_squeezing}
\end{figure}

While one-axis twisting is experimentally more accessible, and underlies many spin-squeezing protocols in cold atoms and trapped ions, two-axis counter-twisting represents the optimal scenario for achieving the fastest squeezing towards the Heisenberg limit [see Fig.~\ref{fig:Ramsey_squeezing}]. For an optimally squeezed state generated by two-axis counter-twisting [see Fig.~\ref{fig:Ramsey_states}(c)], the noise reduces to
\begin{align}
    [\Delta \hat S_z(t)]^2 \approx \frac{1}{2}\;,
\end{align}
where this noise reduction comes at the expense of reducing the signal by half. This overall improves the optimal SNR to
\begin{align}
    S_\omega
    \approx \frac{N^2 t^2}{4}\;.
\end{align}

\subsection{Maximally entangled states and the Heisenberg limit}
Is the above SNR for population difference measurements the ultimate limit for coherently evolving systems? To answer this, we must delve into the concept of the {QFI}, a central quantity in quantum metrology that generalizes the SNR by optimizing it over arbitrary observables. The QFI with respect to a parameter $\omega$ for arbitrary pure quantum states is given by
\begin{align}\label{eq:QFIdef}
    \mathcal{I}_\omega = 4 \left[\langle \partial_\omega \psi(t) | \partial_\omega \psi(t) \rangle - |\langle \partial_\omega \psi(t) | \psi(t) \rangle|^2\right]\;,
\end{align}
where $|\partial_\omega \psi(t)\rangle$ represents the derivative of the quantum state $|\psi(t)\rangle$ with respect to $\omega$. The QFI quantifies how sensitively the quantum state depends on the parameter $\omega$, thereby setting an ultimate theoretical bound on the precision of parameter estimation.

For the Ramsey interferometer, where the collective state rotates with frequency $\omega$, the QFI simplifies to
\begin{align}
    \mathcal{I}_\omega = 4 t^2 (\Delta \hat S_y)^2\;,
\end{align}
where $(\Delta \hat S_y)^2$ is the variance of the collective spin operator $\hat S_y$ in the initial, for now arbitrary, state. Importantly, $(\Delta \hat S_y)^2$ depends on the quantum correlations within the system. The QFI reaches its maximum value—commonly referred to as the {Heisenberg limit}—when the system is prepared in a maximally entangled state [see Fig.~\ref{fig:Ramsey_states}(d)]:
\begin{align}
|\psi\rangle &= \frac{1}{\sqrt{2}} \left(|0\rangle^{\otimes N} + |1\rangle^{\otimes N}\right)\;,
\end{align}
which maximizes the variance of $\hat S_y$. Depending on the field, such states are known under different names: GHZ states in the many-body and quantum information community~\cite{Greenberger1989}, Schrödinger cat states in quantum optics~\cite{PhysRevLett.57.13}, and NOON (N00N) states in interferometry and photonics~\cite{Dowling01032008,PhysRevA.54.R4649}. More generally, they are also described as macroscopic superposition states or maximally path-entangled states. For these states, the QFI attains its maximal value
\begin{align}
\mathcal{I}_\omega = N^2 t^2\;.
\end{align}
This upper bound is four times as large as the SNR obtained from population difference measurements for optimally squeezed states, indicating that population difference measurement is not the optimal strategy. Nevertheless, population difference measurements remain near-optimal, and are much simpler to implement in practice. 

While maximally entangled states offer the highest QFI, their practical use is limited by significant challenges. For these states, the population difference signal $\langle \hat S_z \rangle$ vanishes. In Ramsey interferometry with maximally entangled states, one typically relies on measuring a collective coherence $\hat{\sigma}_x^{\otimes N}$. In practical settings, it is typically more difficult to measure a collective coherence than a population difference (previously used for CSS and squeezed spin states). For arbitrary system size $N$ one obtains on average a measurement signal of 
\begin{equation}
    \langle \hat{\sigma}_x^{\otimes N} \rangle = \cos{(N\omega t})\;.
\end{equation}
The corresponding measurement noise arising from quantum fluctuations is given by
\begin{equation}
    (\Delta \hat{\sigma}_x^{\otimes N})^2 = \sin^2(N \omega t)\;.
\end{equation}
The precision in the estimation procedure when utilizing a maximally entangled state can then be quantified using the SNR as
\begin{align}
    S_\omega = \frac{[\partial_\omega \langle \hat \sigma_x^{\otimes N} \rangle]^2}{(\Delta \hat \sigma_x^{\otimes N})^2} = N^2 t^2\;.
\end{align}
The above result has profound implications for metrology and high-precision measurements. When Ramsey interferometry is performed with CSSs, the QFI scales as $\propto N t^2$, corresponding to the standard quantum limit. By contrast, if the system is prepared in a maximally entangled state, the scaling improves to $\propto N^2 t^2$. This quadratic dependence is commonly referred to as \emph{Heisenberg scaling}, reflecting the collective accumulation of phase across the entire ensemble, which amplifies small changes in the frequency $\omega$.

However, it is important to stress that {Heisenberg scaling is not a fundamental limit}. A system can display quadratic (or even higher than quadratic~\cite{gietka2023overcoming}) scaling while still performing worse than the SQL in absolute terms, depending on prefactors and resources. For this reason, the term \emph{quantum enhancement} should be understood as genuinely surpassing the standard quantum limit, rather than merely achieving a quadratic scaling. In practice, quantum advantage is often associated with better-than-linear scaling, with quadratic scaling representing the ultimate enhancement in idealized scenarios. Yet, this potential benefit comes at a cost: maximally entangled states are extremely fragile to noise, challenging to prepare, and difficult to measure reliably.

\subsection{Limitations in open quantum systems}
Thus far, our discussion has focused on closed quantum systems undergoing coherent evolution---idealized scenarios that, while insightful, are not fully representative of real-world quantum systems. In reality, no quantum system exists in complete isolation. Every atom---or any quantum system---interacts with its surrounding environment. For instance, atoms couple to the electromagnetic vacuum, which contains infinitely many modes of the electromagnetic field. This interaction leads to phenomena such as {spontaneous emission}, where an excited atom irreversibly emits a photon into the vacuum. Spontaneous emission fundamentally limits the lifetime of an atom's excited state, and has profound implications for Ramsey interferometry and quantum metrology whatsoever. 

Paradoxically, while environmental coupling is a source of decoherence and dissipation, it is also what makes quantum measurement possible. Without this interaction, quantum systems would remain entirely isolated, inaccessible to observation. The challenge in quantum metrology lies in carefully balancing the detrimental effects of decoherence with the necessity of coupling for measurement. Various models describe the interplay between quantum systems and their environments, accounting for decoherence and dissipation in different contexts~\cite{DemkowiczDobrzanski2009lossy,noisy2011metrology, wscher2011Natphysnoise,Koodyski2013efficienttools,maccone2014noiseagainst,demko2012elusiveHL,DemkowiczDobrzaski2015limitsint,smirne2016ultimatenoise,demkowicz2017adaptive,noisy2011metrology}. These effects impose constraints on the time over which a quantum system can maintain coherent evolution. For example, in the presence of dissipation the optimal SNR no longer exhibits a simple quadratic dependence on time. Instead, it often takes the form
\begin{align}
     S_\omega= \frac{f_N T}{\Gamma}\;,
\end{align}
where: \( \Gamma \) quantifies the dissipation strength (for instance, the inverse lifetime of an atomic excited state); \( T \) is the optimal evolution time, which often is no longer a freely tunable parameter but is determined by the interplay between quantum dynamics and noise; and \( f_N \) is a function that depends on the number of atoms \( N \), the initial state, and other parameters of the system. Importantly, \( f_N \) is asymptotically a linear function of $N$~\cite{demko2012elusiveHL}. This relationship underscores the central role of decoherence in metrology: the maximum achievable precision is not solely dictated by quantum mechanics but also by how effectively dissipation and noise can be mitigated or leveraged. These aspects will be analyzed further in Sections~\ref{sec:HO} and \ref{sec:ultrastrong}.

\subsection{Precision limits for systems where the total number of excitations is not fixed}
Last but not least, for extremely large systems (\( N \to \infty \)), the dynamics of collective spin operators can be approximated using harmonic oscillator operators. In this regime
\begin{align}
    \hat{S}_z \sim \hat{a}^\dagger\hat{a}\;, 
\end{align}
where \( \hat{a}^\dagger \) and \( \hat{a} \) are the creation and annihilation operators, respectively. Here, the SNR is no longer fundamentally limited by the number of particles \( N \), as harmonic oscillator operators are unbounded. Instead, the relevant figure of merit becomes the number of excitations in the system
\begin{align}
    \langle \hat{n} \rangle = \langle \hat{a}^\dagger \hat{a} \rangle\;.
\end{align}
This reformulation extends the concepts of the {standard quantum limit} and the {Heisenberg limit} to systems where the total number of excitations is not fixed, as photonic systems. In such cases, \( N \) is effectively replaced by \( \langle \hat{n} \rangle \), providing a consistent framework for characterizing the ultimate precision limits of these systems.

\subsection{Quantum-enhanced measurements}

{We can therefore see how quantum coherence and entanglement can be used as a resource for sensing to go beyond the typical precision limits. This is essentially achieved by exploiting the fact that entangled systems collectively respond to small variations in parameters of interest. A potential shortcoming from a practical perspective is the prohibitively large time and resources required to generate squeezed and maximally entangled states~\cite{chu2023strong}. Furthermore, such states are intrinsically sensitive to noise and decoherence, which ultimately degrade their performance below the Heisenberg limit. This motivates the search for alternative platforms which may achieve the Heisenberg limit, without suffering from these prohibitive shortcomings. Critical systems emerge as natural candidates in this regard: many critical systems are ``passive'' in the sense that a many-body system close to criticality can be tuned to its critical point and allowed to respond to a small parameter change. In practice, this is can be implemented in condensed matter systems. 
\section{Minimal critical model: the quantum harmonic oscillator} \label{sec:QPT_HO}
In this section, we introduce the concept of quantum criticality through the simplest possible system: the quantum harmonic oscillator. While phase transitions are typically discussed in the context of many-body systems, we will see that even a single harmonic mode—when suitably parameterized—can exhibit features that mimic the critical behavior of more complex models. Our goal is to build intuition for how nonclassical properties, such as squeezing and enhanced sensitivity to parameters, naturally emerge near critical points. These effects can be understood without the full machinery of many-body physics and provide a pedagogical entry point into the study of quantum phase transitions. In particular, we will explore how the harmonic oscillator—with a tunable frequency—can act as a minimal model for criticality and how its behavior changes as the frequency approaches zero, where effective critical-like phenomena appear. We will also show how this critical behavior can be leveraged in quantum metrology. This is a central idea in critical quantum metrology, and the harmonic oscillator serves as a clean and analytically tractable platform to illustrate it.

\subsection{Harmonic oscillator: closing the energy gap}\label{sub_sec:HO_QPT}
A quantum phase transition typically occurs when the energy gap between the ground state and excited states closes in the thermodynamic limit, though in some cases (e.g. first-order or topological transitions), the relation can be more subtle. A simple model to capture this behavior is a harmonic oscillator with a tunable frequency. The energy spectrum of such a system is linear, meaning that the energy gaps between consecutive levels are constant and proportional to the frequency of the oscillator
\begin{align}
    \Delta E = \hbar \omega\;.
\end{align}
As the frequency $\omega$ approaches zero, the energy gap closes, which mirrors the critical point of a quantum phase transition. To illustrate this, we consider the Hamiltonian of a harmonic oscillator with a tunable frequency
\begin{align}\label{Hsimpleharmonicoscillator}
    \hat H = \frac{\hat p^2}{2} + \frac{\omega^2(g)}{2}\hat x^2\;,
\end{align}
where $\omega(g) = \omega \sqrt{1 - g/\omega}$ is the tunable frequency, and the parameter $g$ controls the oscillator’s behavior: increasing $g$ opens (unsqueezes) the oscillator, while decreasing it closes (squeezes) the harmonic oscillator (see Fig.~\ref{fig:closing_gap} for a schematic illustration). Clearly, when $g = \omega$ the oscillator flattens, and the energy gap vanishes, mimicking the critical point of a quantum phase transition. To describe this in terms of quantum fields, we express the abstract position and momentum operators as
\begin{align}
    \hat x = \sqrt{\frac{1}{2 \omega}}\left(\hat a + \hat a^\dagger\right), \quad \hat p = \sqrt{\frac{\omega}{2}}\left(\hat a - \hat a^\dagger\right)\;,
\end{align}
where $\hat a$ and $\hat a^\dagger$ are the annihilation and creation operators, respectively. Substituting these into Eq.~\eqref{Hsimpleharmonicoscillator}, we obtain 
\begin{align}
    \hat H = \omega \hat a^\dagger \hat a - \frac{g}{4}\left(\hat a + \hat a^\dagger\right)^2\;,
\end{align}
which represents a squeezing Hamiltonian. This Hamiltonian either squeezes or unsqueezes the harmonic oscillator as well as the quantum states, depending on the value and sign of $g$. The eigenstates of the Hamiltonian are related to the original Fock states (or number states) $|n\rangle$ of the harmonic oscillator with frequency $\omega$ through the squeezing operator
\begin{align} \label{eq:eigenstatesHO}
    \hat S(\xi)|n\rangle = \exp\left[\frac{\xi}{2} \left(\hat a^{\dagger 2}- \hat a^2\right)\right]|n\rangle\;,
\end{align}
where $\xi=\left[\log\left(1-g/\omega\right)\right]/4$ is the squeezing parameter. The transformed Hamiltonian, expressed in terms of a new set of annihilation and creation operators $\hat c$ and $\hat c^\dagger$, becomes  
\begin{align}
    \hat S(\xi)\hat H \hat S^\dagger(\xi) = \omega\sqrt{1-g/\omega}\, \hat c^\dagger \hat c\;,
\end{align}
\begin{figure}[t]
      \centering
    \includegraphics[width=1\linewidth]{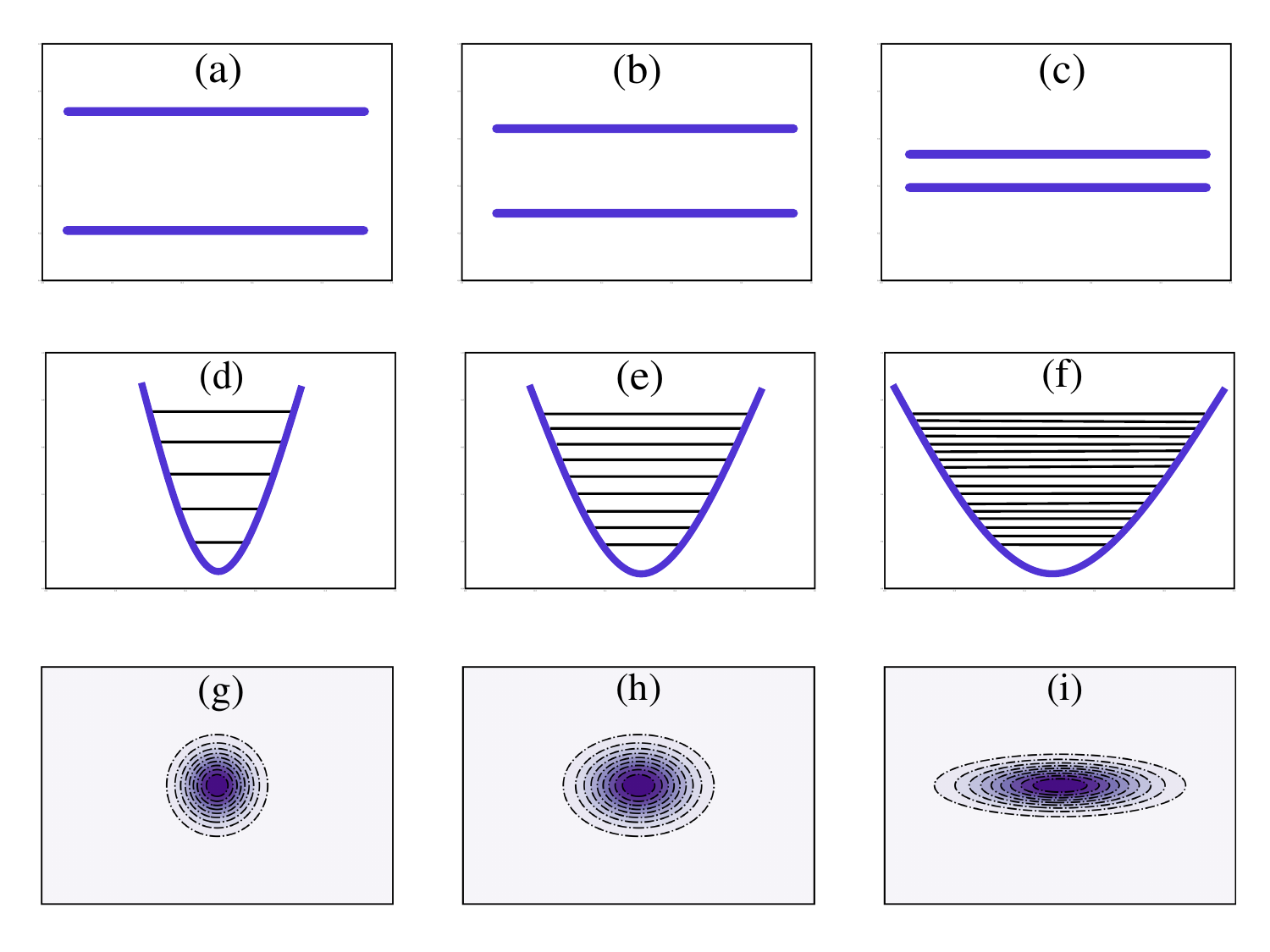}
    \caption{\textbf{Schematic illustration of the connection between energy-gap closing and quantum correlations.} Panels (a)–(c) show the progressive closing of the energy gap in a harmonic oscillator, corresponding to the flattening of its potential (d)–(f). This deformation of the potential strongly modifies the ground state. As the oscillator opens up, the state becomes increasingly squeezed (g)-(i). At the critical point, where the harmonic oscillator is maximally flattened, the ground state exhibits extreme squeezing and heightened sensitivity to the oscillator frequency. This mechanism provides an intuitive picture for the enhanced sensitivity underlying critical quantum metrology.}
    \label{fig:closing_gap}
\end{figure}
where the new operators are defined as 
\begin{align}
    \hat c = \hat S^\dagger(\xi)\hat a \hat S(\xi) = \hat a \cosh \xi  + \hat a^\dagger \sinh \xi \;,
\end{align}
and
\begin{align}
    \hat c^\dagger = \hat a^\dagger \cosh \xi  + \hat a \sinh \xi\;.
\end{align}
This describes the modes of a harmonic oscillator with a renormalized frequency $\omega\sqrt{1-g/\omega}$, encapsulating the behavior near the critical point.

Keep in mind that the definition of harmonic oscillator operators inherently depends on the frequency of the oscillator. For example, in a mechanical harmonic oscillator, the annihilation operator is given by
\[
\hat{a} = \sqrt{\frac{m \omega}{2 \hbar}} \hat{x} + i \sqrt{\frac{1}{2 m \hbar \omega}} \hat{p}\;,
\]
where \(m\) is the mass of the system, \(\omega\) is its angular frequency, \(\hat{x}\) is the position operator, and \(\hat{p}\) is the momentum operator.

\subsection{Quantum Fisher information for an opening harmonic oscillator}\label{sub_sec:QFI_QHO_Opening}

The squeezing associated with opening or closing of the energy gap already hints at the possibility of reducing quantum noise and thus improving measurement precision, in analogy to noise squeezing in Ramsey interferometry. To further explore this, let us calculate the QFI for the ground state of the modified harmonic oscillator previously described, treating $\omega$ as the unknown parameter. Plugging the ground state from Eq.~\eqref{eq:eigenstatesHO} with $n=0$ into the expression for the QFI~\eqref{eq:QFIdef}, we obtain 
\begin{align}\label{eq:QFIho}
    \mathcal{I}_\omega = 2{(\partial_\omega \xi)^2} = \frac{g^2}{8\omega^2(g-\omega)^2}\;.
\end{align}
This result explodes at the critical point $g\sim\omega$ where the harmonic oscillator frequency vanishes as $\omega(g) \to 0$, signaling an infinite QFI. However, this form of the QFI is unconventional; notably, it appears independent of time and the number of excitations. Does this imply that the bounds derived in earlier sections do not apply to critical metrology?  

To clarify this, consider a general Hamiltonian of the form 
\begin{align}
    \hat H = \omega \hat H_\omega + \hat H_c\;,
\end{align}
where neither $\hat H_\omega$ nor $\hat H_c$ depend on $\omega$. For such systems, the QFI is bounded by~\cite{Boixo2008limitQuantumEstimation}  
\begin{align}
    \mathcal{I}_\omega \leq t^2 ||\hat H_\omega ||^2\;,
\end{align}
where the pseudo-norm $||\hat H_\omega || = \lambda_\mathrm{max} - \lambda_\mathrm{min}$ is defined in terms of the maximum ($\lambda_\mathrm{max}$) and minimum ($\lambda_\mathrm{min}$) eigenvalues of $\hat H_\omega$. For finite systems---such as atoms in a Ramsey interferometer, where the eigenvalues are bounded---this bound translates to the Heisenberg limit  
\begin{align}
    \mathcal{I}_\omega = N^2 t^2.
\end{align}
This confirms that the QFI is bounded by the Heisenberg limit even in critical metrology. Consequently, adding extra terms to the Hamiltonian that do not depend on the unknown parameter, regardless of their complexity, cannot breach this limit.  

Interestingly, the QFI from Eq.~\eqref{eq:QFIho} can be linked to the number of excitations in the ground state. For a squeezed vacuum state, the number of excitations is given by  
\begin{align}\label{eq:Nvacum}
    \langle \hat{n}\rangle=\langle \hat a^\dagger \hat a \rangle = \sinh^2 \xi 
    \xrightarrow[]{g\rightarrow \omega} \frac{1}{4}\frac{1}{\sqrt{1-g/\omega}}\;.
\end{align}
Using this, the QFI can be rewritten as 
\begin{align}
    \mathcal{I}_\omega \approx \frac{4 g^2}{\omega^4} \langle \hat n \rangle^4\;,
\end{align}
which seemingly surpasses the Heisenberg limit, $\langle \hat n \rangle^2 t^2$, for times shorter than  
\begin{align}
    t < \frac{2 g}{\omega^2} \langle \hat n \rangle\;.
\end{align}
\subsection{Taking time into consideration}
How is this possible? The answer lies in neglecting the time required to prepare such a state. Using the adiabatic theorem, the preparation time for a state near the critical point, $g \sim \omega$, starting from an initial state deep in the trivial phase, $g \sim 0$, can be estimated as
\begin{align}
T \approx \frac{2}{\gamma \omega} \frac{1}{\sqrt{1 - g/\omega}} = \frac{8}{\gamma \omega} \langle \hat n \rangle\;,
\end{align}
where the condition $\gamma \ll 1$ is imposed to ensure the adiabatic condition. Here, the time $T$ is not a freely tunable parameter as in the case of free evolution; rather, it is a function of the number of excitations $\langle \hat n \rangle$, which grows as one approaches the critical point. This connection between time and excitations is a hallmark of critical systems and plays a central role in critical quantum metrology. Near a quantum phase transition, the energy gap closes in the thermodynamic limit, leading to critical slowing down—the phenomenon where the system's intrinsic timescales diverge as it approaches criticality. As a result, any adiabatic process that attempts to traverse the critical region becomes prohibitively slow, requiring increasingly long times to maintain adiabaticity and avoid diabatic transitions. This divergence is reflected in the scaling of the relaxation time, which typically follows a power-law behavior with the distance to the critical point.

In our context, this means that while the ground-state susceptibility—and hence the QFI—can be significantly enhanced near the critical point, preparing the system in such a highly sensitive state becomes increasingly difficult. The required adiabatic preparation time diverges with $\langle \hat n \rangle$, and any deviation from perfect adiabaticity can lead to excitations that degrade or disturb the metrological performance. This trade-off between sensitivity and preparation time is at the heart of critical metrology. Substituting the adiabatic time into the expression for the QFI yields
\begin{align}
\mathcal{I}_\omega \approx 8\gamma^2 T^2 \langle \hat n \rangle^2\;,
\end{align}
which exhibits a quadratic scaling in both $\langle \hat n \rangle$ and $T$—commonly referred to as Heisenberg scaling. However, this comes with an important caveat: the prefactor is suppressed by $8 \gamma^2$, making the achievable precision substantially worse than the Heisenberg limit. In fact, the QFI surpasses the standard quantum limit only when
\begin{align}
\langle \hat n \rangle > \frac{1}{8\gamma^2}\;.
\end{align}
Given that $\gamma \ll 1$ is required for adiabaticity, the number of excitations needed to exceed the standard quantum limit becomes exceedingly large, which again highlights the practical limitations imposed by critical slowing down.

In summary, while critical points offer the promise of enhanced sensitivity through divergent susceptibilities, this advantage comes at the cost of long preparation times and increasing fragility. These features are emblematic of critical quantum metrology, where the interplay between enhanced sensitivity and slow dynamics sets a fundamental limit on performance.
\begin{figure*}[htb!]
      \centering
    \includegraphics[width=1\linewidth]{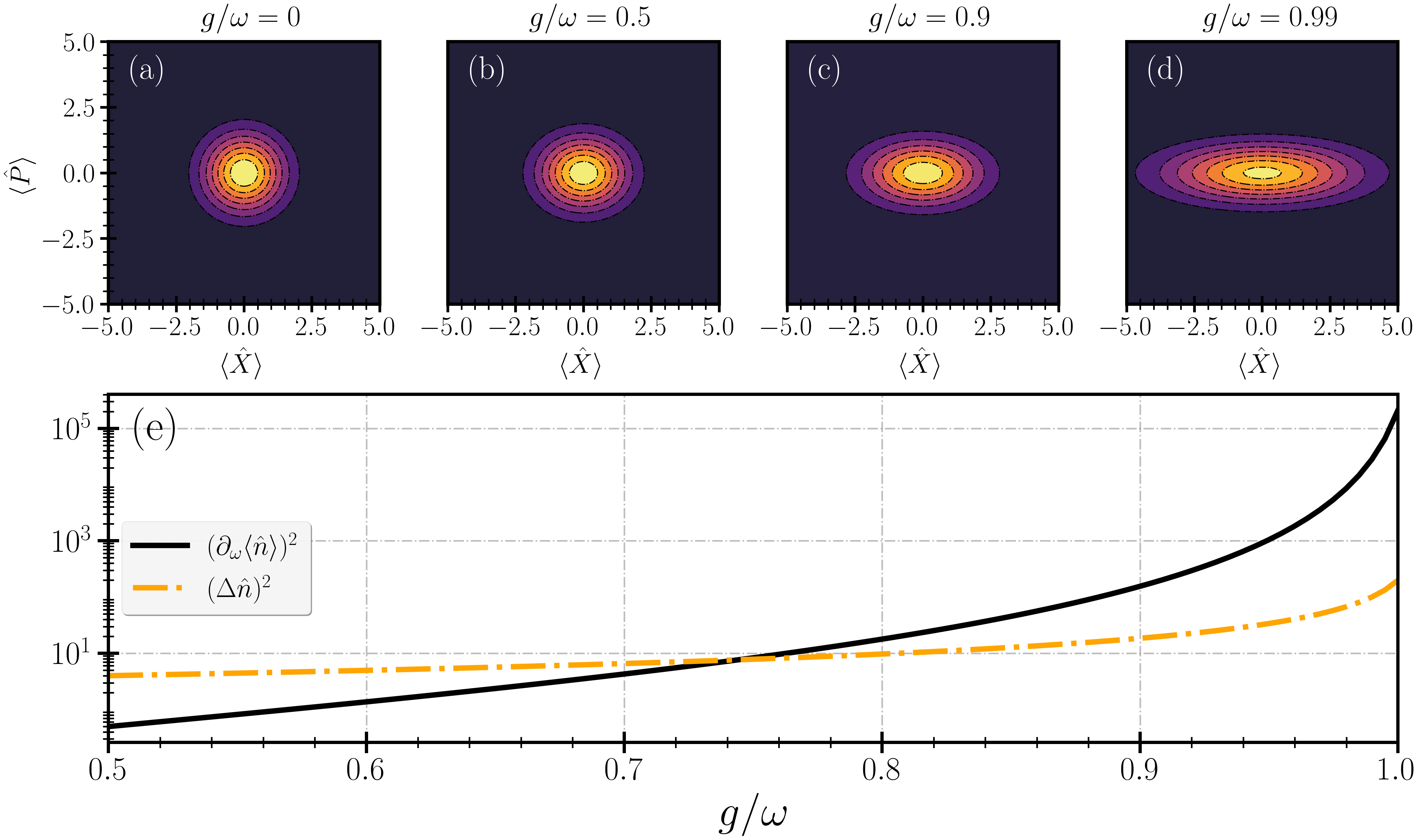}
    \caption{\textbf{Operating mechanism of critical metrology illustrated with an opening harmonic oscillator.} Approaching the critical point $g=\omega$ opens (or squeezes) the harmonic oscillator which in turn squeezes the ground state, as shown in panels (a)–(d), where we display the Husimi Q-function of the ground state. This squeezing increases the signal (here related to $\langle \hat n \rangle$) but also enhances fluctuations $(\Delta \hat n)^2$, as shown in (e). However, near the critical point, the signal grows much faster than the fluctuations, leading to an overall enhancement of the SNR and the QFI. Ideally, one would aim to enhance the signal while simultaneously suppressing the noise.}
    \label{fig:Critical_Metrology}
\end{figure*}

\subsection{Optimal observable}\label{sub_sec:SNR_HO}

So far we have considered only the QFI. Let us now analyze the SNR in the harmonic oscillator model. A natural choice of observables includes the excitation number \( \hat{n} = \hat{a}^\dagger \hat{a} \), or the squeezing represented by the width of the harmonic oscillator, \( (\hat{a} + \hat{a}^\dagger)^2 \). For the excitation number \( \hat{n} \), the fluctuations are given by
\begin{align}
    (\Delta \hat{n})^2 = \langle \hat{n}^2 \rangle - \langle \hat{n} \rangle^2 = 2 \langle \hat{n} \rangle (1 + \langle \hat{n} \rangle)\;,
\end{align}
where \( \langle \hat{n} \rangle = \sinh^2\xi \) is the average number of excitations. Substituting the value of \( \langle \hat{n} \rangle \) from Eq.~\eqref{eq:Nvacum}, we calculate the SNR for measuring the average excitation number:
\begin{align}
     S_\omega = \frac{g^2 }{8\omega^2(g - \omega)^2}\;.
\end{align}
Interestingly, this SNR matches the QFI, indicating that measuring \( \hat{n} \) achieves the optimal precision in this model. 

Next, consider the observable \( (\hat{a} + \hat{a}^\dagger)^2 \), which corresponds to the squeezing of the harmonic oscillator. The expectation value of this observable is
\begin{align}
    \langle (\hat{a} + \hat{a}^\dagger)^2 \rangle = \exp(-2\xi)\;,
\end{align}
and its fluctuations are given by
\begin{align}
    \Delta^2 (\hat{a} + \hat{a}^\dagger)^2 = 2 \exp(-4\xi)\;.
\end{align}
The corresponding SNR is 
\begin{align}
     S_\omega = \frac{g^2 }{8\omega^2(g - \omega)^2}\;,
\end{align}
which, remarkably, is again equal to the QFI.

These results are both elegant and insightful, shedding light on several key aspects of critical metrology. First, they suggest a certain universality in critical metrology: finding optimal or efficient measurement strategies appears to be relatively straightforward in many cases, particularly because, near the critical point, the states are well-approximated by Gaussian distributions. As a result, the majority of information about these states can be captured through the first two moments. This observation points to a profound and intrinsic connection between quantum squeezing and the closure of the energy gap at the critical point, emphasizing the fundamental role of critical phenomena in enhancing measurement precision. 

Second, while critical systems are known to amplify noise (as seen in the increased fluctuations of observables), they also amplify the signal even more strongly, leading to significant improvements in metrological performance (see Fig.~\ref{fig:Critical_Metrology}). This resonates with the broader observation that the precision of critical metrology improves with system size, a trend we have previously discussed. However, an ideal scenario would involve increasing the signal while simultaneously minimizing—or at least maintaining—the noise. Striking this balance represents a compelling challenge and an exciting avenue for future research in critical metrology. Later in Section~\ref{sec:ultrastrong}, we will revisit this concept and explore strategies for optimizing the SNR in critical systems.

\subsection{Finite-size system phase transition}\label{sub_sec:finite_PT}

The energy gap can only fully close in the thermodynamic limit where the system size becomes infinite, a condition that cannot be realized in practical, finite-size systems. As a result, the energy gap remains slightly open, leading to deviations from the idealized harmonic oscillator picture---and Gaussianity---often used to describe the system near criticality. To account for these finite-size effects and the resulting modifications to the harmonic behavior, one can introduce a quartic correction to the harmonic oscillator model. This correction takes the form of an anharmonic term in the Hamiltonian:
\begin{align}
\label{quartic_resonator}
    \hat H = \frac{\hat p^2}{2} + \frac{\omega(g)^2}{2}\hat x^2 + \chi \hat x^4\;,
\end{align}
where $\chi$ represents the degree of nonlinearity introduced by the quartic term. The parameter $\chi$ reflects the influence of finite-size effects and the residual interactions that modify the harmonic potential. Such a Hamiltonian captures the essence of critical phenomena more accurately in realistic scenarios, as the quartic term softens the potential near criticality, leading to asymmetric energy levels and deviations from the purely parabolic profile of a harmonic oscillator. This correction can have significant implications for the system's dynamics, eigenstates, optimal measurements, and the scaling properties near the critical point, making it a crucial component in understanding quantum phase transitions in finite systems and limitations of critical metrology in finite systems.

\subsection{Lessons from the harmonic oscillator and the road ahead}
The harmonic oscillator analysis provides a powerful pedagogical springboard into the broader landscape of quantum critical metrology. While it captures many hallmark features---such as the closing of the energy gap, ground-state squeezing, and enhanced parameter susceptibility---it remains a single-mode, non-interacting system. As such, it lacks essential characteristics of genuine quantum phase transitions: collective behavior emerging from many-body interactions, universal scaling governed by critical exponents, and nontrivial dynamics under parameter ramps through critical points.

To begin addressing these aspects, we now take a modest but conceptually important step toward many-body physics by introducing the Landau-Zener model. Although still involving only two levels, this model captures the essential dynamical signature of criticality and provides a tractable setting to explore the interplay between critical dynamics and metrological sensitivity.  In the next section, we analyze the Landau-Zener problem through the lens of quantum metrology. This serves as a bridge between minimal single-mode models and genuinely many-body systems, such as the transverse-field Ising model, where gap closures occur in the thermodynamic limit and give rise to true quantum phase transitions.

\graphicspath{{./Landau_Zener_Model_Figs}}

\section{Quantum phase transitions, criticality, and critical metrology: elementary Landau-Zener model}\label{sec:QPT}

{Quantum phase transitions are a central concept in condensed matter physics and in the study of many-body quantum systems. Unlike classical phase transitions---such as the freezing of water into ice---that are driven by thermal fluctuations, quantum phase transitions occur at absolute zero temperature and are instead driven by quantum fluctuations. They take place when a control parameter (for instance, a magnetic field or an interaction strength between particles) is tuned, altering the balance between competing phases of the system. At the critical point the quantum fluctuations dominate and the system exhibits dramatic changes: long-range correlations can emerge, susceptibilities may diverge, and characteristic energy scales vanish. This reflects the fact that the system becomes extraordinarily sensitive to even infinitesimal changes in its parameters---a phenomenon known as quantum criticality. This extreme sensitivity due to quantum criticality, suggests that quantum phase transitions might become especially useful in quantum metrology. While critical sensitivity has long been exploited in classical contexts (for instance, in photodetectors or bubble chambers), the nonclassical correlations and enhanced fluctuations that arise near quantum critical points open fundamentally new possibilities for quantum-enhanced metrology by exploiting inherently non-classical features and collective behaviours. 

In this section, we introduce quantum phase transitions through the Landau-Zener model, which serves as a minimal, analytically tractable example capturing the essential features of critical behavior in a driven two-level system. In the subsequent section, we then move on to the transverse-field Ising model—a paradigmatic system exhibiting a genuine quantum phase transition in the thermodynamic limit. These models allow us to explore how criticality emerges in quantum systems and how the associated nonclassical features, such as diverging susceptibilities and long-range correlations, can be harnessed to enhance metrological precision
}

\subsection{Landau-Zener model}

The Landau-Zener (LZ) model describes a simple two-level quantum system driven by an external control field~\cite{zener1932non}. Due to its inherent simplicity, the LZ model does not exhibit genuine quantum criticality. Nevertheless, it serves as a valuable minimal model that captures several essential features commonly found near quantum critical points~\cite{innocenti2020ultrafast, zhang2009directcriticality}, offering useful insights into the mechanisms underpinning critical quantum metrology~\cite{Gietka2021adiabaticcritical, PhysRevA.96.020301}. The Hamiltonian of the model reads:
\begin{equation}
    \label{eq:lz_model_ham}
    \hat{H}_{\text{LZ}} = \frac{\omega}{2} \hat{\sigma}_z - \frac{g(t)}{2} \hat{\sigma}_x\;,
\end{equation}
where $\omega$ denotes the energy-level splitting in the absence of a time-dependent control field $g(t)$, and $\hat{\sigma}_\alpha$ ($\alpha = x, y, z$) are the Pauli matrices. In what follows, we restrict our attention to the equilibrium (adiabatic) regime where $g(t)$ varies slowly enough for the system to remain in its instantaneous ground state, and thus can be treated as a static parameter $g$.

When $g = 0$, the eigenstates are the bare spin states $|\!\downarrow\rangle$ and $|\!\uparrow\rangle$ with energies $\pm \omega/2$, respectively. If, in addition, $\omega = 0$, the system becomes exactly degenerate and a level crossing occurs. However, for any finite $\omega$, this degeneracy is lifted and an avoided crossing~\cite{zener1932non} emerges at $g = 0$, with the energy gap between ground (G) and excited (E) states reaching a minimum value of $\Delta = \omega$. The instantaneous eigenstates in this case are superpositions of the bare spin states, and the associated energy spectrum reads:
\begin{align}
    | \psi_{\text{G}} \rangle &= \cos{\left(\frac{\phi}{2}\right)}~| \!\downarrow \rangle + \sin{\left(\frac{\phi}{2}\right)}~| \!\uparrow \rangle, \quad E_{\text{G}} = -\tfrac{1}{2} \sqrt{g^2 + \omega^2}\;,\\
    | \psi_{\text{E}} \rangle &= \sin{\left(\frac{\phi}{2}\right)}~| \!\downarrow \rangle - \cos{\left(\frac{\phi}{2}\right)}~| \!\uparrow \rangle,  \quad E_{\text{E}} = +\tfrac{1}{2} \sqrt{g^2 + \omega^2}\;,
\end{align}
with the mixing angle defined as 
\begin{equation}
    \tan{\left(\frac{\phi}{2}\right)} = \frac{g + \sqrt{g^2 + \omega^2}}{\omega}\;.
\end{equation}
The energy gap is then given by
\begin{equation}
    \Delta = \sqrt{\omega^2 + g^2}\;.
\end{equation}

Although the LZ model consists of a single spin and lacks true many-body criticality, its behavior near the avoided crossing ($g=0$) closely mirrors key aspects of second-order quantum phase transitions. In particular, the energy gap $\Delta = \sqrt{g^2 + \omega^2}$ plays a role analogous to the excitation gap in many-body systems, which vanishes at the critical point in the thermodynamic limit. In this analogy, the parameter $\omega$ mimics a finite-size effect that prevents an actual gap closure—yet the system still undergoes a qualitative transformation around $g = 0$. 

Far from the avoided crossing ($|g| \gg \omega$), the transverse field dominates and the eigenstates align with the eigenstates of $\hat{\sigma}_x$. In contrast, near $g = 0$, the $\hat{\sigma}_z$ term becomes dominant, and the system closely approximates the bare $|\!\downarrow\rangle$ and $|\!\uparrow\rangle$ basis. As the control field $g$ is varied adiabatically from large negative to large positive values, the ground state evolves smoothly but significantly, interpolating between two qualitatively distinct regimes. This continuous yet abrupt transformation of the ground state is characteristic of a quantum phase transition. The \emph{phases} in this context correspond to regions where either the $\hat{\sigma}_x$ or $\hat{\sigma}_z$ term dominates, while the crossover near $g = 0$ serves as an analogue of a quantum critical point. Importantly, the sharpness of this transition depends on $\omega$. A small $\omega$ results in a narrow avoided crossing and a more abrupt change in the ground state, closely resembling a true quantum phase transition [see Figs.~\ref{fig:lz_back}(a) and~\ref{fig:lz_back}(b)]. Conversely, larger $\omega$ values smooth out the transition and suppress critical-like features. Despite its simplicity, the LZ model thus encapsulates many of the qualitative hallmarks of quantum criticality, making it a powerful pedagogical tool and a stepping stone toward understanding more complex models, such as the transverse-field Ising model.

\begin{figure*}[t!]
    \centering
    \includegraphics[width=1\linewidth]{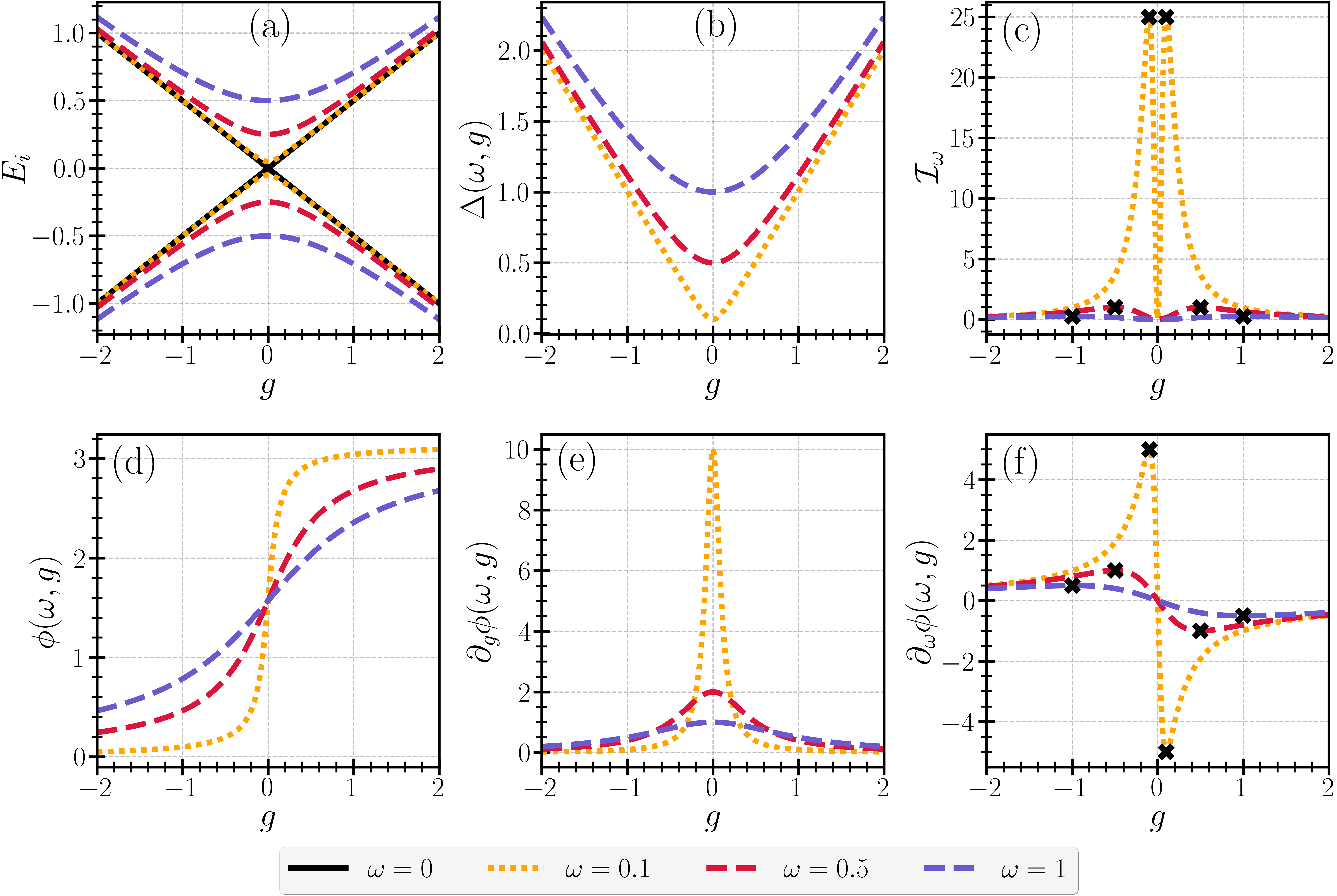}
\caption{ \textbf{Properties of the LZ model.} (a) Energy eigenvalues of the LZ model. (b) Energy gap $\Delta = \sqrt{\omega^2 + g^2}$ of the LZ model. (c) QFI for estimating $\omega$. (d) Geometric change of the angle $\phi$. (e) Rate of change of angle $\phi$ with respect to $g$, showing maximal rate of change when $\omega = g$, where the energy gap closes. (f) Rate of change of angle $\phi$ with respect to $\omega$ illustrating why the QFI does not peak at $\omega = g = 0$ (when the energy gap closes) but rather at a finite value corresponding to the point of maximal curvature of the eigenstate manifold. The black cross in (c) and (f) are to emphasize this point and represent the peak of the QFI and $\partial_\omega \phi$, respectively. }
    \label{fig:lz_back}
\end{figure*}

\subsection{Ground state sensitivity}\label{lz_subsec_gs_sens}

We now explore how the features discussed above influence the sensing capabilities of this simple two-level quantum system. Our metrological task is to estimate the frequency $\omega$, with the transverse field $g$ serving as a tunable control parameter. In the adiabatic regime, where the system remains in its instantaneous ground state, the optimal precision of estimating $\omega$ is quantified by the QFI, which for pure states is given by Eq.~\eqref{eq:intro_pure_state_qfi_def}. For the LZ model, we obtain
\begin{equation} \label{eq:QFI_LZ}
    \mathcal{I}_\omega = \frac{g^2}{\left(g^2 + \omega^2\right)^2}\;.
\end{equation}
Fig.~\ref{fig:lz_back}(c) shows the QFI as a function of $g$ for three different values of $\omega = \{0.1, 0.5, 1\}$, indicated by red, orange, and purple dashed lines, respectively. We observe that the QFI increases significantly as $\omega$ decreases, with the red curve ($\omega = 0.1$) reaching much higher values than the purple one ($\omega = 1$). This enhancement in sensitivity arises because the QFI scales inversely with the energy gap,
\begin{equation}
    \label{eq:LZQ_four}
    \mathcal{I}_\omega \propto \frac{1}{\Delta^4}\;,
\end{equation}
where $\Delta = \sqrt{g^2 + \omega^2}$ is the instantaneous energy gap. As $\omega \to 0$, the gap closes at $g = 0$, leading to a divergence in the QFI. This phenomenon is rooted in the general expression for the QFI in terms of $\omega$-dependent eigenstates $| \psi_n(\omega) \rangle$ and eigenvalues $E_n(\omega)$ of the Hamiltonian $\hat{H}(\omega)$:
\begin{equation}\label{eq:1overdeltasquaredqfi}
    \mathcal{I}_\omega = 4 \sum_{n \neq 0} \frac{|\langle \psi_n(\omega)| \partial_\omega \hat{H}(\omega) | \psi_0(\omega) \rangle |^2}{\left[E_n(\omega) - E_0(\omega)\right]^2} \propto \frac{1}{\Delta^2}\;.
\end{equation}
Thus, on a general level, the closing of the gap enhances parameter sensitivity, making it a metrological resource. In the specific case of the LZ model, the dependence of the eigenstates $|\psi_n(\omega) \rangle$ on $\omega$ leads to an even stronger divergence as noted in Eq.~\eqref{eq:LZQ_four}, further amplifying the QFI. This behaviour mimics a fundamental characteristic of many-body quantum phase transitions, where the gap closes at the critical point.

\subsection{Ground state geometry}
This critical-like behavior is also reflected in the geometry of the ground state. In the LZ model, the instantaneous ground state is parametrized by an angle $\phi(\omega, g)$, which determines its orientation on the Bloch sphere. As $g$ is varied adiabatically, the ground state traces out a continuous trajectory on the sphere, transitioning from alignment with the $-\hat{\sigma}_x$ eigenstate at large negative $g$ to the $+\hat{\sigma}_x$ eigenstate at large positive $g$. Figure~\ref{fig:lz_back}(d) illustrates this geometric evolution for several values of $\omega$. As $\omega \to 0$, the energy gap at $g = 0$ vanishes, and this geometric path becomes increasingly sharp. The transition in the state's orientation becomes nearly discontinuous, indicating that even infinitesimal changes in $g$ or $\omega$ near the avoided crossing cause drastic changes in the quantum state. 

Conversely, far from the crossing ($|g| \gg \omega$), the state aligns with the $\hat{\sigma}_x$ basis, becoming largely insensitive to $\omega$, which corresponds to a suppressed QFI, as seen in Fig.~\ref{fig:lz_back}(c). The sharpness of the ground state transformation can be quantified by the rate of change $\partial_\omega \phi(\omega, g)$, which measures how rapidly the ground state rotates on the Bloch sphere as $\omega$ is tuned. This rate is plotted in Fig.~\ref{fig:lz_back}(f) for various values of $\omega$, showing that it peaks at $\omega = g$ and diverges as $g \to 0$. A divergent $\partial_\omega \phi$ indicates that infinitesimal changes in the control field lead to macroscopic changes in the quantum state. This mirrors the divergence of physical susceptibilities---such as magnetization---observed at quantum critical points in many-body systems. In this analogy, $\partial_\omega \phi$ plays the role of a generalized susceptibility that directly quantifies the fragility of the ground state to control parameter variations. A similar behavior is observed in the derivative with respect to $g$ [see Fig.~\ref{fig:lz_back}(e)], $\partial_g \phi(\omega, g)$, which becomes sharply peaked near $g = 0$ and diverges as $\omega \to 0$. In contrast to $\partial_\omega \phi$, which is relevant when $\omega$ is estimated and $g$ is known, the derivative $\partial_g \phi$ quantifies sensitivity when $g$ is the parameter to be estimated. Both derivatives exhibit critical-like features and serve as geometric witnesses of enhanced quantum sensitivity.

The location of the highest rate of change, maximized at $g = \omega = 0$, highlights the avoided crossing as the point where the state transformation is most dramatic. As $\omega$ decreases, the avoided crossing sharpens, the ground state transition becomes more abrupt, and the peak in $\partial_\omega \phi(\omega, g)$ grows. From a metrological perspective, this is of great importance. Quantum estimation relies on how distinguishable quantum states become under small changes in parameters, and the divergence in $\partial_\omega \phi$ signals precisely such a regime of enhanced sensitivity. This underscores the importance of state geometry as a resource for quantum metrology and connects the enhanced estimation precision near criticality to the structure of quantum state space itself.

\subsection{Geometric tensor and quantum Fisher information}\label{sec_lz_subsec_qgt}
To formalize the notion of parameter sensitivity in quantum states, we introduce the \emph{quantum geometric tensor}~\cite{zanardi2007criticalscaling}. For a pure state $\ket{\psi}$, the quantum geometric tensor is defined as
\begin{equation}
\chi_{\mu\nu} = \langle{\partial_\mu \psi | \partial_\nu \psi}\rangle - \langle{\partial_\mu \psi | \psi}\rangle \langle{\psi | \partial_\nu \psi}\rangle\;.
\end{equation}
Its real part defines a Riemannian metric $g_{\mu\nu} = \mathrm{Re}[\chi_{\mu\nu}]$, which characterizes the infinitesimal Bures distance between nearby quantum states
\begin{equation}
ds^2 = g_{\mu\nu} \, d\lambda^\mu d\lambda^\nu\;,
\end{equation}
where $\lambda^\mu$ with $\mu=(1,2,3,\ldots)$ denotes the set of parameters the quantum state depends on. 

In the case of single-parameter estimation, the QFI is simply related to the metric component~\cite{caves1994statisticaldistance,PhysRevD.23.357,uffink}
\begin{equation}
\mathcal{I}_\lambda = 4 g_{\lambda\lambda} = 4 \, \mathrm{Re}[\chi_{\lambda\lambda}]\;.
\end{equation}
This geometric perspective reveals that the QFI quantifies the \emph{curvature} of the quantum state manifold—it measures how quickly the state changes with respect to the parameter $\lambda$. In the context of the LZ model, we consider estimating the frequency $\omega$. Therefore, the relevant quantity is the rate of change of the ground state with respect to $\omega$, i.e., $\partial_\omega \ket{\phi(\omega,g)}$. In Fig.~\ref{fig:lz_back}(f), this derivative reaches its maximum precisely at $g = \omega$, coinciding with the peak of the QFI (shown in the black crosses respectively). This is no coincidence: the QFI is proportional to the Bures metric, and hence directly reflects the distinguishability between infinitesimally close quantum states. The point $g = \omega$ corresponds to the region of maximal curvature in the ground state manifold, where the trajectory of $\ket{\phi(\omega,g)}$ in Hilbert space bends most sharply. From a geometric point of view, this is the \emph{sweet spot} for parameter estimation—where the sensitivity to $\omega$ is greatest. This interpretation also explains why, in the LZ model, the QFI scales as $1/\Delta^4$ rather than $1/\Delta^2$ [see Eq.~\eqref{eq:1overdeltasquaredqfi}], where $\Delta$ denotes the energy gap. In the LZ model, the eigenstates themselves depend non-trivially on $\omega$, and their curvature in parameter space leads to an \emph{additional} enhancement. That is, the geometry of the Hilbert space contributes a second layer of sensitivity beyond the energetic features alone.

In summary, while the closing of the energy gap at $g = 0$ reflects critical-like behavior and induces a diverging response with respect to the control parameter $g$, the optimal sensitivity to the estimated parameter $\omega$ occurs at $g = \omega$, where the ground state manifold exhibits maximal geometric curvature. This highlights a broader principle in quantum metrology: optimal precision is not solely dictated by spectral properties such as gap closings, but also by the underlying differential geometry of the quantum state space.

\subsection{Time considerations}
Let us now revisit the expression for the QFI at the optimal point \(g = \omega\), where  
\begin{align} 
    \label{eq:LZ_time}
    \mathcal{I}_\omega = \frac{1}{4 \omega^2}\;.  
\end{align}  
Interestingly, this result seemingly surpasses the standard quantum limit, which for a single particle and evolution time \(t\) is given by
\begin{align}
    \mathcal{I}_\omega = t^2\;.  
\end{align}
Indeed, when \(\omega < 1 / (2t)\), the QFI in our setup appears to exceed what would be possible even with the best unitary evolution governed by a time-independent Hamiltonian. How can this enhancement occur in a single-particle system?

The apparent paradox arises because we have not yet accounted for all physical resources. In particular we have omitted the \emph{time} required to prepare the optimal probe state. Specifically, the expression Eq.~\eqref{eq:LZ_time} assumes that the system is initialized in the ground state at the optimal point \(g = \omega\), which depends on the parameter \(\omega\) to be estimated. However, to avoid circular logic in estimation theory, the probe state must initially be independent of the target parameter \(\omega\). Therefore, one must consider a complete protocol in which the system is first prepared in a known ground state---for example, at \(g_0 \gg \omega\), where the ground state is simply \(|\!\downarrow\rangle\)---and then adiabatically brought to the critical region \(g_f \sim \omega\), where the sensitivity is maximal.

To remain in the ground state throughout this process, the control field \(g(t)\) must be changed slowly enough to satisfy the adiabatic condition. According to the adiabatic theorem, the characteristic timescale for adiabatic evolution is inversely proportional to the square of the minimum energy gap encountered during the protocol. In the LZ Hamiltonian, the gap closes as \(\omega \to 0\), leading to a divergent timescale
\begin{align}
    T_{\text{adiabatic}} \propto \frac{1}{\Delta_{\text{min}}^2}\;.
\end{align}
This divergence is a manifestation of the general phenomenon known as \emph{critical slowing down}---near a quantum critical point, the energy gap closes and the system requires increasingly more time to equilibrate. As a result, the time required for faithful adiabatic state preparation diverges at the critical point. To be more specific, the adiabatic ramp can be chosen as  
\begin{align}  
    g(t) =\frac{\omega  (1-\gamma   \omega t)}{\sqrt{\gamma   \omega  t(2-\gamma  \omega t )}}\;,  
\end{align}  
where \(\gamma \ll 1\) is a small dimensionless parameter ensuring adiabaticity. The total time to evolve from \(g_0\) to \(g_f \sim \omega\) is then approximately  
\begin{align} \label{eq:LZtime}
    T = \frac{2-\sqrt{2}}{2 \gamma  \omega } \approx \frac{0.3}{\gamma \omega}\;.  
\end{align}
Using this expression, we can rewrite the QFI at the optimal point \(g_f \sim \omega\) as  
\begin{align}  
    \mathcal{I}_\omega \approx 2.8 T^2 \gamma^2\;.  
\end{align}
This shows that the QFI indeed scales quadratically with the total protocol time, consistent with the standard quantum limit for unitary dynamics. While the large prefactor might suggest an advantage, any attempt to push beyond the standard limit requires increasing \(\gamma\), which would violate the adiabatic condition. Hence, no contradiction arises. The apparent metrological enhancement is neutralized once the full resource cost---especially time---is properly accounted for.

This analysis highlights a central feature of critical metrology: although critical points provide enhanced static sensitivity, they are also accompanied by intrinsically slow dynamics. As the energy gap closes near criticality, the timescale required for adiabatic preparation diverges (critical slowing down). Thus, the apparent gain in sensitivity comes at the cost of longer preparation times. This trade-off encapsulates a deeper principle of critical quantum metrology: while criticality enhances susceptibility to parameter changes, the precision attainable in practice is fundamentally limited by the dynamical constraints imposed by the closing gap.

\subsection{Thermal fluctuations}\label{subsec_sp_thermal}

\begin{figure*}[ht!]
    \centering
    \includegraphics[width=1\linewidth]{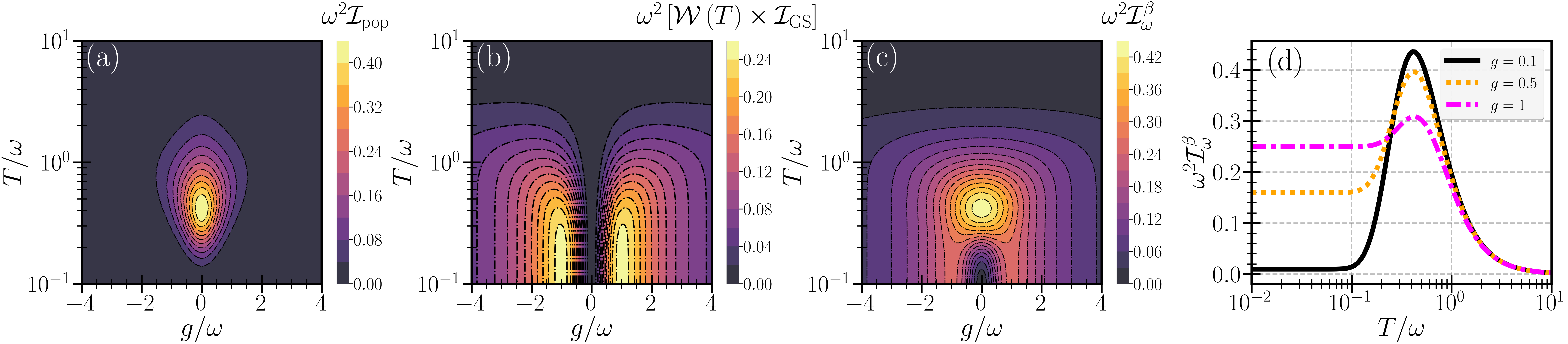}
    \caption{ \textbf{Single-parameter estimation at finite temperature for the LZ model.} (a) Contribution of the thermal populations to probe parameter sensitivity $\mathcal{I}_{\text{pop}}$. (b) Contribution of the temperature-dependent weighting factor $\mathcal{W}(T)$ and ground state sensitivity $\mathcal{I}_{\text{GS}}$. (c) Total finite-temperature QFI for estimating \( \omega \). (d) Comparison of the QFI for several values of \( g \). In all panels we set $\omega=1$.} 
    \label{fig:lz_thermal_fluc}
\end{figure*}

Thus far, our analysis has focused on the ground state properties of the system, assuming a zero-temperature environment. However, in realistic scenarios, isolating a quantum system from its thermal surroundings is often challenging, and residual thermal fluctuations may influence its behavior. These fluctuations are typically viewed as detrimental to quantum resources, introducing decoherence and eroding metrological advantages. Yet, when considering quantum critical metrology—where enhanced sensitivity arises from proximity to a quantum phase transition—the role of thermal effects becomes more subtle~\cite{abiuso2024fundamental}. In the vicinity of a critical point, quantum fluctuations dominate within a finite temperature window. If thermal fluctuations remain small compared to the characteristic energy gap, the system retains critical behavior and associated sensitivity. Conversely, if thermal fluctuations become dominant, the system is driven away from criticality, and metrological advantages may be lost. Understanding this interplay between quantum and thermal fluctuations is essential for assessing the robustness and practical viability of critical metrology. We now explore this interplay concretely using the LZ model.

At finite temperature, the system is no longer in a pure ground state, but is instead described by a thermal Gibbs state
\begin{equation}
    \hat{\varrho} = \sum_n e_n | e_n \rangle \langle e_n | \;,
\end{equation}
where the populations \( e_n \) are given by the normalized Boltzmann weights:
\begin{equation}
    e_n = \frac{e^{-E_n/k_B T }}{\mathcal{Z}} \;, \quad \text{with} \quad \mathcal{Z} = \sum_n e^{-E_n/k_BT } \;,
\end{equation}
where \( \hat{H}|e_n\rangle = E_n |e_n\rangle \) defines the system’s eigenbasis. Throughout the manuscript, we use natural units with \( k_B = 1\). At non-zero temperature \(T\), information about the parameter \( \omega \) can be imprinted both in the thermal populations \( e_n \) and in the structure of the eigenstates \( |e_n\rangle \). 

To quantify the sensitivity of the thermal state to changes in \( \omega \), we examine the QFI \( \mathcal{I}^\beta_\omega \), which at finite temperature decomposes into two contributions:
\begin{align}
    \mathcal{I}_\omega^\beta &= \underbrace{\sum_n\frac{(\partial_\omega e_n)^2}{e_n}}_{\text{Population term}} + \underbrace{2 \sum_{n\neq m}\frac{(e_m - e_n)^2}{e_m + e_n}|\langle e_m|\partial_\omega e_n\rangle|^2}_{\text{Coherence term}} \nonumber \\
    &= \underbrace{\frac{\omega^2\, \text{sech}^2\left(\frac{\Delta}{2T}\right)}{4 T^2 \Delta^2}}_{\mathcal{I}_\text{pop}} + \underbrace{\tanh^2\left(\frac{\Delta}{2T}\right)}_{\mathcal{W}(T)} \underbrace{\left[ \frac{g^2}{(g^2 + \omega^2)^2} \right]}_{\mathcal{I}_\text{GS}}\;,\label{LZ:totalQFI:finiteT}
\end{align}
where \( \Delta = \sqrt{g^2 + \omega^2} \) is the energy gap. This decomposition highlights the distinct physical origins of thermal sensitivity for the LZ model:
\begin{itemize}
    \item \( \mathcal{I}_\text{pop} \): the population term, which quantifies sensitivity arising from changes in the thermal distribution.
    \item \( \mathcal{W}(T) \): a temperature-dependent weighting factor that modulates the contribution from the coherence term.
    \item \( \mathcal{I}_\text{GS} \): the zero-temperature ground state QFI, which encodes the intrinsic sensitivity of the energy eigenbasis.
\end{itemize}

This structure reveals that finite temperature opens new pathways for parameter sensitivity~\cite{PhysRevLett.133.040802,gietka2024tempcqm}. At zero temperature, \( \mathcal{W}(T) \to 1 \) and \( \mathcal{I}_\text{pop} \to 0 \), recovering the ground state QFI. For small but finite temperatures, thermal excitations begin to populate higher eigenstates, introducing additional contributions to sensitivity via both terms. Interestingly, in this intermediate regime, the total QFI can exceed its zero-temperature value. In the high-temperature limit \( T \to \infty \), all populations become uniform, and both terms vanish: \( \mathcal{I}^\beta_\omega \to 0 \). This reflects the loss of information in a maximally mixed state, where the system no longer responds to changes in \( \omega \).

These features are visualized in Fig.~\ref{fig:lz_thermal_fluc}. Panel (a) shows the contribution of the population dependant term of the total QFI, \( \mathcal{I}_{\text{pop}} \), reflecting the degree of information now imprinted on the thermal populations. This quantity peaks at intermediate temperatures \( T \sim \Delta \) and small \( g \), where both population imbalance and coherent transitions are significant. The maximum occurs at $g = 0$ where the crossing occurs and the population change is maximal. Notably, this prefactor vanishes in the high-temperature limit, where thermal populations flatten and coherence is suppressed. Panel (b) illustrates the impact the thermal weighting factor \( \mathcal{W}(T) \) has on the ground state sensitivity \( \mathcal{I}_{\text{GS}} \). In contrast with the population-dependent thermal contribution, here the sensitivity peaks around $\omega = g$ in line with the zero-temperature limit. The sensitivity deteriorates from its maximum as the thermal fluctuations being to dominate.  

In panel (c) the full finite-temperature QFI \( \mathcal{I}^\beta_\omega \) is displayed, revealing a broad sensitivity maximum at \( g \approx 0 \) and \( T \sim \Delta \), precisely where thermal and quantum effects optimally combine. This highlights that temperature does not merely degrade sensitivity; it can in fact enable new metrological channels if harnessed appropriately. Panel (d) further emphasizes this by plotting the temperature dependence of QFI for three values of \( g=\{0.1,0.5,1\} \). In all cases, the QFI approaches the ground state value as \( T \to 0 \), then exhibits an enhancement in the intermediate-temperature regime before decaying to zero at high temperatures.

Taken together, these results demonstrate that optimal sensitivity does not occur at zero temperature. Instead, a finely tuned balance between quantum coherence and thermal activation can yield superior precision. This reveals a key insight in critical metrology: thermal fluctuations, typically considered detrimental, can under specific conditions act as a resource—particularly near critical points, where energy gaps are small and sensitivity is high.

\subsection{Accessible measurements}\label{lz_subsec_accessible}
Thus far, our discussion has revolved around the optimal attainable precision---as quantified by the QFI. This assumes that we are performing measurements in the optimal basis, allowing us to extract maximum information about the parameter of interest $\omega$ and lower bound the precision through the QCRB. However, the optimal measurement might not always be a practically implementable observable, making it important to also consider more feasible measurement strategies. The LZ model is a simple two-level system, and therefore the choice of measurement basis is highly restricted: it consists only of two orthonormal Bloch vectors. 

A natural choice of measurement basis is that formed by the spin-up $|\!\uparrow\rangle$ and spin-down $|\!\downarrow\rangle$ states, which corresponds to the eigenstates of the $\hat{\sigma}_z$ operator. We now examine the metrological performance associated with this observable, analysing its sensitivity at both zero and finite temperature. This allows for a direct comparison with Sections~\ref{lz_subsec_gs_sens} and~\ref{subsec_sp_thermal}, where we considered the optimal sensitivity. We first examine the zero-temperature scenario. For this choice of observable, we obtain on average a measurement signal of
\begin{align}
    \langle \hat{\sigma}_z \rangle = \frac{-\omega}{\sqrt{g^2 + \omega^2}}\;,
\end{align} 
whose corresponding quantum fluctuations are given by
\begin{align}
    (\Delta \hat{\sigma}_z)^2 = 1 - \langle \hat{\sigma}_z \rangle^2 = \frac{g^2}{g^2 + \omega^2}\;.
\end{align} 
The sensitivity can thus be quantified using the SNR as
\begin{align}
     S_\omega=  \frac{\left( \partial_\omega \langle \hat{\sigma}_z \rangle \right)^2}{\left(\Delta  \hat{\sigma}_z \right)^2} = \frac{g^2}{\left(g^2 + \omega^2\right)^2}\;.
\end{align}  
Interestingly, the SNR provides the same precision as the QFI, $S_\omega = \mathcal{I}_\omega$. This implies that the transverse magnetization constitutes an optimal measurement at zero-temperature. Physically, we can understand that at $T = 0$, the Bloch vector points around $\vec{n} = (g,0,\omega)/\Delta$ so changing $\omega$ tilts the state within the ${x}$-${z}$ plane, and $\hat{\sigma}_z$ is capable of reading that out with maximal efficiency. 

A natural question to ask is if such a measurement remains optimal when thermal fluctuations are introduced. As before, the probe state is now described by the canonical Gibbs state $\hat{\varrho} = e^{-\hat{H}/T}/\mathcal{Z}$, rather than a pure ground state as in the zero-temperature case. By measuring the transverse magnetization at finite temperature, we obtain on average a measurement signal of  
\begin{equation}
    \langle \hat{\sigma}_z\rangle_\beta = -\frac{\omega}{\Delta} \tanh{\left(\Delta/2T\right)}\;,
\end{equation}
with corresponding fluctuations
\begin{equation}
    (\Delta \hat{\sigma}_z )^2_\beta = \frac{1}{\Delta^2} \left[g^2 +\omega^2\text{sech}^2\left(\Delta/2T\right)\right]\;.
\end{equation}
This leads to the SNR expression:
\begin{equation}
    S_\omega^\beta = \frac{\text{sech}^2\left(\Delta/2T\right)\left[ \omega^2 \Delta + g^2 T \sinh{(\Delta/T)}\right]^2}{2T^2 \Delta^4\left[g^2 + 2\omega^2 + g^2 \cosh{(\Delta/T)} \right]}\;,
\end{equation}
where $\Delta = \sqrt{\omega^2 + g^2}$ is the energy gap. Although not immediately obvious from the above expression, the resulting SNR is lower-bounded by the QFI of the thermal state 
\begin{equation}
    S_\omega^\beta \leq  \mathcal{I}_\omega^\beta\;.
\end{equation}
This means that through the inclusion of thermal fluctuations, the transverse magnetization no longer constitutes an optimal measurement---despite being optimal at zero-temperature. This highlights the subtle nature of how external factors can influence precision measurements, and that strategies yielding optimal performance at zero temperature do not necessarily extend to finite-temperature regimes. In this case, the optimal measurement is one that fully exploits both the population imbalance and the coherences present in the mixed state.

The transverse magnetization suffers in this regard, as it cannot fully access the geometric component of mixed states in the presence of thermal fluctuations, thereby becoming a suboptimal measurement in this regime. This can be seen explicitly in Fig.~\ref{fig:lz_snr_qfi_thermal}, where we compare the SNR to the QFI. For $T\!\to\!0$, when the system is described by the ground state, and for $T\!\to\!\infty$, when the thermal state becomes maximally mixed, the precision of both strategies is comparable. In contrast, in the intermediate regime, where the QFI is enhanced, the transverse magnetization is not capable of fully exploiting the features necessary to attain optimal precision. 

\begin{figure}[t]
    \centering\includegraphics[width=\linewidth]{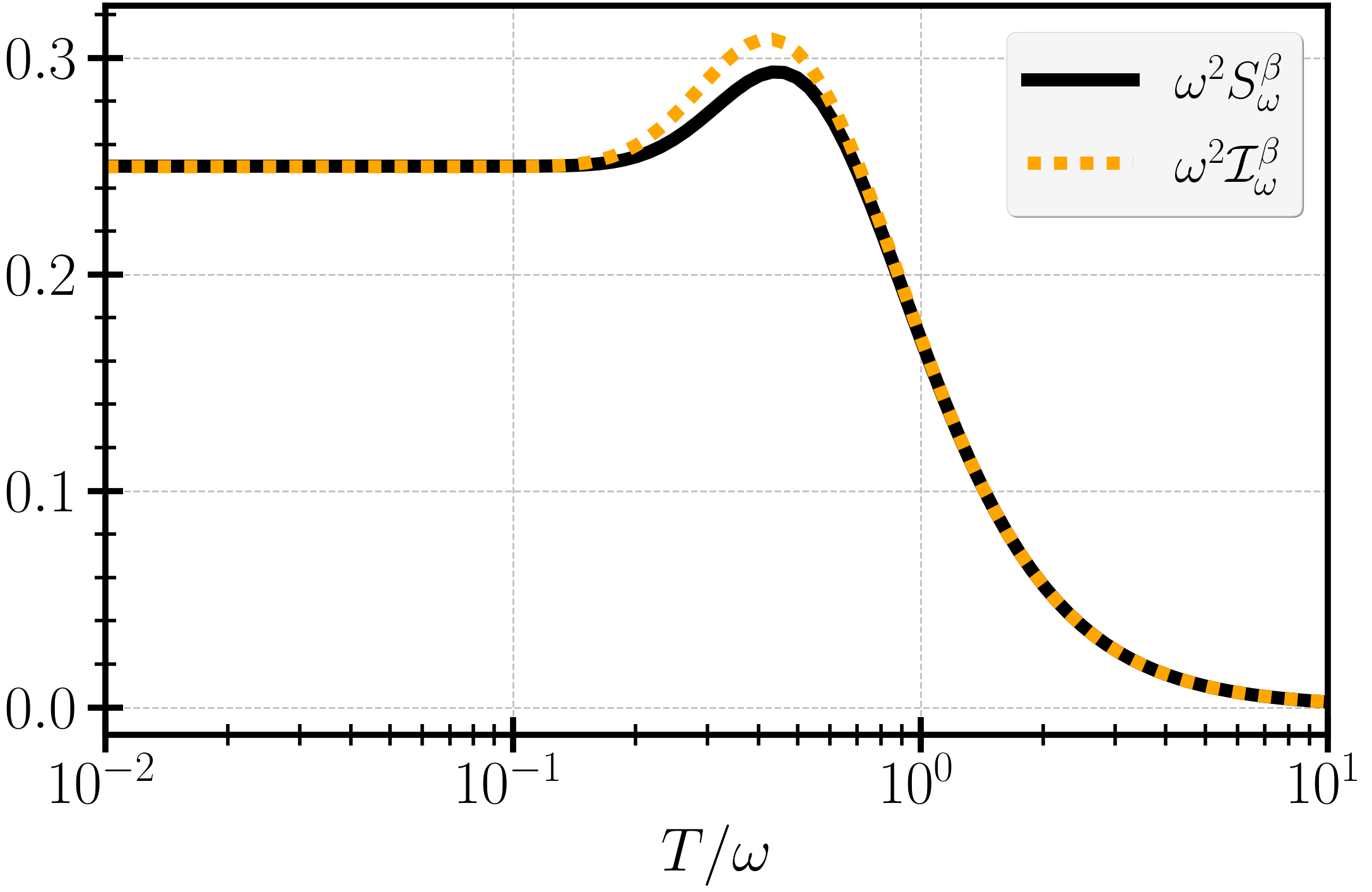}
    \caption{\textbf{Comparison of optimal and experimentally accessible sensitivities in the LZ model at finite temperature.} The QFI \( \mathcal{I}_\omega^\beta\) is shown by the dotted yellow line. The SNR obtained from measurements in the $\hat{\sigma}_z$ basis is shown by the black solid line. The latter constitutes a suboptimal measurement, not capable of fully exploiting the thermal fluctuations. In both cases, the frequency is set to $\omega = 1$.}
    \label{fig:lz_snr_qfi_thermal}
\end{figure}
On a more general footing, an important caveat arises when adopting $\hat{\sigma}_z$ as the measurement observable. We are assuming we have the freedom to choose the measurement basis. However, in many practical scenarios such freedom is not readily available. Often, the only viable option is to measure in the energy eigenbasis. At first glance, this seems problematic since, in that basis, the Hamiltonian takes the diagonal form
\begin{align}
    \hat{H} = \frac{1}{2} \sqrt{g^2 + \omega^2} \hat{\sigma}_z\;.
\end{align}  
This implies that the population-difference signal is always constant---regardless of the parameters values. Equivalently, this is analogous to stating the system is simply in its ground state, which offers no direct information about the parameters. In such situations, one can instead perform Ramsey interferometry, which measures the energy separation between the two levels. In the rotating frame of the driving laser and following the Ramsey sequence, the signal becomes  
\begin{align}
    \langle \hat{\sigma}_z \rangle = \sin(\delta t)\;,
\end{align}  
where $\delta = \omega_l - \sqrt{g^2 + \omega^2}$ is the detuning between the laser frequency $\omega_l$ and the transition frequency. The corresponding SNR is given by 
\begin{align}
     S_\omega = \frac{t^2 \omega^2 \cos^2(\delta t)}{g^2 + \omega^2} \leq \frac{t^2 \omega^2}{g^2 + \omega^2} \leq t^2\;.
\end{align}  
The SNR is maximized when $g = 0$, which stands in contrast to the optimal condition derived from Eq.~\eqref{eq:QFI_LZ}. This discrepancy highlights again an important point: for few-body systems, critical quantum metrology does not inherently provide any enhancement over the standard quantum limit. These results will serve as a baseline for the following sections, where we explore many-body quantum systems that exhibit genuine quantum phase transitions.

\subsection{Nuisance parameters and multiparameter estimation}\label{sec_LZ_MP}

Critical metrology typically assumes that only one parameter is unknown, while all others are precisely controlled. This assumption is crucial. Near critical points, quantum systems become extremely sensitive not just to the parameter of interest, but to any other parameter which can drive a quantum phase transition. In practice, perfect knowledge of all non-target parameters is often unrealistic. Experimental imprecision or calibration limits naturally introduce uncertainty in these additional quantities, which can be viewed as nuisance parameters when there is ambiguity in their characterization~\cite{suzuki2020quantum}. Such imprecisions on the level of non-target and control parameters used to tune the system toward criticality can be especially problematic, as proximity to a critical point is essential for exploiting the system's diverging susceptibility~\cite{PhysRevA.111.052621,abol2021criticalityglobalsensing,xbrk-357g,81vb-hf2f}.  This poses a significant challenge for multiparameter estimation in critical systems. In cases where only two parameters are involved, as in the Landau–Zener model, treating both as unknown is often impractical. Critical metrology relies on having at least one well-controlled, tunable parameter to bring the system close to criticality. Without such a reference parameter, the notion of multiparameter estimation loses meaning, as the system’s extreme sensitivity prevents independent determination of both parameters simultaneously.

In previous sections, we focused on estimating the energy splitting $\omega$, assuming perfect knowledge of the control parameter $g$. To explore the limitations of this assumption, we now treat both $\omega$ and $g$ as unknown parameters. This generalization allows us to study the fundamental limits of multiparameter estimation near criticality. The QFIM for the LZ ground state is then given by:
\begin{equation}
    \hat{\mathcal{I}}_{\text{LZ}} = \begin{pmatrix}
        \mathcal{I}_{\omega\omega} & \mathcal{I}_{\omega g} \\ 
        \mathcal{I}_{g \omega} & \mathcal{I}_{gg}
    \end{pmatrix} = \begin{pmatrix}
        \frac{g^2}{\left( g^2 + \omega^2\right)^2} & -\frac{g \omega}{\left( g^2 + \omega^2\right)^2} \\ 
        -\frac{g \omega}{\left( g^2 + \omega^2\right)^2} & \frac{\omega^2}{\left( g^2 + \omega^2\right)^2}
    \end{pmatrix}\;.
\end{equation}
This matrix is clearly singular, since its determinant vanishes:
\begin{equation}
    \det\left[\hat{\mathcal{I}}_{\text{LZ}}\right] = \mathcal{I}_{\omega \omega} \mathcal{I}_{gg} - \mathcal{I}_{\omega g}^2 = 0\;.
\end{equation}
The singularity of the QFIM has deep implications: it indicates that $\omega$ and $g$ are statistically dependent, meaning that variations in one parameter can be exactly mimicked by variations in the other. Consequently, the measurement outcomes do not contain enough independent information to resolve both parameters simultaneously. In this case, the inverse QFIM, which determines the QCRB for the covariance matrix of any unbiased estimator, does not exist. As a result, there is no unbiased estimation strategy that can achieve finite precision for both parameters at the same time. At best, one can estimate a single \emph{effective} parameter—some linear combination of $\omega$ and $g$—with finite precision~\cite{Rubio_2020,PhysRevResearch.3.033011,proctor2017networkedquantumsensing,Seveso_2020,PhysRevLett.120.080501,frigerio2024overcomingsloppinessenhancedmetrology,tsang2020quantum,suzuki2020quantum,niu2025rolelongrangeinteractioncritical}. This behaviour is a hallmark of critical metrology with renormalization group ($RG$) relevant parameters in the multiparameter setting~\cite{mihailescu2023multiparameter}: the extreme sensitivity of the ground state enhances precision for one parameter, but at the cost of losing identifiability of the others. This is because $RG$-relevant parameters (those that control the universality class, like the control parameter $g$ driving the transition, or an external field breaking symmetry) are imprinted in the quantum-state in a universal way, typically through scaling functions of the generalized form~\cite{PhysicsPhysiqueFizika.2.263,wilson1975renormalization,PhysRevLett.28.240,sachdev1999quantum,DeGrandi2010,gritsev2009universal}
\begin{equation}
    \chi(\omega,N,T) \sim N^\alpha f\big( N^{1/\nu}(g -g_c), N T^z \big)\;,
\end{equation}
with $\chi$ a generalized susceptibility, $f$ a universal scaling function which depends on the system, $N$ the system size, $\alpha$ a critical exponent describing how the susceptibility $\chi$ diverges, $\nu$ the correlation length critical exponent, $T$ the temperature, and $z$ the dynamic critical exponent. 

Interestingly, this limitation is not unique to the LZ model or truly critical systems. A similar issue arises in paradigms such as Ramsey interferometry, where treating the interrogation time~$t$ as an unknown parameter likewise leads to a singular QFIM. In general, the fragility of the ground state—which enables enhanced sensitivity near critical points—also becomes a source of ambiguity when multiple parameters are simultaneously estimated. Understanding and managing this trade-off is central to the design of practical critical quantum sensors. One way to resolve this impasse is to assume that at least one $RG$-relevant parameter is known and controlled~\cite{mihailescu2023multiparameter,mihailescu2025metrologicalsymmetriessingularquantum}. Doing so immediately lifts the singularity of the QFIM and restores a well-posed estimation problem. In the following, we illustrate this strategy for the Landau–Zener model by introducing temperature as a third, known parameter, which regularizes the QFIM and enables meaningful multiparameter estimation. An alternative approach is to reformulate the problem in terms of effective parameters—linear combinations of the original parameters that can be estimated with finite precision even when the full set remains correlated.

\subsection{Effective parameter in multiparameter estimation}
But what is the effective parameter, and how do we identify it? Effective parameters arise when there are functional relationships or constraints between the parameters being estimated. These constraints give rise to what can be termed \emph{metrological symmetries}---situations where only specific combinations of parameters are physically distinguishable through measurement~\cite{mihailescu2025metrologicalsymmetriessingularquantum}. When such symmetries exist, the singularity of the QFIM reflects the fact that we are attempting to estimate more parameters than the quantum state can resolve. In other words, the estimation problem becomes overparameterized at the metrological level. This overparameterization results in a loss of identifiability for individual parameters, as the quantum state only encodes information about certain combinations, not each parameter independently (see Fig.~\ref{fig:twoimages} for a schematic depiction of overparameterization leading to QFIM singularity)~\cite{PhysRevResearch.7.023060,He_2025,frigerio2024overcomingsloppinessenhancedmetrology,taming_singular,PhysRevA.95.052320,Seveso_2020,PhysRevLett.127.110501,PhysRevA.106.022429,conlon2024roleextendedhilbertspace,candeloro2024dimension}. 

In the case of the LZ model, we can intuitively discern the effective parameter encoding on the level of the Hamiltonian. In particular, we note that the energy splitting $\omega$ sets the effective energy scale of the system. By rescaling the Hamiltonian accordingly we obtain
\begin{align}
    \label{eq:lz_reparam}
    \hat{H}'_{\text{LZ}} = \frac{\hat{H}_{\text{LZ}}}{\omega} &= \frac{1}{2}\hat{\sigma}_z - \frac{g}{2 \omega} \hat{\sigma}_x\;,
\end{align}
with the rescaled Hamiltonian $\hat{H}_{\text{LZ}}'$ being dimensionless and $\omega$ being treated as a unit of energy. Correspondingly, we can deduce that the system can be fully specified by the ratio of parameters $\Omega = g/\omega$. It then becomes obvious that $g$ and $\omega$ cannot be independently estimated from pure states. This is because no measurements using such a state can distinguish between different combinations of the parameters, $(\omega,g)$ and $(2\omega,2g)$ for example, as the ratio of the parameters is the same in both instances, and the state depends only on this ratio. Therefore, while true multiparameter estimation is prohibited, the appropriate object which can be estimated in this case is the single effective parameter $\Omega = g/\omega$. 

Using the rescaled Hamiltonian $\hat{H}_{\text{LZ}}'$, the effective single parameter QFI can be computed for the ground state, yielding
\begin{equation}
    \mathcal{I}_\Omega = \frac{1}{\big(1+\Omega^2\big)^2} \equiv \frac{\omega^4}{\big(\omega^2 + g^2\big)^2}\;,
\end{equation}
where in the second equality we express the effective QFI in terms of the original bare parameters. In the more general case, discerning the nature of effective parameters and their corresponding precision is not as straightforward. This is especially true in complicated many-body systems and in systems undergoing phase transitions. In the latter case, parameter dependencies may naturally emerge in the vicinity of critical points, where universal physics is encoded in order parameters that combine the bare Hamiltonian parameters in non-trivial ways. Moreover, emergent Hamiltonian symmetries at criticality—which are absent away from the transition—may impose additional constraints on the system, potentially giving rise to such metrological symmetries.

\begin{figure}[t!]
    \centering
    \includegraphics[width=1\linewidth]{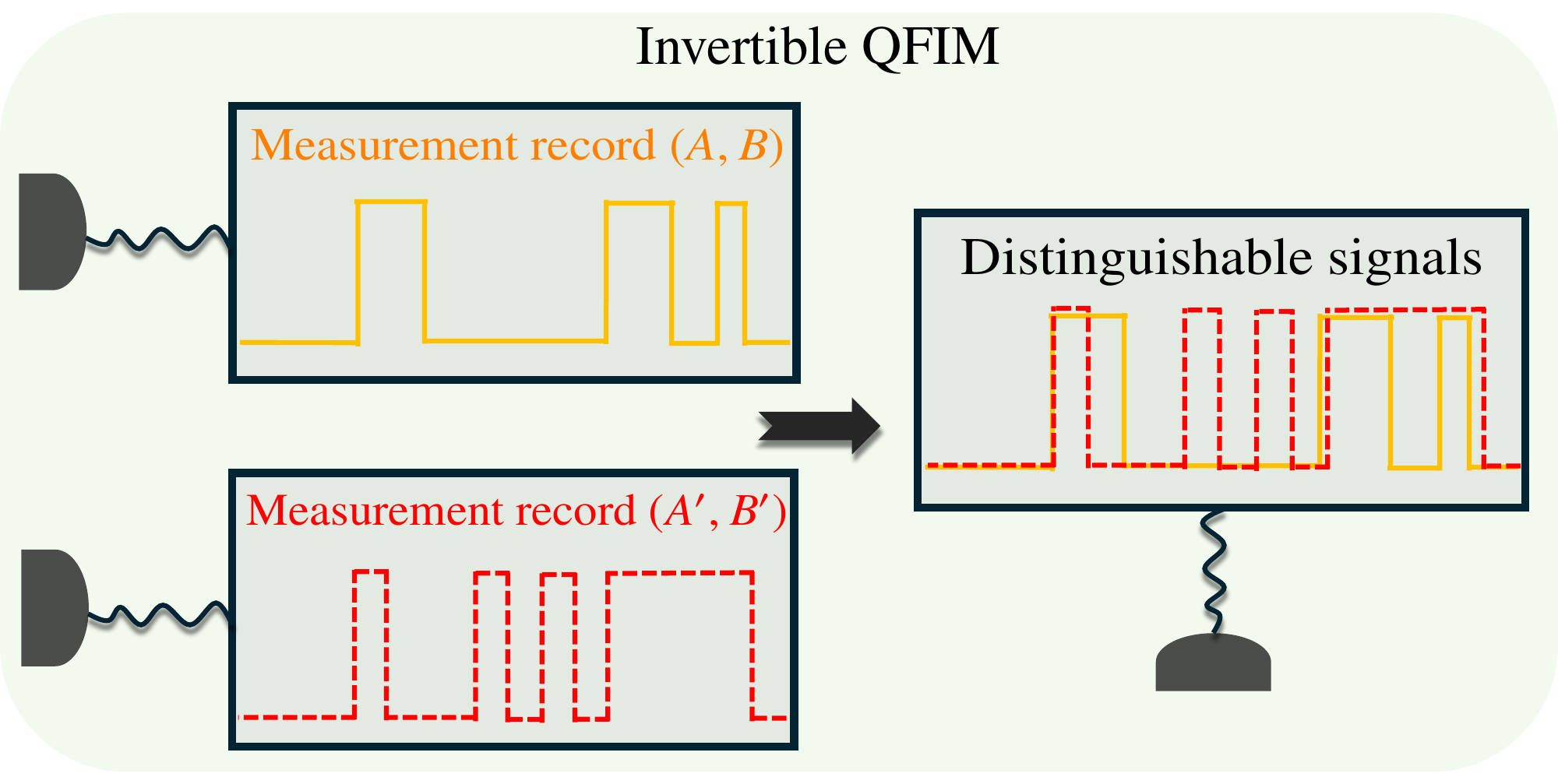}\vspace{-0.5em}
    
    \includegraphics[width=1\linewidth]{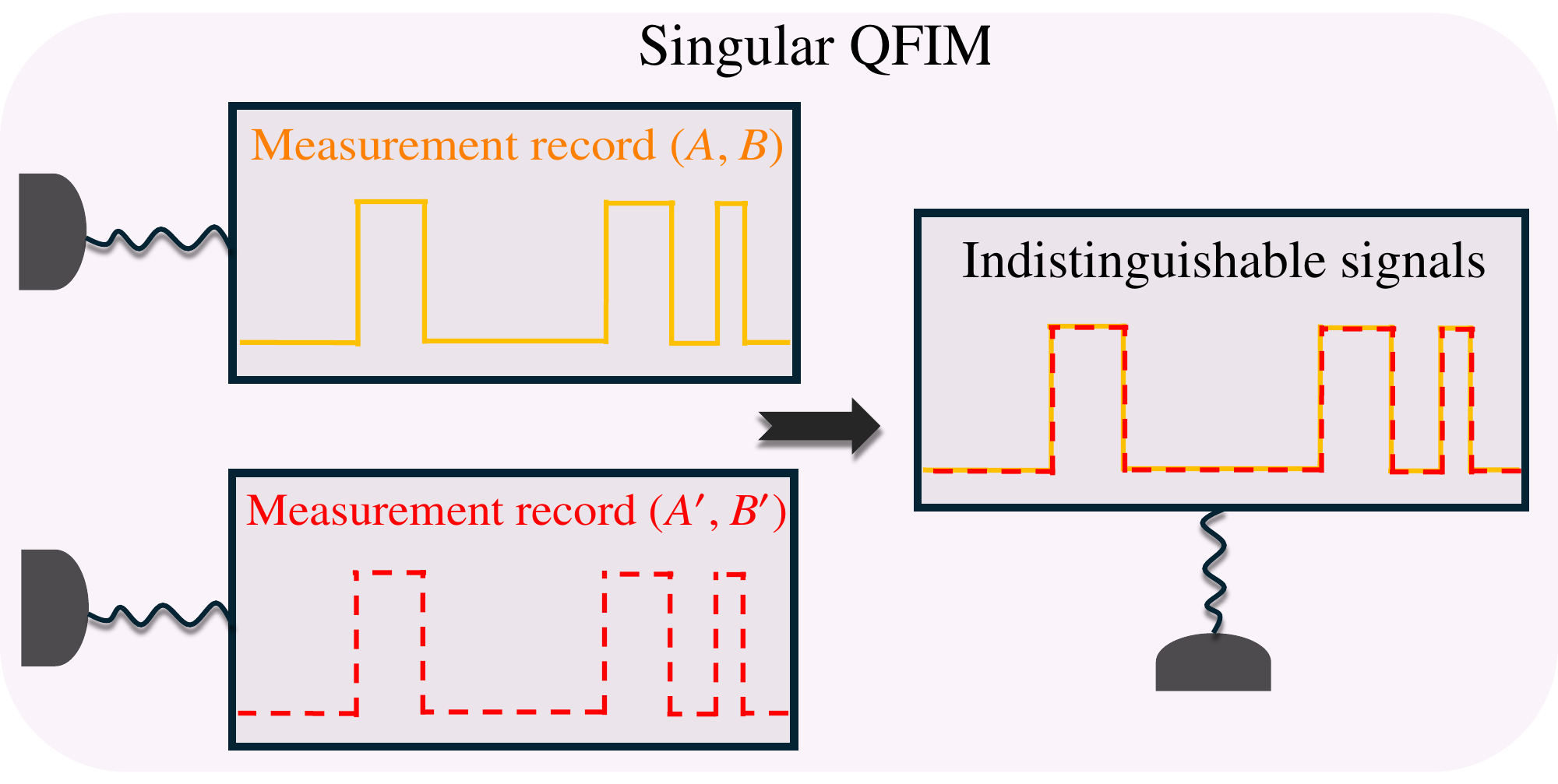}
    
    \caption{\textbf{Depiction of overparameterization leading to QFIM singularity.} When the QFIM is invertible (top), measurement records corresponding to different parameter sets \( (A,B) \) and \( (A',B') \) lead to distinguishable signals, permitting estimation. However, for an overparametrized statistical model, different parameter records \( (A,B) \) and \( (A',B') \) yield indistinguishable signals. This reflects a loss of identifiability, and results in a singular QFIM (bottom).}
    \label{fig:twoimages}
\end{figure}
Since it is not always possible to discern the effective parameter encoding, and thus the attainable precision, from the quantum state alone in complicated many-body critical systems, a more systematic approach is desirable.

\subsection{Deducing effective parameters in multiparameter estimation}

One approach for discerning the effective parameter involves analysing the structure of the \emph{singular} QFIM directly. In particular, diagonalizing the non-invertible QFIM can still yield meaningful insights, especially in analytically tractable models such as the LZ model~\cite{frigerio2024overcomingsloppinessenhancedmetrology,suzuki2020quantum,mihailescu2025metrologicalsymmetriessingularquantum,tsang2020quantum,PhysRevResearch.7.023060,He_2025,candeloro2024dimension,taming_singular,conlon2024roleextendedhilbertspace}. For the LZ model considered, we write the spectral decomposition
\begin{equation}
    \mathcal{D} = \hat{U}^T \hat{\mathcal{I}}_{\text{LZ}}(\vec{\theta}) \hat{U}\;,
\end{equation}
where $\vec{\theta} = (\omega,g)$ is the set of parameters to be estimated, $\hat{U} \in \mathbb{R}^{d\times d}$ is an orthogonal matrix whose columns are the eigenvectors $\vec{u}_a$, and $\hat{\mathcal{D}} = \text{diag}(\vec{G})$ is a diagonal matrix of eigenvalues $G_a$ of the QFIM \( \hat{\mathcal{I}}_{\text{LZ}} \). We note that the QFIM is a real symmetric semi-definite matrix, and can always be diagonalized. Each eigenvalue is given by
\begin{equation}
    G_a = \sum_{i,j} U_{i,a} U_{j,a} \times \hat{\mathcal{I}}_{\theta_i,\theta_j}\;,
\end{equation}
and quantifies the sensitivity along the eigen-direction $\vec{u}_a$ in parameter space. For the LZ model, the QFIM has one vanishing eigenvalue $G_a = 0$, whose corresponding eigenvector $\vec{u}_a$ identifies a redundant direction in parameter space: a combination of parameters along which the quantum state remains invariant. This reflects an over-parametrization at the metrological level. Such a zero mode imposes a constraint of the form $f(\vec{\theta}) = 0$, capturing a fundamental non-identifiability of parameters along that direction. This eliminates a parameter from the problem, thereby reducing the metrological degrees of freedom by one: the remaining effective parameter $\Omega$ which can be deduced. 

Conversely, the non-zero eigenvalue $G_b$ and its associated eigenvector $\vec{u}_b$ identify the effective parameter direction where parameters can be estimated from the state and are given by
\begin{equation}
    \vec{u}_b = \begin{pmatrix}
        \frac{g}{\sqrt{g^2+\omega ^2}} \\ \frac{\omega }{\sqrt{g^2+\omega ^2}}
    \end{pmatrix}\;,\quad \text{with} \quad G_b = \frac{1}{g^2 + \omega^2}\;,
\end{equation}
which provides a non-zero constraint $\hat{\mathcal{I}}(\vec{\theta}) \vec{u}_b = G_b \vec{u}_b$ that can be utilized to discern the effective parameter encoding $\Omega = f(\vec{\theta})$ by inverting the set of non-linear partial differential equations, derived from the condition $\vec{u}_b = \vec{\nabla}_\theta \Omega /|\vec{\nabla}_\theta \Omega|$ in the case of a single zero eigenvalue. By solving the set of partial differential equations, we obtain the following solutions for the effective QFI for the LZ model:
\begin{equation}
    \Omega = \frac{g}{\omega}, \quad \text{and} \quad \Omega =\omega g \;,
\end{equation}
where only the first solution is physical. The above demonstrates that care is essential when deducing the effective parameter encoding, as even in simple cases, unphysical solutions may arise. The presence of multiple solutions also implies potential branch-cut issues when performing numerical computations. Once the functional form of the effective parameter, $\Omega = g / \omega$, has been deduced from the singular QFIM one can, in principle, reparametrize the Hamiltonian or ground state accordingly to find the corresponding effective QFI, as done before in Eq.~\eqref{eq:lz_reparam}. For situations where such a reparametrization is not possible, an alternative is required.

\subsection{Bounding precision of effective parameters in multiparameter estimation}
A widely adopted approach to deal with singular multiparameter estimation and to formulate approximate bounds on the precision is to apply the Moore-Penrose pseudoinverse to the QFIM~\cite{taming_singular,PhysRevLett.127.110501,2001parameterestimationsingular,xavier2004riemannian,PhysRevA.95.052320}. The pseudoinverse of the QFIM $\left[ \hat{\mathcal{I}}(\vec{\theta}) \right]^+$ always exists and is well defined. The interpretation of the pseuodinverse as a lower bound $\text{Cov}[\vec{\theta}] \geq \left[  \hat{\mathcal{I}}(\vec{\theta}) \right]^+$ on the covariance matrix of parameters is known to be overly optimistic compared to the true QCRB~\cite{xavier2004riemannian}. The pseudoinverse of the LZ model can readily be computed in this case
\begin{equation}
    \label{eq:lz_pseudo}
    \left[ \hat{\mathcal{I}}_{\text{LZ}} \right]^+ = \begin{pmatrix}
        g^2 & -\omega g \\ -\omega g & \omega^2
    \end{pmatrix}\;.
\end{equation}
We can then utilize the previously deduced effective parameter encoding $\Omega = g / \omega$ in conjunction with the pseudoinverse in Eq.~\eqref{eq:lz_pseudo} to deduce the true precision of the effective parameter, as opposed to the optimistic one arising from $\text{Cov}[\vec{\theta}] \geq \left[ \mathcal{M} \hat{\mathcal{I}}(\vec{\theta}) \right]^+$~\cite{mihailescu2025metrologicalsymmetriessingularquantum}. This can be achieved by using the linearized error propagation formula
\begin{equation}
    \text{Cov}\left[ \vec{\Omega}\right] = \hat{J} \text{Cov}\left[ \vec{\theta} \right]\hat{J}^T\;,
\end{equation}
with $\vec{\Omega}$ the vector of effective parameters and $ \hat{J} = \nabla \Omega(g,\omega) = (\partial_g \Omega,\partial_\omega \Omega) \equiv (1/\omega,-g/\omega^2)$ the Jacobian. We obtain the following precision from substitution of the above
\begin{align}
    \text{Var}(\Omega) &= \Omega^2 \left[ \frac{\text{Var}(\omega)}{\omega^2} + \frac{\text{Var}(g)}{g^2} - 2 \frac{\text{Cov}(\omega,g)}{\omega g}\right]\;,
\end{align}
where we treat the values $\text{Var}(\omega)$, $\text{Var}(g)$, and $\text{Cov}(\omega,g)$ as elements of the pseudoinverse $\left[ \hat{\mathcal{I}}_{\text{LZ}} \right]^+$. From this, the precision of $\Omega$ can be found to be
\begin{equation}
    \text{Var}(\Omega) = \frac{(g^2 + \omega^2)^2}{\omega^4}\;,
\end{equation}
whose inverse is precisely the effective QFI found by using the reparametrized Hamiltonian $\hat{H}_{\text{LZ}}'$ in Eq.~\eqref{eq:lz_reparam}. We can therefore understand that discerning the effective parameter encoding is crucial in the pseudoinverse approach to retrieve the exact saturable bound, as opposed to the optimistic one~\cite{mihailescu2025metrologicalsymmetriessingularquantum}.

\subsection{Finite temperature multiparameter estimation}\label{subsec_LZ_MP_temp}

In the context of single parameter estimation, we have seen in Section~\ref{subsec_sp_thermal} how finite temperature fluctuations may serve to further enhance the attainable precision beyond what is achievable with the ground state alone~\cite{abiuso2024fundamental,PhysRevLett.133.040802,gietka2024tempcqm}. This improvement arises from two effects: additional information encoded in thermal populations and the mixing of states in the statistical ensemble $\hat{\varrho} = \sum_n e_n |e_n\rangle \langle e_n|$. However, this benefit is limited to low but finite temperatures: in the limit $T\to0$ the precision is governed by the ground state, while at $T\to \infty$ the state becomes maximally mixed and no useful information can be recovered. 

For multiparameter estimation in the LZ model using the ground state, we have seen that estimation of the target parameter $\omega$ is strictly prohibited when there is no information about the control parameter $g$---which may also be the target of a joint estimation. This limitation was signalled by a singular QFIM, which was a direct consequence of an over-parametrization leading to the emergence of a metrological symmetry: the parameters being related by their ratio. This confined us to single parameter estimation of the effective parameter $\Omega = g/\omega$. 
\begin{figure*}[t]
    \centering
    \includegraphics[width=1\linewidth]{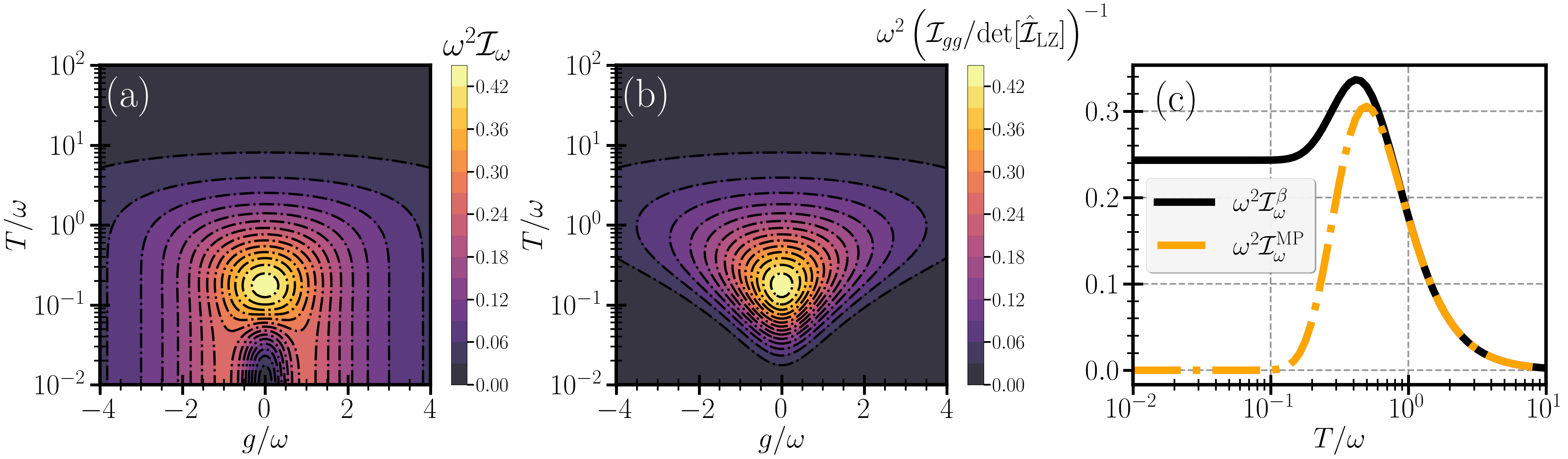}
    \caption{ \textbf{Multiparameter estimation at finite temperature for the LZ model.} (a) Total finite-temperature QFI~\eqref{LZ:totalQFI:finiteT} for estimating \( \omega\) in the single parameter case when the control parameter $g$ is assumed to be known. (b) Total finite-temperature QFI for the joint estimation of \( \omega \) and \( g \) in the multiparameter case. Here the control parameter \( g\) is also treated as an unknown parameter. (c) Comparison of the single and multiparameter QFI shown in the black and dashed orange lines respectively. In all panels we set $\omega = 1$; $g = 0.8$ for panel (c).} 
    \label{fig:lz_MPE}
\end{figure*}

To lift the QFIM singularity and allow for the estimation of our target $\omega$ as an independent parameter, one must break the metrological symmetry. This serves to eliminate the constraints that render the encoding redundant, thereby restoring full rank to the QFIM. One method of achieving this is to add known external controls on the system~\cite{mihailescu2023multiparameter}. Another possibility is to leverage thermal fluctuations: by treating temperature as a known quantity, it may in some circumstances serve as a mechanism for breaking the metrological symmetry, decoupling the parameter dependencies. We illustrate this point by considering the estimation of $\omega$ and $g$ in the LZ model, but this time at finite temperature \(T\). In this case, the QFIM reads
\begin{equation}
    \hat{\mathcal{I}}_{\text{LZ}}^\beta = \begin{pmatrix}
        \mathcal{I}_{\omega\omega}^\beta& \mathcal{I}_{\omega g}^\beta \\ \mathcal{I}_{g\omega}^\beta & \mathcal{I}_{gg}^\beta
    \end{pmatrix}\;,
\end{equation}
with elements given by Eq.~\eqref{eq:intro_sld_elements}
\begin{align}
    \mathcal{I}_{\omega\omega}^\beta &= \frac{(\Delta^2 \omega^2 - 4 g^2 T^2)\text{sech}^2\big( \Delta / 2T\big) + 4 g^2 T^2}{4 T^2 \Delta^4} \;, \\
    \mathcal{I}_{gg}^\beta &= \frac{(\Delta^2 g^2 - 4 \omega^2 T^2)\text{sech}^2\big( \Delta / 2T\big) + 4 \omega^2 T^2}{4 T^2 \Delta^4}\;, \\
    \mathcal{I}_{\omega g}^\beta &= \frac{g \omega \left[ \big(\Delta^2 + 4 T^2 \big)\text{sech}^2\big( \Delta/2T \big) - 4 T^2\right]}{4 T^2 \Delta^4}\;,
\end{align}
where $\Delta = \sqrt{\omega^2 + g^2}$ is the energy gap. As before, diagonal elements of the QFIM relate to the single-parameter sensitivity, and the off-diagonal element ${\mathcal{I}}_{\omega g}^{\beta}$ shows the correlation between the parameters.
 
Introducing temperature as a known parameter, the previous constraint rendering the QFIM of the ground state singular is broken, and the QFIM becomes invertible with $\text{det}\left[\hat{\mathcal{I}}_{\text{LZ}}^\beta\right] = \mathcal{I}_{\omega\omega}^\beta\mathcal{I}_{g g}^\beta - \big(\mathcal{I}_{g\omega}^\beta\big)^2 \neq 0$. Indeed, we have now
\begin{equation}
    \label{eq:lz_det_temp}
    \text{det}\left[\hat{\mathcal{I}}_{\text{LZ}}^\beta\right] = \frac{4 ~\text{csch}^4\big(\Delta /T\big) \sinh^6{\big( \Delta/2T\big)}}{T^2 \Delta^2}\;.
\end{equation}
The above expression reveals an interesting competition: the denominator $T^2 \Delta^2$ suggests a divergence in the determinant as the energy gap closes, $\Delta\!\to\!0$, or temperature vanishes, $T\!\to\!0$, potentially indicating sensitivity is restored in this regime compared to the singular ground state solution. However, this is counteracted by the numerator, specifically by the $\text{csch}^2\big(\Delta / T\big)$ term, which strongly suppresses sensitivity at low temperatures ($T\!\to\!0$), recovering the ground state result of a singular QFIM.  The competition between these two terms determines the overall behaviour of the determinant. In particular, when $ T \ll \Delta$ and $|g| \gg \omega$ the system is described by its ground state properties and the determinant vanishes. In this region the thermal population of the excited states is exponentially suppressed. Conversely in the opposite limit, $T \to \infty$, the thermal state becomes maximally mixed and no information about the parameters is encoded, thus the QFIM becomes singular. 

Between these extremes lies an optimal regime, $T \sim \Delta$ and $|g| \sim \omega$, for which thermal fluctuations break the metrological symmetry and restore full rank to the QFIM.  Therefore, for multiparameter estimation, metrological utility is enhanced in this region by making the estimation of our target $\omega$ possible when the control parameter $g$ is unknown. This result highlights a counter-intuitive yet powerful principle: under certain conditions, thermal noise can become a valuable resource in quantum metrology by unlocking otherwise inaccessible parameter information.

\subsection{Comparison of precision at finite temperature for single and multiparameter cases}

Having shown that thermal fluctuations can break the metrological symmetry and restore full rank to the QFIM within a well-defined regime, we are now in a position to tolerate uncertainty in the control parameter $g$. This allows us to meaningfully compare single and multiparameter estimation strategies for our target parameter $\omega$. In the multiparameter setting, the relevant figure of merit for the estimation precision is the inverse of the QFIM
\begin{equation}
     \left[\hat{\mathcal{I}}^{\beta}_{\text{LZ}}\right]^{-1} = \frac{1}{\text{det}\left[\hat{\mathcal{I}}^{\beta}_{\text{LZ}}\right]}\begin{pmatrix}
         \mathcal{I}_{gg}^\beta & - \mathcal{I}_{\omega g}^\beta \\ -\mathcal{I}_{g\omega}^\beta & \mathcal{I}_{\omega\omega}^\beta
     \end{pmatrix}\;,
 \end{equation}
which enters the QCRB Eq.~\eqref{eq:def_qcrb_mp} for multiparameter estimation. The diagonal elements of this inverse matrix quantify the minimal achievable variance for each parameter, accounting for correlations between them. In contrast, for single-parameter estimation of our target $\omega$ the QCRB simplifies to $\text{Var}(\omega) \geq \left[\mathcal{I}_\omega \right]^{-1}$, the inverse of the variance. To fairly compare the two approaches, we must relate these quantities. This involves relating the single parameter QFI, $\mathcal{I}_\omega$, to the inverse of the corresponding multiparameter variance from the QCRB: \( \left(\mathcal{I}_{gg}^\beta /\text{det}\left[ \hat{\mathcal{I}}_{\text{LZ}}^\beta\right] \right)^{-1} \), which we denote as \( \mathcal{I}_\omega^{\text{MP}}\) to be succinct. 

In Fig.~\ref{fig:lz_MPE}(a) and (b) the precision in estimating \( \omega \) in the single and multiparameter scenarios are shown for a fixed \( \omega = 1\). For both instances at low temperatures, $T \ll \Delta$, the system resides in its ground state and the precision is determined solely by the ground state sensitivity. This means that while in the single parameter case precision in estimating \( \omega \) remains finite---the multiparameter scenario has a vanishing sensitivity due to QFIM singularity.

In the intermediate temperature regimes, $T \!\sim \!\Delta$, the precision becomes enhanced due to thermal fluctuations. For the single parameter case, this manifests in greater precision compared to the ground state, whereas for multiparameter estimation this means a lifting of the QFIM singularity. Furthermore, within this regime, the multiparameter sensitivity is generically \emph{lower} compared with the single-parameter case. This reflects a general feature of quantum estimation theory: introducing uncertainty in additional parameters, here the control parameter $g$, can only degrade, or, at best, maintain the precision for the parameter of interest.

These features can be understood on a more quantitive basis in panel Fig.~\ref{fig:lz_MPE}(c), where the single and multiparameter precision are shown as a function of temperature for $g = 0.8$ and $\omega = 1$. In the limit $T\!\to\!0$, the single-parameter sensitivity remains finite, whereas the multiparameter sensitivity vanishes. When $T \sim \Delta$, thermal fluctuations enhance the single-parameter QFI and restore precision in the multiparameter case. Notably, the multiparameter precision is always lower-bounded by the single-parameter value. 

\subsection{Attainability of multiparameter quantum Cram\'er-Rao bound}
Thus far, we have encountered two key subtleties specific to the multiparameter estimation scenario. First, the QFIM can become singular, indicating that the bare parameters cannot be simultaneously estimated due to underlying dependencies between them. Second, there exists an inherent trade-off between the single and multiparameter estimation strategies. In the multiparameter setting, the presence of additional unknown quantities, while potentially offering a more realistic characterization, can degrade the estimation precision of a target parameter (in our case the energy splitting $\omega$).

Another important issue concerns the attainability of the QCRB. While in single-parameter estimation the bound provided by the QFI is always asymptotically saturable using optimal measurements, this is not generically true in the joint estimation of multiple parameters. When the SLD operators \(\hat{L}_i\) corresponding to different parameters are incompatible—that is, they do not commute—simultaneous optimal measurement is impossible. This non-commutativity induces a fundamental trade-off in precision that cannot be overcome, even asymptotically.~\cite{di2022multiparameter,carollo2020erratum,Carollo_2019,carollo2018uhlmann,bakmou2019quantum,Nat_Incomp,heinosaari2016invitation,Belliardo_2021,ragy2016compatibility,wang2025tighttradeoffrelationoptimal,PhysRevLett.133.040802,Matsumoto_2002,PhysRevA.73.052108,Holevo1976,albarelli2020perspective,PhysRevX.12.011039}. We will unpack these ideas for the presented LZ model. The SLD operators $\hat{L}_\omega$ and $\hat{L}_g$ can be computed using Eq.~\eqref{eq:intro_sld_elements}, and take the concise analytical form:
\begin{equation}
    \hat{L}_{g} = \Big[ \tfrac{g^2}{2 T \Delta^2} + \tfrac{\omega^2}{\Delta^3} \tanh{x} \Big]\hat{\sigma}_x - \left[\tfrac{g \omega}{2 T \Delta^2} + \tfrac{g \omega}{\Delta^3} \tanh x\right]\hat{\sigma}_z - \alpha \mathbb{I}\;,
\end{equation}
\begin{align}
    \hat{L}_\omega &= \left[ \tfrac{g}{2 T \Delta^4}\left( \Delta^2[g \tanh^2{x} + \omega ~\text{sech}^2x] - 2T\omega \Delta \tanh{x} \right)\right]\hat{\sigma}_x \nonumber\\ 
    &-\Big[\tfrac{\omega}{2T \Delta^2}\left( g \tanh^2{x} + \omega ~\text{sech}^2x\right) + \tfrac{g^2}{\Delta^3}\tanh{x}\Big] \hat{\sigma}_z - \alpha \mathbb{I}\;,
\end{align}
where $x = \Delta / 2T$ and $\alpha = (g \tanh{x})/2T \Delta$. From these expressions we may easily compute the commutator
\begin{equation}
\left[\hat{L}_\omega,\hat{L}_g\right] =\left[ i \big(T \Delta\big)^{-1} \tanh{x}\right] \hat{\sigma}_y\;.
\end{equation}
Since \([\hat{L}_\omega,\hat{L}_g]\neq0\), there is no measurement basis that is optimal for both parameters simultaneously.This implies a fundamental quantum precision trade-off, as the information is encoded along non-orthogonal directions in Hilbert space that effectively ``interfere'' with each other. This interference prohibits a single measurement (or single-copy probe) from saturating the QCRB~\cite{Holevo1976,PhysRevA.105.062442,10480691,PhysRevX.12.011039}. Importantly, non-commutativity does not always imply incompatibility. In the LZ model we find that a particularly interesting condition holds: 
\begin{equation}
    \text{Tr}\left( \hat{\varrho}_{\text{LZ}} \left[\hat{L}_\omega,\hat{L}_g\right] \right) = 0\;.
\end{equation}
This reveals a physically important point: although the SLD operators do not commute, their non-commutativity averages out in the quantum state $\hat{\varrho}_{\text{LZ}}$, and thus does not manifest in the estimation statistics. In other words, the quantum incompatibility cost is zero. As a consequence, the ultimate precision limit---given by the HCRB---is asymptotically attainable. While a single-copy measurement is limited by the SLD non-commutativity, access to many copies of the state enables collective measurements that bypass this limitation. These joint POVMs effectively separate the interfering information encoded in $\omega$ and $g$, allowing the full QFI to be extracted.

\graphicspath{{./Ising_Figs/}}

\section{Quantum phase transitions, criticality, and critical metrology: many-body Ising model}\label{sec:Ising_Model}
The Landau–Zener model has served as a minimal framework for understanding the fundamental principles by which quantum critical systems acquire enhanced sensing capabilities. Having established these operating principles, we now turn to many-body quantum systems capable of supporting bona fide quantum phase transitions. For this purpose, we introduce the general form of a spin-chain Hamiltonian
\begin{equation}\label{eq:intspins}
    \hat{H} = \omega \sum_{i=1}^{N} \hat{\sigma}_z^{(i)} - \sum_{i\neq j }^N g_{ij}\hat{\sigma}_x^{(i)} \hat{\sigma}_x^{(j)} \;,
\end{equation}
where $\omega$ is the transverse field (the parameter to be estimated) and $g_{ij}$ denotes the interaction strength between spins $i$ and $j$. While Eq.~\eqref{eq:intspins} encompasses a broad class of interacting spin models, in what follows we restrict our analysis to two limiting cases. By first considering uniform nearest-neighbour interactions, $g_{ij} = g$, we obtain the one-dimensional \emph{transverse field Ising model} (TFIM):
\begin{equation}
    \hat{H}_{\text{TFIM}} = \omega \sum_{i=1}^{N} \hat{\sigma}_z^{(i)} - g \sum_{i=1}^N \hat{\sigma}_x^{(i)} \hat{\sigma}_x^{(i+1)} \;.
\end{equation}
This model exhibits a quantum phase transition when $g\!=\! \omega$ between two phases:
\begin{itemize}
    \item \textbf{Paramagnetic phase ($|\omega| > |g|$):} In this phase, spins align along the transverse field direction $z$ with short-range correlations. Connected spin-spin correlations decay exponentially with distance, reflecting the absence of long-range order. Here, correlations between the spins are small, and the ground state is close to a product state. This is also known as the disordered phase. 
    \item \textbf{Ferromagnetic phase ($|\omega| < |g|$):} In this phase, spins tend to align along one of the two directions $\pm x$, reflecting the $\mathbb{Z}_2$ spin-flip symmetry of the Hamiltonian. For a finite system, the true ground state remains symmetry-preserving and is given by a superposition of the two ferromagnetic configurations. In the thermodynamic limit, however, spontaneous symmetry breaking occurs, and the system selects one of the ordered states. Connected correlations $\langle \hat{\sigma}_x^{(i)} \hat{\sigma}_x^{(i+r)} \rangle - \langle \hat{\sigma}_x^{(i)} \rangle \langle \hat{\sigma}_x^{(i+r)} \rangle$ saturate to a finite value as $r \to \infty$, signalling long-range order. This phase is also known as the ordered phase and supports a degenerate ground state manifold in the thermodynamic limit.

\end{itemize}
The critical point is defined at $g_c \equiv  \omega$. At criticality, the correlation length diverges, the excitation gap closes, and the system becomes maximally susceptible to small parameter changes. It is in this regime where critical metrology offers its strongest advantage. We will explore these concepts in more detail in the following. 

The TFIM represents a class of systems which is exactly solvable via a Jordan-Wigner transformation by mapping spins to spinless fermions. A detailed discussion of the exact solution can be found in the following useful references~\cite{jwt,LIEB1961407,PFEUTY197079,sachdev1999quantum,duncan2025tamingquantumsystemstutorial,10.21468/SciPostPhysLectNotes.82}. After Fourier transforming to momentum space, the Hamiltonian decouples into independent momentum-space pairs $(k,-k)$. These can be viewed as elementary \emph{quasiparticle} modes that are always created or annihilated in pairs of opposite momentum. Within each such sector, the relevant states are the $|0\rangle$ (no quasiparticle excitations) and $\hat{c}^\dagger_k\hat{c}_{-k}^\dagger |0\rangle$ (quasiparticle pair with opposite momenta). This effectively reduces the problem to a two-level system for every momentum-$k$ mode. To make this structure explicit, we introduce the fermionic spinors
\begin{equation}
    \hat{\Psi}_k^{\phantom{\dagger}} = \begin{pmatrix}
        \hat{c}_k^{\phantom{\dagger}} \\ \hat{c}^{\dagger}_{-k}
    \end{pmatrix}\;, \qquad \hat{\Psi}^\dagger_k = \begin{pmatrix}
        \hat{c}^{\phantom{\dagger}}_k & \hat{c}_{-k}^{\phantom{\dagger}}
    \end{pmatrix}\;.
\end{equation}
By utilizing the pseudo-spin Pauli matrices, the Hamiltonian of each momentum space block $\hat{H}_k$ can be written into a more intuitive and familiar form
\begin{equation}
    \hat{H}_k = 2 \hat{\Psi}_k^{\dagger} \left[ (\omega - g \cos{k})\hat{\sigma}_z + (g\sin{k})\hat{\sigma}_x \right] \hat{\Psi}_k^{\phantom{\dagger}}\;.
\end{equation}
In this language, the dynamics of each $(k,-k)$ block resembles a LZ type model whose pseudo-spin representation encodes the presence or absence of quasiparticle excitations. Here, the allowed momenta are given by:
\begin{equation}
    k = \frac{\pi(2n+1)}{N}\; \quad \text{with}\; \quad n = 0,1,2,...,\frac{N}{2}-1\;,
\end{equation}
and $\hat{H}_{\text{TFIM}} = \sum_{k>0} \hat{H}_k$. Each $k$-block behaves as a two-level system in the $(k,-k)$ subspace. The TFIM can therefore be decomposed into a set of independent LZ models with an effective field $\omega_\text{eff} = \omega - g\cos{k}$ and coupling $g_{\text{eff}} = g \sin{k}$. We can understand the connection to the LZ model intuitively by noting that the momentum index $k$ labels the Bogoliubov quasiparticle modes emerging from the Jordan-Wigner and Fourier transformations. This provides a one-to-one mapping between the collective spin dynamics of the chain and the independent fermionic quasiparticle excitations in momentum space. 

By solving the $2\times2$ eigenvalue problem of the pseudo-spin Hamiltonian, the quasiparticle dispersion is obtained as:
\begin{equation}
    \epsilon_k = -2 \sqrt{g^2 + \omega^2 - 2g \omega \cos{k}}\;,
\end{equation}
which represents the energy of the fermionic Bogoliubov modes labelled by momentum $k$. The many-body excitation spectrum of the chain can thus be entirely determined by these single-particle dispersions. At the critical point $g = \omega$, the system undergoes a quantum phase transition and the gap between the ground and first excited state vanishes in the thermodynamic limit. Specifically, the minimum of the dispersion occurs as $k\!\to\!0$, and the finite-size scaling of the gap is given by
\begin{equation}
    \Delta(\omega,g) \approx \frac{4 \pi \omega}{N}\;.
\end{equation}
The $1/N$ scaling of the gap is a hallmark of the critical point in a finite-size system, reflecting how the system size itself sets the dominant length scale when the correlation length diverges. In this true many-body system, the closing of the gap emerges collectively from the full set of momentum-$k$ modes. In contrast to this, the LZ model provides a simplified analogue: its single avoided crossing mimics the role of these momentum-resolved level crossings, thereby capturing the essence of the true many-body critical behaviour.  In this picture, each momentum block behaves like an effective two-level system whose ground state can be written as 
\begin{equation}
    |\psi_{GS}\rangle_k = \cos{\left(\frac{\theta_k}{2}\right)}|0\rangle_k + \sin{\left(\frac{\theta_k}{2}\right)}|1\rangle_k\;,
\end{equation}
with mixing angle determined by 
\begin{equation}
    \tan{\left(\frac{\theta_k}{2}\right)} = \frac{g \sin{k}}{g\cos{k}-\omega}\;.
\end{equation}
The parameter $\theta_k$ thus plays the role of the Bloch-sphere polar angle, encoding how the pseudo-spin representing the $(k,-k)$ sector is oriented with respect to the effective field. The many-body ground state for an $N$-site system is then simply the tensor product 
\begin{equation}
    |\psi_{GS}\rangle = \bigotimes_{k>0}|\psi_{GS}\rangle_k\;
\end{equation}
of the ground state of each momentum space block $\hat{H}_k$. This illustrates how the global state is built from a coherent superposition of quasiparticle configurations across all momentum modes.

\subsection{Criticality as a resource for sensing}

We here seek to explore the underlying characteristics of critical systems which make them an appealing platform in the context of quantum sensing. Quantum criticality is driven by quantum fluctuations, as opposed to thermal fluctuations in classical phase transitions. At zero temperature changes in a Hamiltonian parameter, denoted $g$, causes a fundamental change in the ground state properties of the system. Notably, at or near the close vicinity of the critical point $g \equiv g_c$, the quantum fluctuations begin to dominate. This is accompanied by non-analytical behaviour in the ground state which is reflected in physical observables and qualitative changes in the nature of ground state correlations. At finite temperature, there is a delicate interplay between thermal and quantum fluctuations. 

In such cases, there exists a finite ``quantum critical'' regime where quantum fluctuations still dominate before thermal effects become significant. This interplay has important consequences for the robustness and practicality of quantum critical sensors. Furthermore, quantum criticality and entanglement are closely connected~\cite{Osterloh2002,PhysRevA.66.032110,PhysRevLett.90.227902,PhysRevB.76.184517,RevModPhys.80.517}, although no unified theory links them fully. At first glance, the genuine multipartite entanglement structure of ground states in quantum critical systems makes them attractive from a metrological perspective. For example, we have previously encountered the utility of leveraging entangled states in the context of metrology through Ramsey interferometry using entangled GHZ states in Section~\ref{sec:ramsey_example}. In that instance, entanglement was an explicit resource facilitating enhanced sensing capabilities. However, as we will motivate in the following, it is the confluence of the entanglement structure in conjunction with the non-analytical behaviour near quantum critical points that underlies their utility as quantum sensors. 

Putting these ideas on a more rigorous footing, we will explore how the dramatic changes in ground state properties of critical systems can be directly translated into high metrological precision to quantities of interest. This will be exemplified through the paradigmatic TFIM. Concretely, by utilizing a purely information theoretic quantity, we will motivate how small changes of the form 
\begin{equation}
    \omega' = \omega + \delta \omega\
\end{equation}
in the traverse field $\omega$, manifest in large deviations of the ground state $| \psi_{GS}(\omega) \rangle$, with $\delta\omega$ representing a small perturbation. For this purpose, we will consider the \emph{fidelity} of pure states: a geometric object quantifying the overlap between two quantum states $|\psi_{GS}(\omega) \rangle$ and $| \psi_{GS}(\omega') \rangle$, defined as~\cite{uhlmann1976transition}
\begin{equation}
    \label{eq:gs_fid}
	F(\omega,\omega' ) = | \langle \psi_{GS}(\omega)| \psi_{GS}(\omega') \rangle| \;.
\end{equation}
The fidelity measures the closeness between the two quantum states. If the state is left unchanged then $F(\omega,\omega') = 1$, whereas for orthogonal states $F(\omega,\omega') = 0$. Contextually, the fidelity serves as a useful diagnostic tool for quantifying the variations $\delta \omega$ in the ground state properties across the transition point~\cite{zanardi2008informationgeometry,zanardi2006ground,zanardi2007criticalscaling,PhysRevA.78.032309,abasto2008fidelity,zanardi2008criticalityresource,damski2013exact,PhysRevA.77.032307}. 

Using the exact solution of the TFIM, the ground state fidelity takes the concise form~\cite{Zhou_2008}
\begin{equation}
	\label{eq:ising_exact_fid}
	F(\omega,\omega') = |\langle \psi_{GS}(\omega)| \psi_{GS}(\omega') \rangle | = \prod_{k>0} \cos(\theta_k - \delta \theta_k')\;,
\end{equation}
for arbitrary system size $N$, and where the product runs over momentum the blocks of the Hamiltonian $\hat{H}_k$. Fig.~\ref{fig:ising_fid}(a) shows the ground state fidelity~\eqref{eq:ising_exact_fid} landscape as a function of the detuning from the critical point, $\delta g_c = g - g_c$, and system size $N$, for fixed $\omega = 1$ and $\delta \omega = 0.01$. The colour scale highlights two qualitatively different regimes: the extended yellow regions correspond to fidelity values very close to unity, indicating that the ground state remains largely unchanged under the small parameter shift. In contrast, the black/purple valley centred at the critical point, $\delta g_c = 0$, marks a sharp loss in the fidelity (the ground state changes markedly) in the quantum-critical region. As the system size $N$ increases, this valley deepens and narrows, reflecting a growing instability of the ground state to perturbations at criticality. We further remark that in the thermodynamic limit, the fidelity can drop to zero irrespective of how small or large the deviation $\delta \omega$ is. This phenomena is known as the Anderson orthogonality catastrophe~\cite{PhysRevLett.18.1049}. 

These effects are isolated in Fig.~\ref{fig:ising_fid}(b) by plotting the fidelity at criticality $F(\omega = g)$ as a function of system size $N$. For small $N$ the fidelity remains close to unity, meaning that the ground states $|\psi_{GS}(\omega)\rangle$ and $|\psi_{GS}(\omega')\rangle$ are nearly indistinguishable. As the system size is increased, however, the fidelity decays---reflecting that the same small perturbation $\delta \omega$ produces ground states that are increasingly distinguishable. This dramatic decay in fidelity with system size at criticality reflects the profound change ground states undergo during a quantum phase transition with respect to small parameter variations $\delta \omega$. Intuitively these effects seem favourable in the context of metrology. This is because rapid rates of change with respect to quantities of interest indicate regions where parameter estimability is high. 

\begin{figure}[t]
    \centering
    \includegraphics[width=1\linewidth]{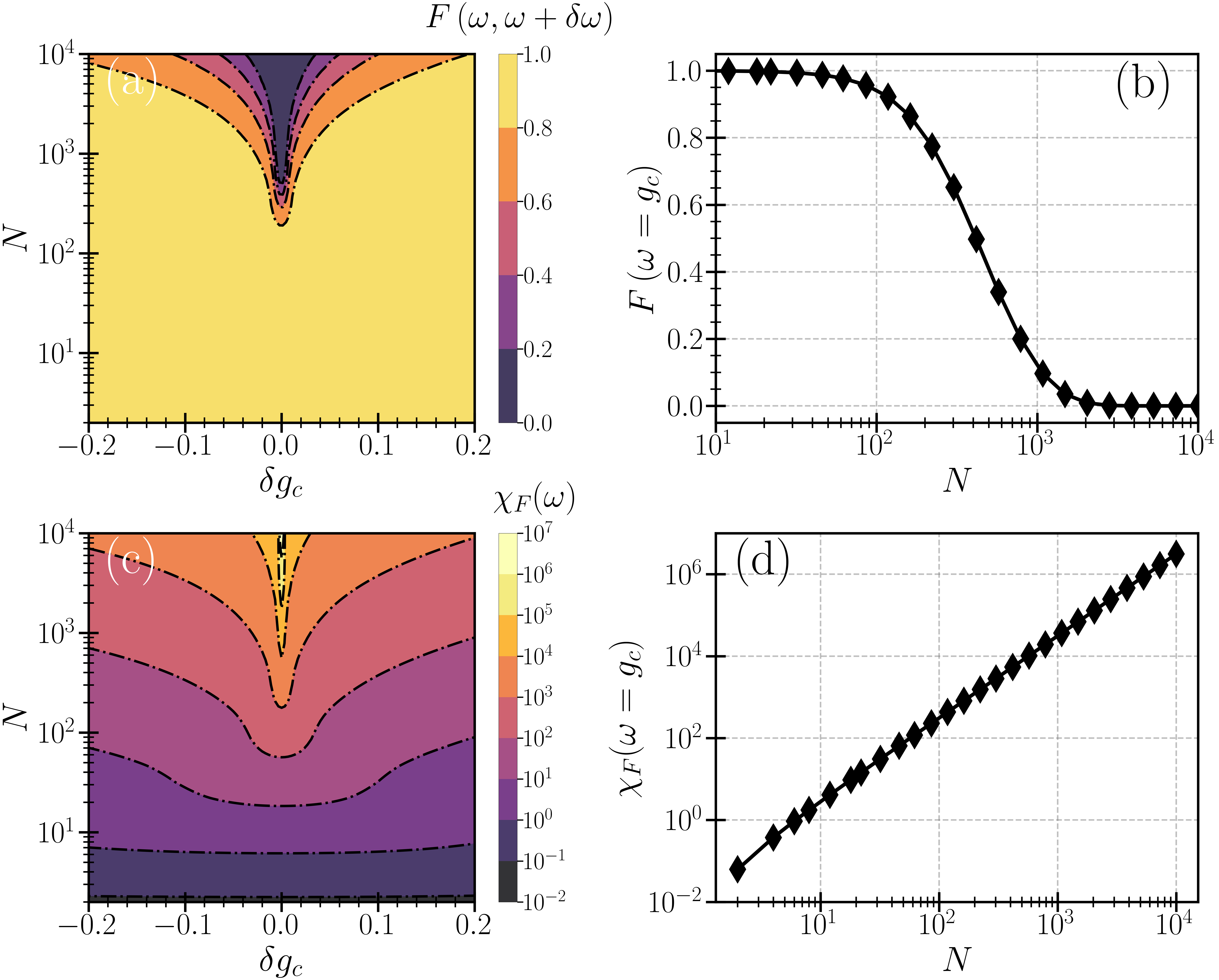}
    \caption{\textbf{Criticality as a metrological resource in the TFIM.} (a) Ground state fidelity $F(\omega,\omega+\delta \omega)$ for $\delta \omega = 0.01$ versus system size $N$ and detuning from critical point $\delta g_c = g_c - 1$, with $\omega = g_c = 1$. The sharp dip at $\delta g_c = 0$ reflects maximal state distinguishability. Note the log-scale on the $y$-axis. (b) Fidelity at criticality $g_c = \omega = 1$ decaying rapidly with system size $N$ (log-scale), signalling the ground state dramatical changing. (c) Fidelity susceptibility $\chi_F(\omega)$ (color in log-scale) as a function of system size $N$ (log-scale) and detuning from criticality $\delta g_c$. $\chi_F(\omega)$ diverges at the critical point marking the fact that the ground state changes dramatically. (d) Scaling of the fidelity susceptibility $\chi_F(\omega)$ at the critical point $g_c = \omega = 1$. It shows a power-law scaling $\chi_F(\omega = g_c) \propto N^\mu$, illustrating the diverging nature at the critical point.}
    \label{fig:ising_fid}
\end{figure}
\subsection{Fidelity susceptibility}\label{sub_sec_fid_sup}

The connection between the dramatic loss of fidelity at the quantum critical point and the indication that these features are useful for metrology can be made rigorous by considering ground state properties which are differentiable in parameter space. In particular, the fidelity between two infinitesimally close ground states, for which $\delta \omega \ll 1$, can be expanded as~\cite{zanardi2006ground,zanardi2008informationgeometry,zhang2009directcriticality,damski2013exact,PhysRevB.85.024301,Zhou_2008,PhysRevB.88.064304}
\begin{equation}
	F(\omega,\omega + \delta \omega ) = 1 - \frac{(\delta \omega)^2}{2} \chi_F(\omega) + ...\;,
\end{equation}
where $\chi_F$ denotes the \emph{fidelity susceptibility} of pure states, defined as
\begin{align}
	\chi_F(\omega) &= - \frac{\partial^2 F}{\partial(\delta \omega)^2}\biggr\rvert_{\delta \omega =0} \\
	\label{eq:fid_suc_def2}
	&= \langle \partial_\omega \psi | \partial_\omega \psi \rangle - \langle \partial_\omega \psi|\psi\rangle \langle \psi | \partial_\omega \psi \rangle\;.
\end{align}
The fidelity $F(\omega,\omega')$ is a geometric measure of how different the ground states are in parameter space, whereas the fidelity susceptibility $\chi_F(\omega)$ represents how sensitive the ground state is to infinitesimal variations in $\omega$. The fidelity susceptibility behaves like a response function in linear response theory. Concretely, it can be interpreted as a generalized susceptibility that quantifies the ground state’s response to an infinitesimal change in the parameter $\omega$ through its state overlap, rather than through changes in expectation values~\cite{PhysRevE.76.022101,PhysRevLett.132.193801,PhysRevA.78.032309,Zhou_2008,zanardi2008informationgeometry}. 

This is important as in quantum critical theory, linear response formalism is the standard tool to study how observables react to weak perturbations, with susceptibilities encoding correlations and critical scaling. In this sense, the fidelity susceptibility plays the role of a geometric analogue, quantifying the response of the ground state itself to infinitesimal parameter changes. For TFIM, the fidelity susceptibility can be computed directly
\begin{equation}
	\chi_F(\omega) = \sum_{k>0} \left( {\partial_\omega \theta_k}\right)^2\;,
\end{equation}
where 
\begin{equation}
	{\partial_\omega \theta_k} = \frac{g \sin (k)}{2 \left(g^2-2 g \omega  \cos (k)+\omega ^2\right)}\;.
\end{equation}
We illustrate the behaviour of the fidelity susceptibility in Fig.~\ref{fig:ising_fid}(c), where it is shown as a function of the detuning from criticality, $\delta g_c = g-g_c$, and the system size $N$. In this analysis, the transverse field is fixed at $\omega = 1$. The color-scale spans several orders of magnitude on a logarithmic scale: from dark purple (low susceptibility) of order $10^{-2}$ to bright yellow (high susceptibility) of order $10^7$, marking the large changes in the fidelity susceptibility. Several key features are immediately visible:
\begin{itemize}
    \item \textbf{Sharp peak at criticality:} For $\delta g_c = 0$, the fidelity susceptibility becomes more strongly enhanced compared to outside of the critical region, forming a ridge that becomes significantly sharper with increasing system size $N$.
    \item \textbf{Broadening away from criticality:} For finite detuning ($|\delta g_c|>0$), the susceptibility grows much slower, indicating that the system is far less sensitive to parameter variations away from criticality.
    \item \textbf{Scaling with size:} As the system size $N$ increases from $10^1$ to $10^4$ (note the log-scale), the maximum value of $\chi_F(\omega)$ grows dramatically, following contour lines that become tight at $\delta g_c = 0$. This signals the onset of a divergence in the thermodynamic limit. 
\end{itemize}
Physically, this behaviour reflects the fact that at criticality $\delta g_c = 0$, the ground state undergoes significant rearrangements in response to even infinitesimal parameter variations. Analogously, this indicates that criticality amplifies the \emph{distinguishability} between nearby ground states. This is what makes the fidelity susceptibility a popular tool for probing quantum phase transitions~\cite{zanardi2006ground,Zhou_2008,doi:10.1142/S0217979210056335,zanardi2007criticalscaling,zanardi2008criticalityresource,zanardi2008informationgeometry,abasto2008fidelity,PhysRevA.78.032309,PhysRevB.77.245109,gritsev2009universal,DeGrandi2010,PhysRevB.81.012303}.

We can quantify the divergence of the fidelity susceptibility at the critical point by performing a scaling analysis. Setting $g = 1$ for simplicity, we find that 
\begin{align}
	\chi_F &= \frac{1}{4} \sum_k \frac{\sin^2(k)}{\left[1 - \cos(k)\right]^2} \\
	&\eqsim \frac{N}{2} \int_{\pi/N}^{\pi(N-1)/N} \frac{1}{k^2} dk \\
	&\propto N^2\;.
\end{align}
From this it is evident that $\chi_F(\omega)$ scales quadratically with system size $N$. Importantly, the fidelity susceptibility is not merely a heuristic measure of the ground state sensitivity to infinitesimal variations, but it can also be related to \emph{universal properties} of a quantum critical point. Its scaling encodes universal properties of the system such as the correlation length exponent $\nu$—which governs the divergence of the correlation length, $\xi \sim |g - g_c|^{-\nu}$—and spatial dimension $d$ when evaluated at the transition point.  In particular it has been shown~\cite{gu2008fidelity,damski2013exact,PhysRevB.77.245109,doi:10.1142/S0217979210056335} that close to a quantum critical point, the fidelity susceptibility follows a universal power-law behaviour
\begin{equation}
    \chi_F(\lambda) \propto N^{2/\nu -d}\;,
\end{equation}
where the exponents depend only on the universality class of the transition, and not on the microscopic details of the Hamiltonian. The connection between the universal scaling functions of critical systems and the fidelity susceptibility is of direct consequence in metrology. This is because the fidelity susceptibility can in fact be seen as a specific component of the quantum geometric tensor, previously introduced in Section~\ref{sec_lz_subsec_qgt}. Utilizing these connections, one finds that
\begin{equation}
	\mathcal{I}_\omega = 4 \chi_F(\omega)\;,
\end{equation}
which follows directly from Eq.~\eqref{eq:intro_pure_state_qfi_def} and Eq.~\eqref{eq:fid_suc_def2}. This shows that the universal scaling of the fidelity susceptibility near criticality directly determines the scaling of the QFI. For the TFIM, this connection implies that at the critical point
\begin{equation}
    \mathcal{I}_\omega \propto N^2\;,
\end{equation}
demonstrating how abstract geometric structures and universal critical exponents manifest as a concrete metrological resource~\cite{zanardi2008criticalityresource}. This scaling is verified and demonstrated in Fig.~\ref{fig:ising_fid}(d). This connection illustrates how features that characterize universality in many-body physics directly determine the degree of quantum-enhanced sensitivity achievable in parameter estimation. 

This makes the rigorous connection between the dramatic changes ground states undergo at quantum phase transitions, diverging susceptibilities, and ensuing metrological utility. Concretely, the fidelity susceptibility precisely captures the inherent change in ground state properties that emerge near the quantum critical point. Furthermore, $\chi_F$ directly encodes universal scaling features of a given phase transition. In turn, through the QCRB, the QFI describes the fundamental attainable precision in parameter estimation, and is directly related to the fidelity susceptibility. This therefore suggests that optimal measurement strategies utilizing critical quantum probes inherently leverage the diverging nature of critical systems. From this simple illustrative system, we have shown—through a geometric perspective—how features of critical systems naturally give rise to Heisenberg (quadratic) scaling in a metrological context. There exists an intrinsic connection between divergent susceptibilities and their role as metrological amplifiers in critical systems. It is important to note, however, that the arguments presented here apply strictly to the zero-temperature case. In more general settings, one expects a delicate interplay between thermal and quantum fluctuations, which jointly determine the metrological performance of quantum critical sensors, as illustrated in the LZ model discussed in Section~\ref{subsec_sp_thermal}.

The coveted Heisenberg scaling of precision in quantum measurements can be attained in both the Ramsey interferometric scheme utilizing entangled GHZ states (as we have seen in {Section~\ref{sec:ramsey_example}}) and in the TFIM at the quantum critical point. Nevertheless, the underlying physical principles facilitating the attainability of this precision are fundamentally different. To clarify this distinction, one may regard the Ramsey protocol and the TFIM as paradigmatic examples of regular quantum metrology and critical quantum metrology, respectively. While the GHZ-based approach achieves enhanced sensitivity through maximally entangled probe sates and collective phase accumulation, critical quantum metrology explicitly leverages the diverging ground state properties and long-range correlations near quantum critical points as the driving metrological resource. The appealing aspect of critical quantum metrology is that such schemes are, for the most part, basically unencumbered by the practical necessities required to prepare highly entangled states explicitly and, furthermore, are more robust against decoherence effects which deteriorate any perceptible quantum advantage. 

\subsection{Quantum Fisher information}
Quantum critical points are characterized by dominant quantum fluctuations, the emergence of universality, and long-range correlation length scales. Furthermore, critical points also mark the regime where the ground state geometry changes profoundly. In the TFIM case, we have motivated how, from a geometric perspective, this extreme sensitivity to infinitesimal variations naturally translates into metrological utility---illustrated through the intrinsic relationship between fidelity susceptibility and QFI for pure states. By exploiting the tensor product structure of the TFIM, ground state $|\psi_{GS}(\omega)\rangle = \bigotimes_{k>0}|\psi_{GS}(\omega)\rangle_k$, we can directly compute the QFI to quantitively reveal its scaling with system size and divergence at the critical point, independently of the fidelity susceptibility. We obtain
\begin{align}
    \mathcal{I}_\omega = \sum_{k} 
    \frac{g^2 \sin^2{k}}{\left(g^2 + \omega^2 - 2g\omega\cos{k}\right)^2}\;.
\end{align}
This summation arises from the contributions of different momentum modes to the total QFI. At the critical point ($g = \omega$), the QFI simplifies significantly, and its elements coalesce into a triangular sequence. The total QFI at criticality is found to scale as
\begin{align}\label{eq:qfiISING}
    \mathcal{I}_\omega = \frac{N^2 + N}{8\omega^2}\;.
\end{align}
This scaling, which is quadratic in the system size $N$, corroborates our findings from Section~\ref{sub_sec_fid_sup}. Such a scaling is characteristic of critical many-body systems and is significantly better than the linear scaling observed in non-interacting systems. 

\subsection{Energy gap and adiabaticity in the transverse field Ising model}
In both the harmonic oscillator and the Landau–Zener models, we have seen that the QFI is not just a formal quantity. The QFI is deeply tied to the physical structure of the system---in particular to the energy spectrum. Endeavouring to extract the predicted QFI sensitivity requires that one must pay close attention to the size of the energy gap between the ground and the first excited state. This is because near a critical point the gap closes. This on the one hand leads to the dramatic enhancement of the QFI, but on the other hand increases the difficulty in preparing the ground state. In these examples, we have assumed that the system evolves adiabatically, and have relied on the fact that the system remains in its instantaneous ground state as the system parameters are changed. This assumption implies that timescale becomes prohibitively long as the gap shrinks. The result of this is a tension between the promise of criticality for metrology and the practical challenge of state preparation---which is a central theme in critical quantum metrology with prepared critical states. For this reason, we must keep the following aspect firmly in mind: while the closing of the energy gap near the critical point is the very source of enhanced sensitivity, it also represents the main obstacle in accessing those optimal states in the laboratory.

The instantaneous energy gap for the TFIM, as a function of $N$, is given by
\begin{align}
    \epsilon = 2 \sqrt{\omega^2 + g^2 - 2\omega g \cos\left(\frac{2\pi}{N}\right)}\;.
\end{align}
At the critical point ($g = \omega$) this reduces to
\begin{align}
    \epsilon \approx \frac{4 \pi \omega}{N}\;.
\end{align}
Thus, the gap scales inversely with $N$, implying that the time required to adiabatically evolve the system increases as the system size grows. Using techniques similar to the LZ model, the total adiabatic evolution time from $g = 0$ can be calculated as
\begin{align}
    T = \frac{1}{4 \gamma \omega }\left(\cot(k) - \frac{\csc(k) \left[g - \omega \cos(k)\right]}{\sqrt{g^2 - 2g\omega \cos(k) + \omega^2}}\right)\;,
\end{align}
where $k = 2\pi/N$. In the limit of large $N$ and at the critical point $g=\omega$, the adiabatic evolution time becomes
\begin{align}
    T = \frac{N}{8 \pi  \gamma  \omega }\;.
\end{align}
This result allows us to rewrite the QFI as
\begin{align}
    \mathcal{I}_\omega = 8 \pi^2 \gamma^2 T^2 + \frac{\pi \gamma T}{\omega}\;,
\end{align}
which does not exhibit an explicit scaling with the number of spins $N$. This is an interesting and pedagogically valuable result. If we examine the QFI for the TFIM, as given in Eq.~\eqref{eq:qfiISING}, it initially appears promising with the Heisenberg scaling even though the system exhibits only nearest neighbour interactions.

Ultimately, what matters in metrological protocols is the precision per unit of time. This again highlights the critical importance of carefully considering both resource contributions and time constraints when interpreting the QFI in critical metrology. Ignoring these factors can lead to misleading conclusions about the efficiency of different metrological strategies. At an extreme, a protocol that requires a very long time will naturally accumulate a large QFI compared to a faster protocol, simply due to the time dependence of the QFI. In such cases, the apparent gain in sensitivity might be illusory. A metrological strategy that achieves high precision only in the infinite-time limit is of limited practical value. Factoring out time can reveal that a protocol with a lower QFI may, in fact, be more efficient. This consideration becomes especially crucial in systems where the protocol duration inherently scales with system size.

\subsection{Accessible measurements}
The QFI quantifies the ultimate achievable sensitivity to changes in a parameter and assumes optimal measurement which extracts this precision. Independent of the specific measurement employed, you cannot surpass this bound. In the TFIM, the super-extensive scaling of the QFI at the quantum critical point originates from the diverging correlation length and accompanying non-analytic changes in the ground state. At criticality, the fidelity susceptibility diverges, reflecting that even an infinitesimal shift in the target parameter produces a macroscopically distinct state. This dramatic ground state restructuring translates directly into enhanced metrological utility, as discussed in Section~\ref{sub_sec_fid_sup}. 

A natural question therefore arises: \emph{Which measurements at criticality provide access to this enhanced sensitivity?} In practice, not all observables constitute measurements that are equally responsive to the diverging correlation length scales near a quantum phase transition. Indeed, some operators couple directly to critical modes, and thus their expectation values or fluctuations exhibit non-analytical behaviour reflecting the underlying universality class. However, other observables remain finite and analytic across the transition, and are therefore effectively ``blind'' to the singular features of the ground state. Implementing such measurements may render the sensor agnostic to the very critical fluctuations that provide the metrological advantage offered by critical quantum systems.

Focusing on the choice of measurement, we will explore below how the observable probed plays an intrinsic role to extracting the metrological potential of critical systems. This highlights a key point. Designing a sensor with exceptional intrinsic sensitivity is not sufficient on its own, and it is equally essential to implement measurement strategies that can fully harness the underlying features responsible for that sensitivity.

\subsection{Local measurement}
In the LZ model, discussed in Section~\ref{sec:QPT}, we considered the sensitivity achievable when measuring the magnetization $\hat{\sigma}_z$. While this strategy was optimal in the context of ground state measurements, the LZ model lacks key features of true many-body quantum criticality---most notably, long-range correlations. To address this, we now extend the analysis to a genuine many-body setting and examine the metrological performance of a local observable across a quantum phase transition. The average magnetization along the field direction $z$ is given by~\cite{barouch1970statistical,PFEUTY197079,PFEUTY197079TFIM}
\begin{equation}
    \langle \hat{\sigma}_z^{(i)} \rangle = -\frac{2}{N}\sum_k \frac{\omega - g\cos{k}}{\sqrt{g^2 + \omega^2 - 2g \omega \cos{k}}} \;.
\end{equation}
We note that $\langle \hat{\sigma}_z^{(i)} \rangle $ is even under the $\mathbb{Z}_2$ symmetry and therefore this choice of observable does not directly probe the symmetry breaking. Instead, it reflects the degree of spin alignment with the field direction. Importantly, this observable is finite on either side of the transition and varies smoothly across the critical point. In the thermodynamic limit, a closed form solution can be obtained at the critical point $g=\omega$
\begin{equation}
    \label{eq:ising_sz}
    \langle \hat{\sigma}_z^{(i)}\rangle = \frac{1}{\pi} \int_0^\pi \frac{\omega - g\cos{k}}{\sqrt{g^2 + \omega^2 - 2g \omega \cos{k}}} dk = \frac{2}{\pi}\;.
\end{equation}
In the paramagnetic phase, $\langle \hat{\sigma}_z^{(i)}\rangle$ is large and approaches its maximum value of $-1$ as the transverse field begins to dominate, whereas in the ferromagnetic phase $\langle \hat{\sigma}_z^{(i)}\rangle$ decreases (though remains finite) as $\omega\!\to\!0$. At the critical point, $\omega = g$, the observable remains finite and analytic.

For a finite system, the first-order expansion captures the finite-size scaling of the magnetization
\begin{equation}
    \langle \hat{\sigma}_z^{(i)}\rangle \propto \frac{4}{N}\;.
\end{equation}
We are interested in observing the consequences of this behaviour in terms of the offered metrological capacity of this local measurement scheme. We will again consider the SNR, defined as
\begin{equation}
    S^{\text{loc}}_\omega = \frac{|\partial_\omega \langle \hat{\sigma}_z^{(i)}\rangle|^2 }{\text{Var}(\hat{\sigma}_z^{(i)})}\;.
\end{equation}
To compute the SNR, we first require the magnetic susceptibility:
\begin{equation}
    \chi_z = \partial_\omega \langle \hat{\sigma}_z^{(i)} \rangle = -\frac{2}{N} \sum_k \frac{g^2 \sin^2{k}}{(g^2 + \omega^2 - 2 g \omega \cos{k})^{3/2}}\;,
\end{equation}
which is non-analytic at the critical point $\omega = g$. Performing a first order expansion of the magnetic susceptibility, we can obtain its finite sized scaling at criticality:
\begin{equation}
    \chi_z(k) \approx \frac{2}{N \omega k } + \mathcal{O}(k^2)\;,
\end{equation}
where higher order terms are neglected. Integrating this expression reveals the scaling with $N$:
\begin{equation}
    \int_{\pi/N}^\pi \frac{2}{N \omega k} dk = \frac{2\log(N)}{N \omega}\;,
\end{equation}
which shows a weak logarithmic divergence at the critical point. Using this expression, coupled with the fact that $\text{Var}(\hat{\sigma}_z^{(i)}) = 1 - \langle \hat{\sigma}^{(i)}_z \rangle^2$, we can obtain the SNR analytically [Shown in Fig.~\ref{fig:ising_scaling} (a) in red circles]. By combining these two results, we obtain the scaling of the SNR with system size
\begin{equation}
    S^{\text{loc}}_\omega = \frac{|\partial_\omega \langle \hat{\sigma}_z^{(i)}\rangle|^2}{\text{Var}(\hat{\sigma}_z^{(i)})} \propto\frac{\left[\log{N}\right]^2}{4\omega^2}\;.
\end{equation}
The extractable precision of the magnetization shows a precision scaling which is significantly weaker than the optimal QFI. This demonstrates clearly that the presence of a quantum critical point does not, on its own, guarantee enhanced sensitivity. The choice of measurement itself must be carefully chosen to probe the relevant diverging length scale and symmetry-breaking sectors of the system. In this case, the weak logarithmic singularity of the signal derivative, $\partial_\omega \langle \hat{\sigma}_z^{(i)} \rangle$, is directly reflected in the attainable precision. This behavior arises because $\langle\hat{\sigma}_z^{(i)}\rangle$ is even under the $\mathbb{Z}_2$ symmetry and thus does not couple directly to the symmetry-broken modes, remaining finite and smooth across the transition. As a result, the susceptibility exhibits only a weak (logarithmic) singularity, and fails to capture the divergent correlation length. Another important consideration is the fact that such a local measurement has a reduced capacity for probing the correlation and entanglement accrued within the ground state---a direct consequence of not capturing the correlation length scale.  

In general, the precision is governed by how sharp the measurement observable changes near the critical point, and how much variance it has. The considered local quantity, \( \langle \hat{\sigma}_z^{(i)} \rangle \), is invariant to the $\mathbb{Z}_2$ spin-flip symmetry and is agnostic to the diverging length scale. Therefore, it has a reduced capacity for accessing the inherent critical like features. This highlights a general aspect of quantum metrology. One must not only consider the ultimate sensitivity promised by a system or probe, but also the practical accessibility of the measurement. The achievable precision reflects a delicate balance between the choice of observable, the resources required to implement it, and the extent to which it captures the underlying critical behaviour.

\subsection{Collective measurement}
In contrast with the local measurement scheme, we now employ a collective measurement strategy. Collective measurements consist of observables acting on the entire ensemble of particles, rather than on a single subsystem. In particular, we consider the collective magnetization
\begin{equation}
    \langle \hat{S}_z\rangle = \left\langle \frac{1}{2} \sum_i\hat{\sigma}_z^{(i)}\right\rangle\;,
\end{equation}
which takes a concise analytical form by utilizing the definition in Eq.~\eqref{eq:ising_sz} of \(\langle \hat{\sigma}_z^{(i)}\rangle \). This collective measurement retains the $\mathbb{Z}_2$ spin-flip symmetry and therefore remains agnostic to the symmetry broken phases. Importantly, its fluctuations involve two-point correlators
\begin{align}
    \label{eq:ising_coll_fluc}
    \text{Var}( \hat{S}_z ) &=  \frac{1}{4}\sum_{i,j} \left( \langle \hat{\sigma}_z^{(i)} \hat{\sigma}_z^{(j)} \rangle - \langle \hat{\sigma}_z^{(i)}\rangle \langle \hat{\sigma}_z^{(j)} \rangle \right)  \;,
\end{align}
whose correlations do not vanish under the $\mathbb{Z}_2$ symmetry. Therefore, the fluctuations are symmetry-resolving in contrast to the fluctuations of the local scheme. By expanding the square in Eq.~\eqref{eq:ising_coll_fluc}, the fluctuations can be written as 
\begin{equation}
    \text{Var}(\hat{S}_z) = \frac{N}{4} + \frac{N}{4}\sum_{r \neq 0}^{N} \langle \hat{\sigma}_z^{(0)} \hat{\sigma}_z^{(r)} \rangle - \frac{1}{4}\langle \hat{S}_z\rangle^2\;,
\end{equation}
with \(r = j - i\) being the separation distance for two spins in the chain. The two-point functions for a finite sized system at zero temperature are given by~\cite{barouch1970statistical,PFEUTY197079,PFEUTY197079TFIM,PhysRev.149.380,LIEB1961407,PhysRevA.4.2331}
\begin{equation}
    \langle \hat{\sigma}_z^{(0)} \hat{\sigma}_z^{(r)} \rangle = \langle \hat{\sigma}_z^{(i)}\rangle^2 - G_r \times G_{-r}\;,
\end{equation}
with 
\begin{equation}
    G_r= -\frac{4}{N} \sum_k \frac{1}{\epsilon_k} \left[ \cos{(rk)(g\cos{(k)} - \omega)}- g\sin{(rk)} \sin(k) \right]\;.
\end{equation}
The metrological precision in estimating our target \( \omega \) can be numerically quantified using the SNR as done before. We find through a numerical fit that for the collective measurement \( \langle \hat{S}_z \rangle \) the precision scales with system size as
\begin{equation}
    S^{\text{col}}_\omega \propto N^{4/3}\;,
\end{equation}
as shown by the blue crosses in Fig.~\ref{fig:ising_scaling} (a). The superior scaling of the SNR for the collective magnetization compared with a local measurement stems from the fact that this choice of observable leverages correlations across the entire system. A local operator such as \( \hat{\sigma}_z^{(i)} \) is symmetry-suppressed and therefore largely insensitive to the diverging correlation length near the critical point (its SNR grows only logarithmically with system size). By contrast, the variance of the collective operator contains contributions from \emph{all} two-point correlators, \( \langle \hat{\sigma}_z^{(i)} \hat{\sigma}_z^{(j)} \rangle \), which near criticality extend over long distances, permitting super-extensive scaling with \( N \).  Physically, this reflects that at criticality the system behaves coherently as a whole, so summing over all spins amplifies the sensitivity to parameter changes much faster than independent local probes could. Collective observables therefore provide a natural way to harvest critical fluctuations for metrological gain. The behaviour of the various observables compared to the optimal measurement for a fixed probe size of $N = 10$ is isolated and compared for clarity in Fig.~\ref{fig:ising_scaling} (b).  

\begin{figure}[t!]
    \centering
    \includegraphics[width=1\linewidth]{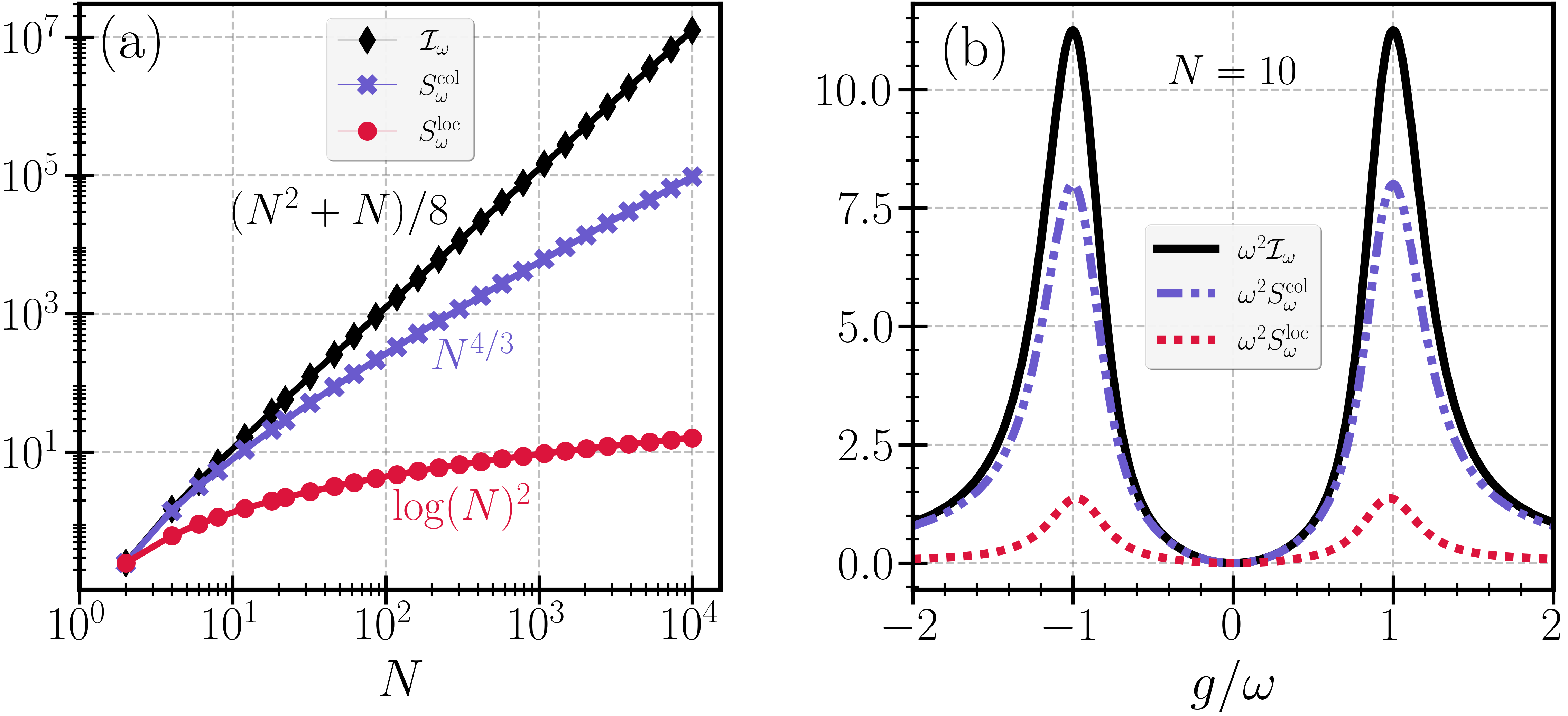}
    \caption{\textbf{Measurement strategies and attainable precision in the TFIM.} (a) Scaling of the optimal QFI, $\mathcal{I}_\omega$, (black diamonds) with system size $N$, compared to the SNR from collective spin measurements $S^{\text{col}}_\omega$ (yellow crosses) and local spin measurements $S^{\text{loc}}_\omega$ (red circles), at criticality $ g_c =\omega = 1$. While the QFI exhibits maximal $N^2$ Heisenberg scaling, collective measurement retain a sub-Heisenberg quantum advantage ($N^{4/3}$), and the local measurement displays only a weak $\log(N)^2$ growth. (b) Behaviour of the various observables for a fixed system size $N = 10$ as a function of the control parameter $g$, for $\omega = 1$. The QFI upper bounds all achievable precisions, with accessible measurements showing reduced but still criticality-enhanced sensitivity near $g_c = \omega = 1$.}
    \label{fig:ising_scaling}
\end{figure}

At the same time, it is important to recognize that the TFIM is not fully symmetric. Its ground state resides in a restricted symmetry sector rather than in the fully symmetric subspace spanned by Dicke states. As a result, collective spin operators, while extremely effective at capturing global correlations, may miss more subtle, symmetry-breaking features of the state. One may thus view them as coarse-grained probes---ideally suited for exploiting universal critical fluctuations, but not guaranteed to detect every microscopic detail. In practice, this suggests a trade-off. Collective measurements offer the strongest scaling advantages, while more tailored observables would be required to fully resolve the structure of the many-body state.

\graphicspath{{./LMG_Figs}}

\section{All-to-all interacting Lipkin-Meshkov-Glick model: return of the harmonic oscillator}\label{sec:LMG_Model}

In the previous section, we examined the nearest-neighbor interacting TFIM, where each spin interacts only with its immediate neighbors. This locality of interactions is what gives rise to the well-known one-dimensional critical behavior, and it illustrates how correlations spread gradually through the system as the critical point is approached. We now shift our focus to a very different scenario, an all-to-all interacting spin model, in which every spin couples equally to every other spin. At first glance, increasing the number of interactions might suggest that correlations can build up more efficiently, especially near a phase transition. Indeed, the high degree of symmetry in such models changes both the structure of the ground state and the nature of criticality. In this section, we will explore how moving from local to global interactions reshapes the landscape of critical metrology, and what new insights it offers compared to the TFIM.

The Lipkin-Meshkov-Glick (LMG) model is a paradigmatic many-body system introduced in nuclear physics to study collective interactions among spins~\cite{LIPKIN1965188}. It has since become a fundamental model in quantum physics due to its rich phase structure and applicability in various fields, including condensed matter physics, quantum information, and metrology. The model describes \(N\) interacting spins under a collective Hamiltonian given by  
\begin{align}
\hat{H}_{\text{LMG}} = \omega \hat{S}_z -\frac{g}{N} \left(\hat{S}_x^2 - \zeta\hat{S}_y^2\right) \;,
\end{align}
where \(\hat{S}_x, \hat{S}_y, \hat{S}_z\) are collective spin operators, \(g\) controls the strength of the spin-spin interactions, \(\zeta\) governs the anisotropy between the interaction components, and \(\omega\) represents the energy separation between the individual spin levels. Depending on the values of the interaction parameters, the system undergoes a quantum phase transition between a symmetric paramagnetic phase and a symmetry-broken ferromagnetic phase at zero temperature.

The LMG model serves as an ideal testbed for exploring quantum criticality and its implications for metrology~\cite{paris2014cqmLMG,vuletic2023LMG}. Near the critical point, the system exhibits long-range correlations and enhanced sensitivity to external perturbations, making it a prime candidate for parameter estimation tasks. Its analytical tractability in certain regimes and experimental realizations in systems such as trapped ions, superconducting qubits, cavity QED~\cite{Li2022LMG,muniz2020LMG}, and Bose-Einstein condensates~\cite{oberthaler2010lmg} make it an invaluable tool for understanding the interplay between quantum criticality and metrological precision~\cite{PhysRevA.106.062442}. Consequently, the LMG model stands at the forefront of theoretical and experimental studies in critical quantum metrology~\cite{chu2021dynamic, gietka2022speedup, gietka2024tempcqm,PRXQuantum.6.030309}. 

For clarity, we consider an isotropic case for $\zeta = 0$, simplifying the Hamiltonian to  
\begin{align}
    \hat{H}_{\text{LMG}} = \omega\hat{S}_z -\frac{g}{N} \hat{S}^2_x\;,
\end{align}
which can be expressed in the form of Eq.~\eqref{eq:intspins} by replacing $g_{ij}$ with $2g/N$. This isotropic version of the LMG model retains its essential features while exhibiting greater symmetry~\cite{vidal2007LMGtl,lmg2004vidal}. To express the Hamiltonian in terms of bosonic modes, we employ the Holstein-Primakoff transformation
\begin{align}
    \hat{S}_z &= \hat{a}^\dagger \hat{a} - \frac{N}{2}\;, \\
    \hat{S}_- &= \sqrt{N - \hat{a}^\dagger \hat{a}} \, \hat{a}\;, \\
    \hat{S}_+ &= \hat{a}^\dagger \sqrt{N - \hat{a}^\dagger \hat{a}}\;,
\end{align}
where \(\hat{a}\) and \(\hat{a}^\dagger\) are bosonic annihilation and creation operators satisfying \([\hat{a}, \hat{a}^\dagger] = 1\). Neglecting constant terms and performing the following approximation
\begin{align}
    &\hat S_-^2 \approx \hat a\left(N-\hat a^\dagger \hat a\right)\hat a\;,\\
    &\hat S_+^2 \approx \hat a^\dagger\left(N-\hat a^\dagger \hat a\right)\hat a^\dagger\;,
\end{align}
the Hamiltonian becomes 
\begin{align}
    \hat{H} &= \omega \hat{a}^\dagger \hat{a} - \frac{g}{4} \left(\hat{a} + \hat{a}^\dagger\right)^2  \\
    &\quad + \frac{g}{4 N}\left( \hat a^\dagger \hat a^\dagger \hat a\hat a^\dagger +  \hat a \hat a^\dagger\hat a \hat a + 2 \hat a^\dagger \hat a \hat a^\dagger  \hat a\right)\;,
\end{align}
where the second line captures the nonlinear interaction terms with an anharmonicity \(\chi = g/4N\). These nonlinear contributions describe deviations from ideal harmonic behavior and become significant for finite system sizes or near criticality. In the thermodynamic limit (\(N \to \infty\)), the nonlinear terms become negligible, and the Hamiltonian simplifies to
\begin{align}
    \hat{H} \approx \omega \hat{a}^\dagger \hat{a} - \frac{g}{4} \left(\hat{a} + \hat{a}^\dagger\right)^2\;,
\end{align}
which corresponds to the opening harmonic oscillator Hamiltonian previously encountered in Section~\ref{sec:QPT_HO}. However, in the finite-component regime, nonlinear effects persist and can profoundly influence metrological properties. In the following, we will retain the full Hamiltonian, including nonlinear terms, to investigate their impact on critical metrology. Although this approach requires numerical calculations, it provides deeper insight into the role of anharmonicity and finite-size effects near the critical point.

\subsection{Quantum Fisher information for Lipkin-Meshkov-Glick model}\label{sub_sec:LMG_QFI}
As usual, we start with the calculation of the QFI. In the large \(N\) limit (\(N \gg 1\)), the QFI with respect to \(\omega\) can be approximated as~\cite{gietka2022speedup} 
\begin{align}\label{eq:qfiLMG}
    \mathcal{I}_\omega \approx \frac{N^{4/3}}{\omega^2}\;.
\end{align}
Surprisingly, the scaling with the number of spins is actually worse here than in the Ising model QFI from Eq.~\eqref{eq:qfiISING}, even though the interactions are stronger and more global in nature. This counterintuitive result suggests that stronger interactions do not automatically guarantee better metrological performance. However, the picture changes once we take time into account. If the ground state is prepared via adiabatic evolution, the Lipkin–Meshkov–Glick model turns out to be more resource-efficient. When the adiabatic preparation time is properly included, the effective QFI becomes~\cite{gietka2022speedup}
\begin{align}
    \mathcal{I}_\omega \approx \gamma^2 N^{2/3} T^2\;,
\end{align}
where \(\gamma \ll 1\), and \(T\) is the adiabatic time needed to ramp the system from $g=0$ to the vicinity of the critical point. This interesting result can be traced back to the closing of the energy gap with increasing system size, which scales as $N^{-1/3}$~\cite{lmg2004vidal,PhysRevLett.95.050402}—significantly slower than in the TFIM. Consequently, the evolution time scales as $T^2 \sim N^{2/3}$. Importantly, the resulting precision always remains below the standard quantum limit for any system size $N$. In the following sections, we discuss strategies to boost the quantum Fisher information and explore whether optimized protocols can overcome this limitation.

\subsection{Excited state quantum phase transition}\label{sub_sec:Excited_state_LMG}

Previously, we observed that temperature can enhance the QFI by allowing incoherent occupation of excited states, leading to the conclusion that the ground state is not necessarily optimal for metrological purposes. In this section, we explore a different route, instead of relying on thermal effects, we investigate the role of excited state quantum phase transitions~\cite{Cejnar_2021_ESQPT}. Unlike traditional quantum phase transitions, which occur in the ground state, excited state quantum phase transitions manifest as abrupt changes in the structure or properties of excited states. By leveraging these transitions, we aim to uncover novel strategies for enhancing metrological precision. To this end, we first consider the thermodynamic limit of the LMG model which is simply a squeezed harmonic oscillator (see Section~\ref{sec:QPT_HO}). The $n$-th excited state of such harmonic oscillator takes the form of a squeezed Fock state
\begin{align}
    |n_\xi\rangle = \hat{S}(\xi) |n\rangle\;,
\end{align}
where $\hat{S}(\xi) = \exp\left[\frac{\xi}{2} (\hat{a}^{\dagger 2} - \hat{a}^2) \right]$ is the squeeze operator, and 
\(
    \xi = \frac{1}{4} \log\left(\frac{\omega}{\omega - g}\right)
\)
is the squeezing parameter. Substituting this excited state into the expression for the QFI yields
\begin{align}
    \mathcal{I}_\omega = 2(n^2 + n + 1) \left( {\partial_\omega \xi} \right)^2\;.
\end{align}
This result shows that the QFI for a squeezed Fock state is $(n^2 + n + 1)$ times larger than that of the corresponding squeezed vacuum (ground) state. The enhancement can be intuitively understood by noting that squeezed Fock states exhibit stronger non-classical correlations compared to squeezed vacuum. However, there is a trade-off. Squeezed Fock states are no longer Gaussian states, which complicates the task of finding optimal measurement strategies. This limitation suggests that while excited states offer enhanced metrological performance, practical implementation may require sophisticated measurement schemes.  Last but not least, it is much harder to manipulate an excited state, especially when the system is open and can lose energy to the environment.

Another limitation comes from the finiteness of the quantum system. While the result in the thermodynamic limit suggests exciting the system as much as possible, this is clearly not feasible for a finite system. In particular, the maximally excited state of a non-interacting spin system is Gaussian (the fully inverted ensemble). Intuitively, the optimal excited state should lie in the middle of the spectrum, corresponding to a Dicke state with maximal Bloch sphere radius. However, since the harmonic oscillator approximation breaks down near the middle of the spectrum, the true optimal excited state lies well below it. Numerical calculations (up to 1500 spins) indicate that the optimal state is approximately the $N/4$-th excited state, for which the QFI is enhanced by a factor of $N^{2/3}/6$, yielding
\begin{align}
    \mathcal{I}_\omega \approx \frac{\gamma^2}{6} N^{4/3}T^2\;,
\end{align}
which surpasses the standard quantum limit once the number of spins reaches a few thousand (assuming $\gamma \approx 0.01$). 

The QFI and SNR for various excited states of the LMG model are shown in Fig.~\ref{fig:LMGexcited}. It can be clearly seen that the QFI increases for low-lying excited states and decreases again for higher excitations. The optimal point for the ground state does not have to coincide with that of the excited states [see Fig.~\ref{fig:LMGexcited}(a)]. This gain in precision, however, comes with a trade-off since the optimal excited states are highly non-Gaussian and exhibit complex correlations, which makes them harder to probe experimentally. Unlike ground states or coherent spin states where simple collective measurements suffice, extracting metrological information from excited states may require access to higher-order observables or even full probability distributions [see Fig.~\ref{fig:LMGexcited}(b)]. In the worst-case scenario one might need to perform quantum state tomography, which becomes prohibitively costly for macroscopic systems due to the exponential growth of the required measurement settings with system size. A complimentary approach to excited state quantum phase transitions is to leverage transitions which manifest across the full spectrum~\cite{li2025nonequilibriumcriticalityenhancedquantumsensing}.

\begin{figure*}
    \centering
    \includegraphics[width=0.8\linewidth]{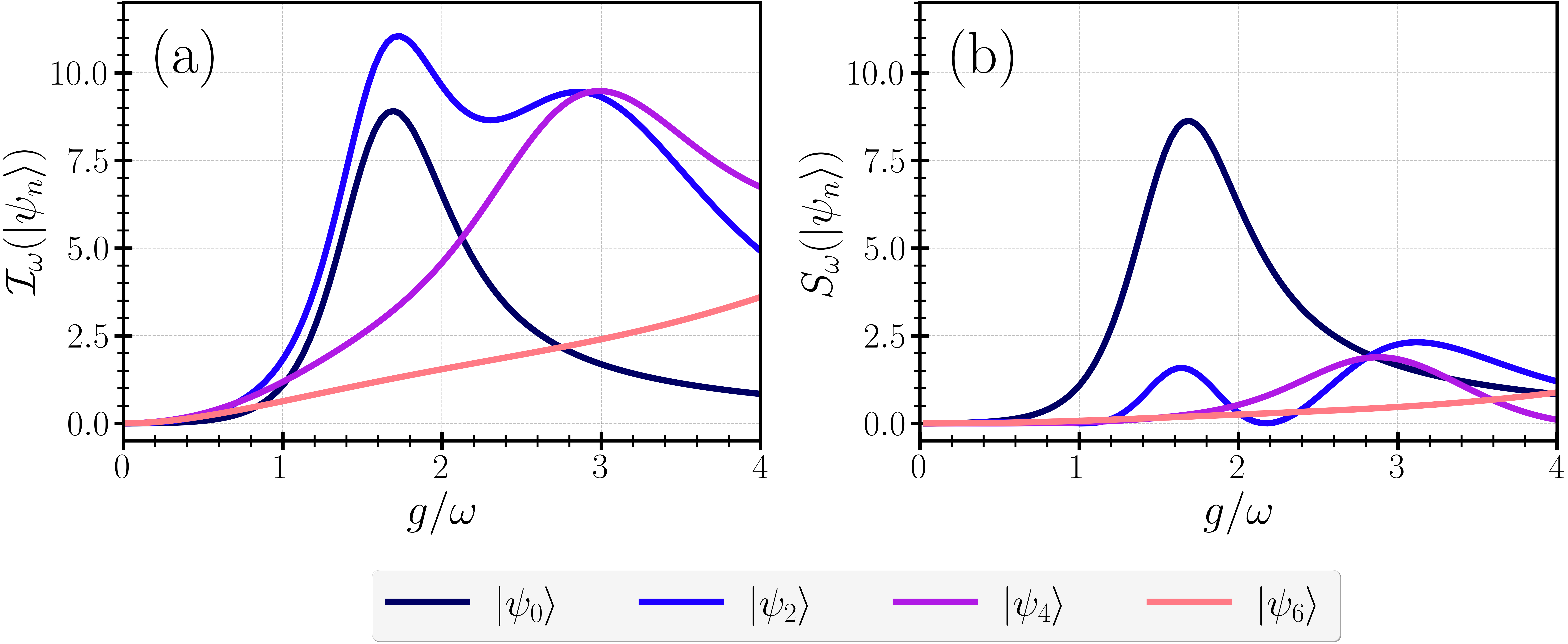}
   \caption{\textbf{Excited-state quantum phase transition in the LMG model for \( N = 10 \).} (a) QFI for the ground state (black) and the 2nd (blue), 4th (violet), 
and 6th (pink) excited states. The QFI initially increases for low-lying excited states but decreases 
again at higher excitations. 
(b) Corresponding SNR for measurements of 
\(\langle \hat S_z \rangle\) (equivalently \(\langle \hat S_x^2 \rangle\)). For the ground state, the 
SNR closely follows the QFI, reflecting its nearly Gaussian character (exact in the thermodynamic limit). 
In contrast, for excited states the SNR is substantially lower than the QFI, as their non-Gaussian nature 
requires more sophisticated measurement strategies to access the available metrological information.}

    \label{fig:LMGexcited}
\end{figure*}

\subsection{Dynamical sudden quench protocol}\label{sub_sec:Dyanical_LMG}
If ground states are not necessarily optimal and excited states can significantly enhance the QFI, one may naturally ask: why use an adiabatic ramp at all? Rapidly ramping the interaction parameter from zero to the vicinity of the critical point could not only boost the QFI by generating non-adiabatic excitations but also reduce the total protocol duration~\cite{chu2021dynamic}. To explore this possibility, let us once again return to the thermodynamic limit of the LMG model, which maps to a harmonic oscillator that progressively opens up as the coupling strength increases. The QFI as a function of time can be calculated in this case as
\begin{align}
    \mathcal{I}_\omega = 4 (\Delta \hat h_\omega)^2\;,
\end{align}
where
\begin{align}
\begin{split}
\hat h_\omega  = &{t}\,\hat a^\dagger \hat a -\frac{g}{2 \omega^2}\frac{\cos\left(\sqrt{\delta}\omega t\right)-1}{ \delta}\frac{\hat a^2 - \hat a^{\dagger 2}}{i} \\
&+\frac{g }{\omega^2}\frac{\sin\left(\sqrt{\delta} \omega t\right)-\sqrt{\delta} \omega t}{\sqrt{\delta}\delta}\left(\hat a^2 +\hat a^{\dagger 2}\right)  \\
&- \frac{g^2}{\omega^3}\frac{\sin\left(\sqrt{\delta} \omega t\right)-\sqrt{\delta} \omega t}{\sqrt{\delta}\delta}\frac{\left(\hat a + \hat a^\dagger\right)^2}{2}
\end{split}
\end{align}
is the operator which imprints the information about the unknown parameter onto the initial state [with $\delta =  4(1-{g}/{\omega})$ ]. For an initial ground state and $g\approx \omega$, the QFI becomes 
\begin{align}
    \mathcal{I}_\omega \approx \frac{2  }{\omega^2} \left(\frac{\sin\left(\sqrt{\delta} \omega t\right)-\sqrt{\delta} \omega t}{\sqrt{\delta}\delta}\right)^2\;,
\end{align}
which exhibits an oscillatory behavior with a frequency given by twice the natural frequency of the squeezed harmonic oscillator $2 \omega \sqrt{1-g/\omega}$. Intuitively, when the harmonic oscillator suddenly opens, the wavefunction begins to expand, generating squeezed excitations and thereby increasing the QFI. If $g < g_c$, this squeezing eventually stops as the wavefunction hits the potential, and a reverse process of unsqueezing sets in, which reduces the QFI. However, at the critical point $g = g_c$, the effective harmonic potential vanishes, and the wavefunction undergoes perpetual squeezing, resembling the expansion of a free particle wavefunction. As a consequence, the QFI becomes
\begin{align}
    \mathcal{I}_\omega \approx \frac{\omega^4 t^6   }{18 }=  \frac{8}{9} {\langle \hat n \rangle^2 t^2}\;,
\end{align}
where $\langle \hat n \rangle = \omega^2 t^2/4$ is the instantaneous number of excitations. Note how close to the Heisenberg limit it is.

While this perpetual squeezing sounds ideal, there are faster and more efficient methods for generating squeezing which means that the instantaneous number of excitations can grow faster than quadratically with $t$. In previous sections, we briefly introduced the two-axis counter-twisting Hamiltonian, known for producing optimal squeezing. In the thermodynamic limit $N \to \infty$, this Hamiltonian reduces to
\begin{align}
    \hat H = \chi\left(\hat S_y^2 - \hat S_x^2\right) \approx \chi\left(\hat p^2 - \hat x^2\right)\;,
\end{align}
which effectively describes an inverted harmonic oscillator~\cite{PhysRevE.104.034132}. In this regime, the oscillator's frequency becomes imaginary. Interestingly, this transition occurs naturally when $g > g_c$, meaning that the system is quenched beyond the critical point of the quantum phase transition. But how can a Hamiltonian for an inverted harmonic oscillator make sense when it is unbounded from below and has no ground state? The inverted harmonic oscillator is an idealization valid only in the thermodynamic limit $N \to \infty$. Physically, quenching beyond the critical point corresponds to transforming the harmonic potential into a double-well potential. If the bottom of such a double-well potential is infinitely low---as a consequence of taking the $N \to \infty$ limit and neglecting the nonlinear terms---squeezing can continue indefinitely. However, in any finite system, the wavefunction will eventually bounce back from the bottom of the double-well, limiting the amount of achievable squeezing and thereby constraining the growth of the QFI.

Under ideal conditions, quenching as far as possible beyond the critical point appears optimal for maximizing both the QFI and the SNR. The dynamics of the inverted harmonic oscillator naturally generate Gaussian states, which are easier to measure optimally. In this regime, the QFI becomes
\begin{align}
    \mathcal{I}_\omega \approx \frac{2 g^3 }{\omega^5} \frac{\sinh^2\left(\sqrt{|\delta|} \omega t\right)}{|\delta|^3} \approx \frac{ g^3 }{2 \omega^5} \frac{\exp\left(2\sqrt{|\delta|} \omega t\right)}{|\delta|^3}\;,
\end{align}
where the exponential term reflects the enhanced squeezing dynamics of the inverted harmonic oscillator and the faster generation of excitations.

For the special case of $g = 2 \omega$ (see Ref.~\cite{Gietka2022understanding} for a general case), the Hamiltonian becomes
\begin{align}
    \hat H = - \frac{\omega}{2} \left(\hat a^2 + \hat a^{\dagger 2}\right)\;,
\end{align}
whose time evolution operator is the squeeze operator
\begin{align}
    \hat U(t) = \exp\left[  \frac{i \omega t}{2} \left(\hat a^2 + \hat a^{\dagger 2}\right)\right] = \exp\left[  \frac{1}{2} \left(\xi^*\hat a^2 - \xi \hat a^{\dagger 2}\right)\right] \;,
\end{align}
with the squeezing parameter $\xi = \omega t \exp(-i \pi/4)$. This implies that for $g = 2 \omega$, the wavefunction evolves through the eigenstates of the harmonic oscillator with $g \leq g_c$. For this specific choice of $g$, the number of excitations as a function of time is given by
\begin{align}
    \langle \hat n \rangle = \sinh^2( \omega t) \approx \frac{\exp(2 \omega t)}{4}\;.
\end{align}
The corresponding QFI becomes
\begin{align}
    \mathcal{I}_\omega \approx \frac{1}{2 \omega^2} \frac{\exp\left(4 \omega t\right)}{8} \approx \frac{\langle \hat n \rangle^2}{\omega^2}\;,
\end{align}
while the time required to reach this regime is
\begin{align}
    T \approx \frac{\log\left(4 \langle \hat n \rangle\right)}{2 \omega}\;.
\end{align}
This allows us to rewrite the QFI as
\begin{align}
    \mathcal{I}_\omega \approx \frac{4 \langle \hat n \rangle^2 T^2}{\log\left(4 \langle \hat n \rangle\right)^2}\;,
\end{align}
which no longer exhibits Heisenberg scaling.

It is important to emphasize that the absence of Heisenberg scaling does not imply an inferior strategy. Ultimately, it is the absolute value of the QFI after a given time that determines the efficiency of a metrological protocol, rather than its scaling behavior. This distinction is particularly significant in critical metrology, where the number of excitations is inherently linked to the required time. Expressing both QFI expressions as functions of time yields (quenching to the critical point)
\begin{align}
    \mathcal{I}_\omega \approx \frac{\omega^4 t^6}{18}\;,
\end{align}
and (quenching beyond the critical point) 
\begin{align}
    \mathcal{I}_\omega \approx \frac{\exp(4 \omega t)}{16 \omega^2} = \frac{32}{5}\frac{\omega^4 t^6}{18} + \frac{1}{16 \omega^2}\sum_{n \neq 6} \frac{\left(4 \omega t\right)^n}{n!}\;,
\end{align}
which clearly demonstrates the superiority of the quenching beyond the critical point strategy. This again highlights a potential pitfall of focusing solely on scaling with respect to excitations: optimizing the scaling behavior does not necessarily lead to optimal QFI.

However, while this approach appears promising, it presents several significant challenges. First, quenching beyond the critical point involves injecting substantial energy into the system, often by increasing laser power or another control parameter. This sudden energy increase can destabilize the system and cause it to deviate from the ideal two-level model, inadvertently coupling to additional energy levels. As a result, the system may become chaotic, rendering it unsuitable for precise sensing applications. Second, the fast dynamics induced by the abrupt quench introduce severe constraints on system control. Maintaining coherence and accurately tracking the evolving quantum state become more challenging, particularly when measurement hardware lacks the necessary bandwidth or precision. These rapid dynamics also increase the likelihood of decoherence and heating, which degrade system performance. Finally, the inherent non-adiabatic nature of the quench leads to unpredictable excitations and complex dynamics, making it harder to design optimal measurement strategies. In finite systems, the wavefunction will eventually reach the bottom of the double-well potential, setting an upper bound on the achievable squeezing and thus the QFI.

Despite these challenges, quenching beyond the critical point remains a promising avenue for metrology, particularly if experimental limitations can be mitigated. Advanced techniques, such as optimal control~\cite{duncan2025tamingquantumsystemstutorial} and reservoir engineering, may help overcome some of these obstacles and unlock the full potential of critical quantum sensing strategies. 
\graphicspath{{./HO_Figs/}}

\section{Driven-dissipative Kerr resonator}\label{sec:HO}

Up to this point, we have focused on toy and paradigmatic models of quantum phase transitions. While these considerations have been extremely insightful, they omit certain unavoidable ingredients—most notably, coupling to the environment, which leads to dissipation and relaxation. In this section, we turn to open quantum systems, which can display critical behavior arising from the interplay of external driving and dissipation. Such driven–dissipative phase transitions are formally defined as non-analyticities in the system’s steady-state manifold as a function of its physical parameters \cite{Minganti2018Spectral}. 

Our primary aim here is to illustrate, through a concrete example, how dissipation modifies the fundamental scaling of estimation precision in critical quantum sensing. It has been formally proven that, under very general assumptions and regardless of the considered sensing strategy, the Heisenberg scaling cannot be achieved for arbitrarily long times~\cite{noisy2011metrology, demko2012elusiveHL, maccone2014noiseagainst}, and the quantum advantage asymptotically comes down to a constant. More specifically, the QFI is bound to grow at most linearly, unless the generator of the evolution cannot be expressed as a linear combination of Lindblad operators~\cite{preskill, PhysRevLett.131.090801}, a condition that is rarely met in relevant scenarios~\cite{Gorecki2025}. 

Accordingly, we will show that critical protocols based on the system's steady state can achieve at most a linear scaling of the QFI. However, we will then show how the dynamical critical properties can be exploited to achieve the Heisenberg scaling for finite evolution times. In particular, in Section~\ref{HeisenbergToDissipative}, we analyse the transition between the Heisenberg and the dissipative regimes. Finally, in Section~\ref{Dissipative_dynamical_protocol}, we show how to identify the optimal working regime of a dissipative critical protocol for finite resources.

To highlight the key differences and challenges that arise in open quantum systems, we once again consider a minimal model that is both conceptually simple and highly relevant from an experimental point of view~\cite{beaulieu2025observation,beaulieu2025Criticality}, that is, a parametrically pumped single-mode Kerr resonator~\cite{PhysRevA.94.033841,PhysRevX.10.021022,c91r-8t3h}, described by the Hamiltonian
\begin{equation}
\label{Kerr_Hamiltonian_t}
    \hat H = \omega_r \hat a^\dagger \hat a + \frac{\epsilon}{2}\left(e^{-i\omega_p t} \hat a^2 + e^{i\omega_p t} \hat a^{\dagger 2}\right) + \chi \hat a^\dagger \hat a^\dagger \hat a \hat a\;,
\end{equation}
where $\omega_r$ is the resonance frequency, \(\chi\) is the Kerr nonlinearity, and \(\epsilon\) is the intensity of the two-photon pump of frequency \(\omega_p\). 
Nonlinear quantum-optical oscillators represented a key testbed for the study of finite-component first-order dissipative phase transitions, both from the experimental~\cite{Chen2023,Brookes2021,FinkNatPhys18,RodriguezPRL17} and the theoretical viewpoints~\cite{Ciuti2016, Simone2020, Minganti2023,Simone2023}. Similar parametric systems have also been considered in sensing applications, such as optical parametric oscillators ~\cite{Gu2025,roques2023biasing,choi2025observing}. Here, we consider in particular the case of parametric (two-photon) pump, where a second-order dissipative phase transition has recently been observed~\cite{beaulieu2025observation,beaulieu2025Criticality}.

By moving into a frame rotating at the frequency $\omega_p/2$, the Hamiltonian of Eq.~\eqref{Kerr_Hamiltonian_t} can be rewritten as
\begin{align}
\label{Kerr_Hamiltonian}
    \hat H = \omega \hat a^\dagger \hat a + \frac{\epsilon}{2}\left( \hat a^2 +  \hat a^{\dagger 2}\right) + \chi \hat a^\dagger \hat a^\dagger \hat a \hat a\;, 
\end{align}
with $\omega = \omega_r - \omega_p/2$. We assume the resonator to be coupled to a zero-temperature Markovian environment, so that the time evolution of the system state \(\hat \varrho\) is governed by the Lindblad equation~\cite{breuer2002theory},
\begin{equation}
\label{Lindblad}
\partial_t \hat \varrho = \mathcal{L}[\hat \varrho] =  -i \left[ \hat H, \hat \varrho \right] + \Gamma \mathcal D [ \hat \varrho]\;,
\end{equation}
where $\mathcal{L}[\cdot]$ is the Liouvillian superoperator and $\Gamma$ denotes the dissipation rate. 
The first term, involving the commutator \([H,\hat \varrho]\), describes the unitary evolution of the system, while the second term, \(\mathcal{D}[\hat \varrho]\) accounts for non-unitary dissipative dynamics. The dissipator is expressed in terms of the bosonic ladder operators \(\hat{a}\) as
\begin{equation}
\label{dissipator}
\mathcal D [\hat \varrho] = 2 \hat a \hat \varrho \hat a^\dagger - \hat a^\dagger \hat a \hat \varrho - \hat \varrho \hat a^\dagger \hat a\;.
\end{equation}
 Our analysis can be easily generalized to include finite-temperature baths~\cite{alushi2024optimality}. We point out that this model is characterized by a $\mathbb{Z}_2$ symmetry, given by the parity of the photon number $\hat \Pi = e^{i \pi \hat a^\dagger \hat a}$. This operator commutes with the system Hamiltonian and remains invariant under the Lindbladian evolution given by Eq.~\eqref{Lindblad}. However, it does not commute with the jump operator (single-photon decay) appearing in Eq~\eqref{dissipator}. This situation is called \emph{weak} symmetry, which implies that the steady state must be an eigenstate of $\hat \Pi$, even if parity is not necessarily conserved during the system dynamics.
 
\subsection{Gaussian approximation: steady-state solutions}
\label{sub_sec:HO_steadystate}
In order to study analytically the phenomenology of the model from Eq.~\eqref{Kerr_Hamiltonian}, we now consider the limit of weak nonlinearity \(\chi\to 0\) and consider only the Gaussian part of the model and its steady-state solution. The saturation effect of the quartic potential has been discussed, for example, in Ref.~\cite{alushi2024collectivequantumenhancementcritical}. Under this Gaussian approximation, the model becomes quadratic and is fully characterized by its first-moments vector \(\textbf{v}\) and covariance matrix \(\bf\Sigma\). For single-mode bosonic systems these objects are defined as \cite{Serafini}
\begin{align}
   {\bf v}&=\begin{pmatrix}
        \langle \hat x\rangle\\
       \langle \hat p\rangle
    \end{pmatrix}\,, \label{fmv1}\\
     {\bf \Sigma}&= \begin{pmatrix}
        2\langle \hat x^2\rangle-2\langle \hat x\rangle^2& \langle\{\hat x,\hat p\}\rangle-2\langle \hat x\rangle\langle \hat p\rangle\\
        \langle\{\hat x, \hat p\}\rangle-2\langle \hat x\rangle\langle \hat p\rangle&2\langle \hat p^2\rangle-2\langle \hat p\rangle^2
    \end{pmatrix} \label{covmatr1} \;,
    \end{align}
with \(\hat{x}=(\hat{a}+\hat{a}^\dagger)/\sqrt{2}\) and \(\hat{p}=-i(\hat{a}-\hat{a}^\dagger)/\sqrt{2}\). From the Lindblad equation~\eqref{Lindblad}, we can derive dynamical equations for the expectation value of any arbitrary quantum operator $\hat{\mathcal{O}}$: 
\begin{equation}
\label{Lindblad_operator}
\partial_t \langle \hat{\mathcal{O}} \rangle = 
- i \left\langle [\hat{\mathcal{O}}, \hat H] - 2\Gamma\hat a^\dagger \hat{\mathcal{O}} \hat a 
- \Gamma  \{ \hat a^\dagger \hat a, \hat{\mathcal{O}} \}\right\rangle\;,
\end{equation}
where we have used the relation $\partial_t \langle \hat{\mathcal{O}} \rangle = \text{Tr}\left[ \hat{\mathcal{O}}\  \partial_t \hat \varrho \right]$. Setting \(\partial_t\langle \hat{\mathcal{O}}\rangle=0\) allows us to obtain the steady-state solution. The first-moments vector vanishes due to the parity symmetry, while the steady-state covariance matrix is given by
\begin{align}
\label{eq:covsteady}
{\bf \Sigma}_{ss}&= \frac{1}{\epsilon_c^2-\epsilon^2} \begin{pmatrix}
   \epsilon^2_c-\omega\epsilon     & -\Gamma \epsilon \\
  -\Gamma \epsilon      &  \epsilon^2_c+\omega\epsilon 
    \end{pmatrix}\;,
\end{align}
which corresponds to a squeezed thermal state. Notice that this solution is physical only when the pump intensity \(\epsilon\) is below the critical point \(\epsilon_c\equiv\sqrt{\omega^2+\Gamma^2}\), i.e., when \(\epsilon<\epsilon_c\). Beyond threshold, the quartic potential must be taken into account to find converging solutions.  The steady-state number of photons \(N_{ss}\) is
\begin{equation}
\label{photnum_SS}
N_{ss}=\frac{\Tr[{\bf\Sigma}_{ss}]-2}{4}=\frac{\epsilon^2}{2(\epsilon_c^2-\epsilon^2)}\;,
\end{equation}
which indeed diverges as soon as the criticality is approached, i.e., in the limit \(\epsilon\to\epsilon_c\). The Kerr nonlinearity \(\chi\) regularizes the model by eliminating the divergences, and ensures that the model is well defined for any values of the pump intensity $\epsilon$. This means that it plays the role of the quartic potential in the toy model of Eq.~\eqref{quartic_resonator}. 

In proximity of the critical point and in the limit \(\chi\to0\), the system~\eqref{Kerr_Hamiltonian} undergoes a second-order driven-dissipative phase transition due to the spontaneous breaking of the \(\mathbb{Z}_2\)-symmetry. We note that the validity of the Gaussian approximation is closely related to the number of photons in the system and the value of the Kerr parameter. When the number of photons is sufficiently high, i.e., as \(\epsilon\to\epsilon_c\), the Kerr nonlinearity can no longer be neglected and the quartic terms in Eq.~\eqref{Kerr_Hamiltonian} must be taken into account. A simple condition for the validity of Eq.~\eqref{photnum_SS} has been derived in~\cite{alushi2024collectivequantumenhancementcritical}.

An effective Gaussian model can also be constructed for $\epsilon>\epsilon_c$, and it becomes increasingly accurate further away from the critical point. In this regime, the steady state is well approximated by a classical mixture of two equiprobable displaced squeezed thermal states $\hat \varrho_{\pm}$~\cite{Ciuti2016, paraoanu2023parametriccritical}. This result can be understood by noting that, for small $\chi$, the model~\eqref{Kerr_Hamiltonian} is well approximated by a double-well potential, and that the low-energy physics can be captured via a quadratic expansion around each minimum. The centers of these minima correspond to semiclassical equilibrium positions, which can be found by applying a displacement operation $\hat U$ such that $\hat U^\dagger \hat{a} \hat U = \hat{a} + \alpha$, with $\alpha$ the displacement parameter. These positions are determined by replacing $\hat{\varrho} \to \hat{U}^\dagger \hat{\varrho} \hat{U}$ and imposing that the linear terms in $\hat{a}$ and $\hat{a}^\dagger$ vanish, leading to two degenerate solutions characterized by an optimal displacement $\alpha_{\rm opt}$ with magnitude $|\alpha_{\rm opt}| = \left(\sqrt{\epsilon^2 - \Gamma^2} - \omega\right)/(2\chi)$ and phase $\arg(\alpha_{\rm opt})=\left[\arcsin\left(\Gamma/\epsilon\right) \pm \pi\right]/2$. We note that, in this manuscript, we will not consider the dynamics beyond the threshold; the detailed structure of the effective Hamiltonian and further corrections in this regime are therefore omitted. The validity of the Gaussian approximation is still determined by the Kerr nonlinearity $\chi$: the smaller $\chi$, the larger the region where the approximation holds.

\subsection{Static protocol}\label{sub_sec:HO_static}
Building on the previous results, we now consider a static sensing protocol for the frequency estimation. This protocol is dubbed \emph{static} as it is based on the steady-state properties. The static protocol follows three key steps:
\begin{itemize}
\item The system is initialized in its steady state with \(\epsilon=0\) (no pumping).
\item The drive is turned on near the critical point, allowing the system to evolve freely according to the Lindblad equation.
\item The system is measured at a chosen time \(t\), which must be long enough for the system to achieve its steady state.
\end{itemize}

Let us first analyze the metrological performance in terms of the scaling, with respect to fundamental resources, of the QFI calculated over the steady-state manifold. As shown in the previous subsection, the steady state of the considered model is Gaussian. We can then use a general formula for the QFI that holds for arbitrary single-mode Gaussian state manifolds. For the estimation of a single parameter \(\omega\), such formula can be computed from first-moments vector \(\bf v\) and covariance matrix \({\bf \Sigma}\) \cite{Serafini}: 
\begin{equation}\label{QFIGaussian}
    \mathcal{I}_\omega=\frac{1}{2}\frac{\Tr{\left({\bf\Sigma}^{-1}{\partial_\omega\bf\Sigma}\right)^2}}{1+\mu^2}+\frac{2(\partial_\omega\mu)^2}{1-\mu^4}+2 (\partial_\omega{\bf v})^{\rm T}{\bf \Sigma}^{-1}\partial_\omega {\bf v}\;,
\end{equation}
where \(\mu=1/\sqrt{\det\left({\bf \Sigma}\right)}\) is the purity of the Gaussian state. We focus on the estimation of the parameter \(\omega\) in Eq.~\eqref{Kerr_Hamiltonian_t}, assuming that all the other Hamiltonian and dissipative parameters are known. We consider the case of \emph{local} parameter estimation---as defined in Section~\eqref{sec:parameter_estimation_theory}---and so we assume that \(\omega=\omega_0+\delta\omega\), where \(\omega_0\) is the prior information over the parameter and \(\delta\omega\) is a small frequency shift we aim to estimate. In this case, the QFI in Eq.~\eqref{QFIGaussian} is evaluated by taking the derivatives and then evaluating the resulting function at $\omega=\omega_0$.  Accordingly, we derive the QFI for the estimation of \(\omega\) by substituting the steady-state solutions of Eq.~\eqref{eq:covsteady} into Eq.~\eqref{QFIGaussian}, yielding   
\begin{equation}\label{QFIss}
  \mathcal{I}_\omega^{ss} =  \frac{1}{2\epsilon_c^2 - \epsilon^2}
 \left[ 2N_{ss} + \frac{8\omega_0^2}{\epsilon^2} N^2_{ss} \right]\;.  
\end{equation}
Notice that, if \(\omega_0\neq0\), when $\epsilon\to \epsilon_c$ the QFI achieves a quadratic scaling with respect to the number of photons, i.e., \(\mathcal{I}_\omega^{ss}\sim 8\omega_0^2N^2_{ss}/\epsilon^4\). However, this quadratic scaling does not take into account the time $T_{ss}$ required to reach the steady state. As discussed in Section~\ref{sub_sec:HO_dynamics}, it can be lower bounded by the inverse of the smallest eigenvalue of the dynamics matrix $T_{ss} \gtrsim \lambda_-^{-1}$. We have that $ \lambda_-^{-1}\sim\Gamma/\epsilon_c(\epsilon_c-\epsilon)$, in the limit $\epsilon\to\epsilon_c$. We then find
\begin{equation}
\mathcal{I}_\omega^{ss} \sim \frac{2 \omega_0^2}{\epsilon^2 \Gamma} T_{ss} N_{ss}\;,
\end{equation}
which shows that, in agreement with fundamental bounds~\cite{Gorecki2025}, for long times the QFI in the dissipative case is asymptotically linear in time and system size.

As emphasized throughout this manuscript, the QFI alone does not provide information about the optimal measurement strategy needed to saturate the upper bound that it sets on the SNR. Therefore, we have to check whether the optimal scaling of Eq.~\eqref{QFIss} can actually be saturated with an experimentally feasible measurement. In our case, homodyne detection turns out to be the optimal measurement scheme, as it can be analytically proven by direct inspection. We evaluate the classical Fisher Information for homodyne detection and show that indeed it saturates the QFI. Recall that implementing homodyne detection involves measuring the quadrature \(\hat x(\varphi)=\left(e^{-i\varphi}a+e^{i\varphi}a^\dagger\right)/\sqrt{2}\), with \(\varphi\in[0,2\pi)\). For single-mode Gaussian states the Fisher information for homodyne detection is given by
\begin{equation}
\mathcal{F}_\omega=\frac{4S(\varphi)\left[\partial_\omega\langle x(\varphi)\rangle\right]^2+\left[\partial_\omega S(\varphi)\right]^2}{2S^2(\varphi)}\;,
\end{equation}
where \(S(\varphi)=\cos^2\left(\varphi\right){\Sigma_{11}}+\sin^2\left(\varphi\right){\Sigma_{22}}-\sin\left(2\varphi\right){\Sigma_{12}}\), \(\Sigma_{ij}\) are the elements of the system's covariance matrix, and the derivatives are evaluated at \(\omega_0\). The explicit expression for the steady-state homodyne Fisher information is~\cite{paraoanu2023parametriccritical}
\begin{equation}
\mathcal{F}_\omega^{ss}=\frac{\epsilon^2\left[\left(\Gamma^2-\omega_0^2-\epsilon^2\right)\cos(2\varphi)+2\omega_0\epsilon+2\omega_0\Gamma\sin(2\varphi)\right]^2}{2(\epsilon_c^2-\epsilon^2)^2\left[\epsilon_c^2-\epsilon\left(\omega_0\cos(2\varphi)-\Gamma\sin(2\varphi)\right)\right]^2}\;.
\end{equation}
One can immediately verify that, for \(\omega_0\neq0\) and in the limit \(\epsilon\to\epsilon_c\), the homodyne Fisher information approaches the QFI, i.e., \(\mathcal{F}_\omega^{ss}\to8\omega_0^2N_{ss}^2/\epsilon^4\). Thus, homodyne detection is the optimal measurement strategy.

\subsection{Gaussian approximation: dynamics}
\label{sub_sec:HO_dynamics}
So far, our discussion has been restricted to steady-state properties, where the system has already relaxed under the combined action of drive and dissipation. However, it is equally important—and often more practically relevant—to ask how the metrological performance evolves during the transient dynamics, before the system settles into its steady state. In what follows, we explore this time-dependent behaviour and assess the potential of dynamical protocols for critical quantum sensing. Under Gaussian approximation, analytical solutions for the system's dynamics can be found by solving the Langevin equations~\cite{Milburn}:
\begin{equation}
\label{in_out_eqs}
\dot{\hat{\textbf{a}}}(t)=A\hat{\textbf{a}}-\sqrt{2\Gamma}\hat{\textbf{b}}(t)\;,
\end{equation}
where \(\hat{\textbf{a}}=(\hat{a}(t),\hat{a}^\dagger(t))^{\rm T}\) is the vector of the bosonic modes and \(A\) is the dynamics matrix defined as
\begin{equation}
    A=\begin{pmatrix}
        -i\omega-\Gamma&&-i\epsilon\\i\epsilon&&i\omega-\Gamma
    \end{pmatrix}\;.
\end{equation}
Equation~\eqref{in_out_eqs} describes the evolution of the resonator field $\hat{a}$, where the quantum noise of the Markovian environment is described by the input mode \(\hat{\textbf{b}}(t)=(\hat{b}(t),\hat{b}^\dagger(t))^{\rm T}\). Since we are working at zero temperature and we have already included the parametric driving as an Hamiltonian term, we assume the input mode to be the vacuum, such that \(\langle \hat{b}^\dagger(t)\hat{b}(t')\rangle=\delta(t'-t)\). Exact solutions to the linear system of first-order differential equations~\eqref{in_out_eqs} can be written as~\cite{alushi2024optimality}:
\begin{align}
    \hat{a}(t)&=c_1(t)\hat{a}(0)+c_2(t)\hat{a}^\dagger(0)-\sqrt{2\Gamma}\int_{0}^{t}c_1(t-\tau)\hat{b}(\tau)d\tau\nonumber\\
    &-\sqrt{2\Gamma}\int_{0}^{t}c_2(t-\tau)\hat{b}^\dagger(\tau)d\tau\;.\label{Langevin solution}
\end{align}
The coefficients \(c_1,\,c_2\) are defined as
\begin{align}
    c_1(x)&=\left(\frac{1}{2}-\frac{i\omega}{2\sqrt{\epsilon^2-\omega^2}}\right)e^{-\lambda_+x}+\left(\frac{1}{2}+\frac{i\omega}{2\sqrt{\epsilon^2-\omega^2}}\right)e^{-\lambda_-x}\;,\\
    c_2(x)&=\frac{-i\epsilon}{2\sqrt{\epsilon^2-\omega^2}}\left(e^{-\lambda_+x}-e^{-\lambda_-x}\right)\;,
\end{align}
where \(\lambda_{\pm}=\Gamma\pm\sqrt{\epsilon^2-\omega^2}\) are the eigenvalues of the dynamics matrix \(-A\). To derive the time-dependent covariance matrix of the system, \({\bf{\Sigma}}(t)\), we can express the position and momentum operators as functions of time using Eq.~\eqref{Langevin solution}. We emphasize that solution from Eq.~\eqref{Langevin solution} is valid only below the critical point, i.e., when \(\epsilon<\epsilon_c\). Under Gaussian approximation, analytical time-dependent solutions can also be found in the regime \(\epsilon>\epsilon_c\). Analogously to the treatment discussed in Section~\ref{sub_sec:HO_steadystate}, a Gaussian description beyond the critical point can be found by considering quantum fluctuations around semiclassical solutions. Since our main interest lies in understanding the behavior as the critical point is approached, in the following we will focus on the case \(\epsilon<\epsilon_c\).

\begin{figure*}
    \centering
    \includegraphics[width=0.9\textwidth]{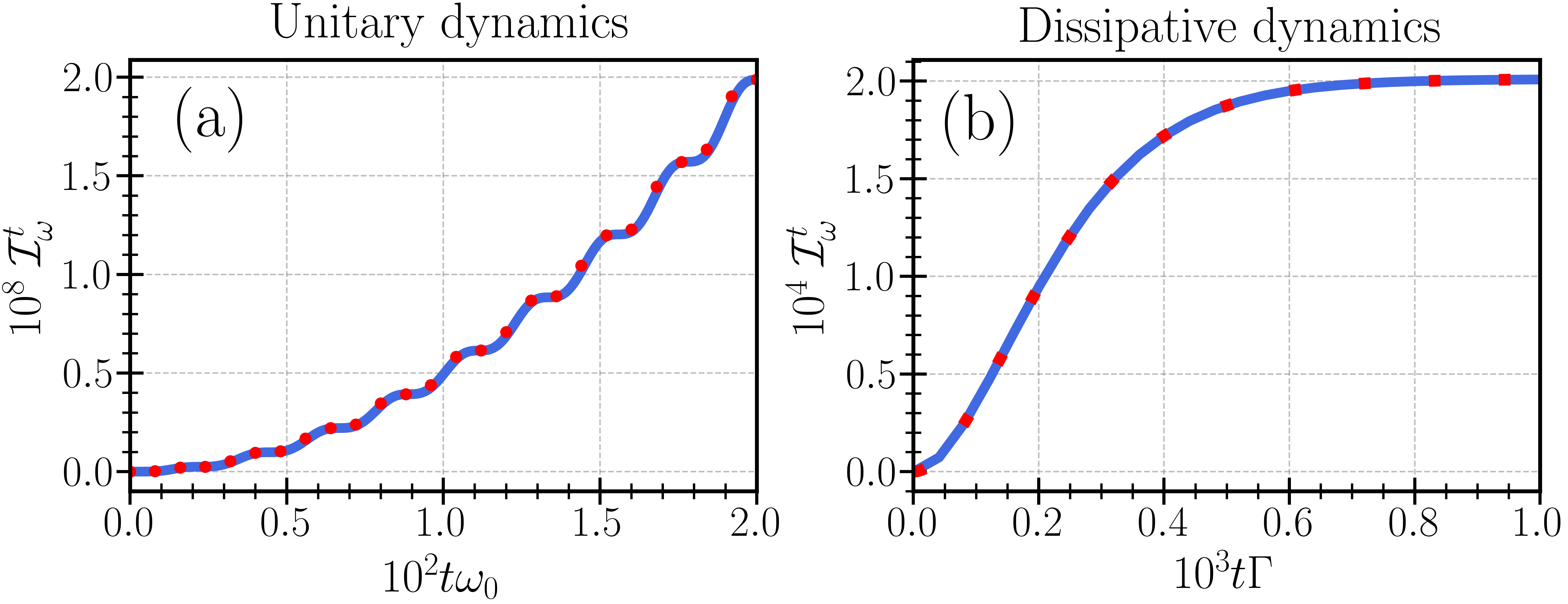}\caption{\textbf{Optimal measurement strategy of dynamical protocols.} We compare the QFI (blue line) and the Fisher information of the homodyne detection (red dots), optimized over the quadrature angle. Panel (a) shows the unitary regime (\(\Gamma=0\)) with \(\epsilon=0.99\omega_0\), while (b) the dissipative regime with \(\omega_0=\Gamma,\,N_{\rm max}=100,\,\epsilon=\epsilon_{\rm opt}=\sqrt{2N_{\rm max}/(1+2N_{\rm max})}\epsilon_c\). The optimized homodyne detection is shown to saturate the QFI.}

    \label{DynamicQFIvsHomodyne}
\end{figure*}

Before proceeding, let us briefly examine the dynamics of the photon number inside the cavity, assuming the system is initialized in the vacuum. The average number of photons at time \(t\) is given by \(N(t)=\langle \hat{a}^\dagger(t)\hat{a}(t)\rangle\) where \(\hat{a}(t)\) is defined in Eq.~\eqref{Langevin solution}. Depending on the value of the pump strength \(\epsilon\), two distinct behaviours of \(N(t)\) emerge.  For \(\epsilon<\omega\), the average photon number is given by
\begin{align}
\label{before_exceptional}
 N(t)&=\frac{\epsilon^2}{2(\epsilon_c^2-\epsilon^2)}\bigg\{1-e^{-2\Gamma t}\Big[\cos\left(2t\sqrt{\omega^2-\epsilon^2}\right)\nonumber\\\quad&+\frac{\Gamma}{\sqrt{\omega^2-\epsilon^2}}\sin\left(2t\sqrt{\omega^2-\epsilon^2}\right)\Big]\bigg\}\;.
\end{align}
In this regime, \(N(t)\) remains bounded and stays finite as long as \(\epsilon<\omega\). In contrast, for \(\epsilon>\omega\),  the photon number evolves according to
\begin{align}
N(t)&=\frac{\epsilon^2}{2(\epsilon^2_c-\epsilon^2)}\bigg[1-\left(\frac{\Gamma}{\sqrt{\epsilon^2-\omega^2}}+1\right)e^{-2\lambda_-t}\nonumber\\\quad&+\left(\frac{\Gamma}{\sqrt{\epsilon^2-\omega^2}}-1\right)e^{-2\lambda_+t}\bigg]\;.
\end{align}
In this case, the growth of the photon number is unbounded  and diverges as \(\epsilon\to\epsilon_c\). For any finite value of the Kerr parameter $\chi$, the nonlinearity will ultimately saturate the photon growth.

\subsection{Dynamical protocol: from Heisenberg to dissipative regime}
\label{HeisenbergToDissipative}
Having established an analytical description of the system’s time-dependent dynamics, we are now in a position to return to our central task: parameter estimation. The transient evolution not only enriches our understanding of critical behavior but also provides a natural stage to investigate how information about the parameters of interest is encoded in the system at different times. In the following, we shift our focus back to estimation theory and analyze how these dynamical features can be harnessed for quantum-enhanced sensing. The time-dependent solutions given in Eq.~\eqref{Langevin solution} allow us to compute the instantaneous QFI, \(\mathcal{I}_\omega^t\). The latter has to be interpreted as the QFI of a protocol where the system is evolved until time $t$ and then measured. In particular, by substituting Eq.~\eqref{Langevin solution} into Eq.~\eqref{covmatr1} and using the general expression for the QFI of single-mode Gaussian states of Eq.~\eqref{QFIGaussian}, we can directly evaluate \(\mathcal{I}_\omega^t\). Although its full expression is too lengthy to report here, in the following we will analyse how it scales with the available resources in different regimes of interest. 

We begin by examining the simpler case where we set \(\Gamma=0\), and the resonator evolves following a unitary dynamics. Consequently, the critical point simplifies to \(\epsilon_c=\omega\). The covariance matrix of the system in this regime is given by
\begin{align}
    {\bf \Sigma}(t)&=\begin{pmatrix}
        \frac{\omega+\epsilon\cos\left(2t\sqrt{\omega^2-\epsilon^2}\right)}{\epsilon+\omega}&&\frac{\epsilon\sin\left(2t\sqrt{\omega^2-\epsilon^2}\right)}{\sqrt{\omega^2-\epsilon^2}}\\\frac{\epsilon\sin\left(2t\sqrt{\omega^2-\epsilon^2}\right)}{\sqrt{\omega^2-\epsilon^2}}&&\frac{-\omega+\epsilon\cos\left(2t\sqrt{\omega^2-\epsilon^2}\right)}{\epsilon-\omega}
    \end{pmatrix}\;,
\end{align}
and the corresponding average number of photons is \(N(t)=\epsilon^2\sin^2\left(\sqrt{\omega^2-\epsilon^2}t\right)/(\omega^2-\epsilon^2)\). Thus, in the limit \(\epsilon\to\epsilon_c\), the QFI scales as
\begin{equation}
\mathcal{I}_\omega^t\sim\left[2N(t)+\frac{8}{9}N^2(t)\right]t^2\;,
\end{equation}
with \(N(t)\sim\omega_0^2t^2\). Therefore, in the unitary case (\(\Gamma = 0\)), a dynamical protocol can achieve a genuine quadratic scaling with respect to both the photon number and the protocol duration. However, for any finite value of the nonlinearity \(\chi\) [see Eq.~\eqref{Kerr_Hamiltonian_t}], the photon number cannot grow indefinitely: there exists a maximum value \(N_{\rm max}\) beyond which saturation effects set in. Since the QFI increases monotonically with time, the optimal interrogation time \(T_{\rm opt}\) is therefore determined by the condition \(N(T_{\rm opt}) = N_{\rm max}\), which yields 
\[
T_{\rm opt} \approx \frac{\sqrt{N_{\rm max}}}{\omega_0}\;.
\]
Regarding measurements, Fig.~\ref{DynamicQFIvsHomodyne}(a) demonstrates that homodyne detection saturates the QFI in unitary dynamical protocols, making it an optimal measurement strategy in this setting.

Finally, let us incorporate dissipative effects by considering the case \(\Gamma \neq 0\). We begin with the short-time regime, where the impact of dissipation can be treated perturbatively. Specifically, we expand the leading-order behavior (for \(\epsilon \to \epsilon_c\)) of the single-shot QFI in powers of \(\Gamma\) for short evolution times~\cite{alushi2024collectivequantumenhancementcritical}:
\begin{equation}
    \mathcal{I}_\omega^t \sim \frac{8}{9}N^2(t)t^2 
    - \frac{32}{27}N^3(t)t^3\Gamma 
    + o(\Gamma^2)\;.
\end{equation}
The first \(\Gamma\)-independent term corresponds to the unitary case and yields Heisenberg scaling. The second term represents the leading correction due to dissipation. By comparing these contributions, one can identify the timescale beyond which the advantageous scaling of the unitary dynamics is lost. This analysis shows that Heisenberg scaling in both photon number and time is preserved only for sufficiently short protocol durations, namely
\[
t \ll \frac{3}{4 N_{\rm max} \Gamma}\;.
\]
Importantly, the extent of this unitary regime decreases inversely with photon number, implying that for any finite value of \(\Gamma\) the Heisenberg limit cannot be sustained asymptotically. Nevertheless, in realistic experimental conditions there may exist parameter regimes where quadratic scaling remains practically relevant.

\subsection{Optimization of dynamical protocol in the dissipative regime}
\label{Dissipative_dynamical_protocol}

Lastly, we consider the full dissipative regime, and identify the optimal protocol with respect to fundamental resources. While in the unitary regime the QFI increases indefinitely with time, the presence of dissipation causes it to eventually saturate to its finite steady-state value, \(\mathcal{I}_\omega^{ss}\) given in Eq.~\eqref{QFIss}. Specifically, for any finite value of \(\Gamma>0\), a transient regime emerges, bridging the unitary and the steady-state dynamics. During this transient phase, the QFI continues to increase with time—hence, the longer this regime lasts, the higher the attainable QFI before saturation. The duration of the transient regime is governed by two distinct timescales, determined by the inverse of the Liouvillian eigenvalues \(\lambda_\pm=\Gamma\pm\sqrt{\epsilon^2-\omega^2}\). Specifically, \(\rm Re\,\lambda_+^{-1}\) defines the effectively unitary timescale—beyond which the dynamics deviate from purely unitary evolution—while \(\rm Re\,\lambda_-^{-1}\) sets the timescale for reaching the steady state. Notably, in the unfavourable regime \(\epsilon<\omega\) the eigenvalues \(\lambda_\pm\) are complex conjugate of each other and so \(\rm Re \,\lambda_+^{-1} =\rm Re\,\lambda_-^{-1}=\Gamma^{-1}\). As a result, the transient regime effectively vanishes, and the QFI quickly saturates. This is consistent with the fact that, for \(\epsilon<\omega\), the photon number is bounded and cannot grow indefinitely [see Eq.~\eqref{before_exceptional}]. On the other hand, in the limit \(\epsilon\to\epsilon_c\) the steady-state timescale diverges as \(\lambda_-^{-1}\sim\Gamma/\epsilon_c(\epsilon_c-\epsilon)\), meaning that the duration of the transient regime is stretched by the critical slowing down, and the QFI can grow proportionally. This situation is depicted in Fig.~\ref{DynamicQFI}.

An asymptotic analysis in the limit \(\epsilon\to\epsilon_c\) shows that the QFI scales as \(\mathcal{I}_\omega^t\sim[2\omega_0^2/\epsilon^2(2\epsilon_c^2-\epsilon^2)]N^2(t)\) during the unitary regime, i.e., for \(t\lesssim \lambda_+^{-1}\). In the transient regime, \(\lambda_+^{-1}\lesssim t\lesssim \lambda_-^{-1}\), the scaling improves until  saturating at the steady-state \(\mathcal{I}_\omega^{ss}\sim[8\omega_0^2/\epsilon^2(2\epsilon_c^2-\epsilon^2)]N^2_{ss}\). In a single-shot realization of the critical estimation protocol—where time is not treated as a resource—the QFI maintains a \emph{quadratic} scaling with the photon number during all its evolution. Notice that the QFI rate can be maximized with respect to \(\omega_0\). In fact, it is straightforward to verify that the optimal choice is \(\omega_0=\Gamma\). Moreover, the optimal strategy in this regime is to perform the measurement at steady state, i.e., once the QFI saturates around \(T_{\rm opt}\approx\lambda_-^{-1}\). When we impose a constraint on the maximum number of photons allowed in the resonator, we can select the optimal pump strength such that \(N_{ss}=N_{\rm max}\). This yields \(\epsilon_{\rm opt}=\sqrt{2N_{\rm max}/(1+2N_{\rm max})}\epsilon_c\). Therefore, under the photon number constraint and with \(\omega_0=\Gamma\), the QFI achieves the scaling
\begin{equation}
    \mathcal{I}_\omega^{\rm opt}\sim\frac{2N^2_{\rm max}}{\Gamma^2}\;.
\end{equation}
This optimization has been done considering a single repetition of the protocol, assuming that the main constraint limiting the estimation precision is the cut-off on the photon number.

Let us now take into account time  and assume that one can repeat the protocol \(\mathcal{M}\) times to achieve a higher total QFI. In this multi-repetition strategy, both the number of photons and the total protocol duration time \(T\) are treated as resources. Since the QFI is additive for independent probes, the total QFI is given by \(\mathcal{I}_{\rm tot}(t)=\mathcal{M}\mathcal{I}_\omega^t\). The optimal approach here is to divide the total time \(T\) into \(\mathcal{M}\) equal intervals and perform each measurement at a time \(t=T/\mathcal{M}\). This results in \(\mathcal{I}_\omega^{\rm tot}(t)=\mathcal{I}_\omega^t\, T/t\), which is maximized when the ratio \(\mathcal{I}_\omega^t/t\) is maximized. As discussed previously, the single-shot QFI reaches its maximum value around \(T_{\rm opt}\approx\lambda_-^{-1}\), where the system reaches its steady-state. In the optimal scenario, i.e., when \(\omega_0=\Gamma\) and \(\epsilon=\epsilon_{\rm opt}\), the optimal measurement time simplifies to 
\begin{equation}
T_{\rm opt}\approx\frac{1+2N_{\rm max}+\sqrt{2N_{\rm max}(1+2N_{\rm max})}}{2\Gamma}\;,
\end{equation}
which scales as \(T_\omega^{\rm opt}\sim 2N_{\rm max}/\Gamma\). Given that the single-shot QFI scaling is \(\mathcal{I}_\omega\sim 2N_{\rm max}^2/\Gamma^2\), the total QFI under the optimal repetition strategy scales as
\begin{equation}
    \mathcal{I}_\omega^{\rm tot}\sim\frac{N_{\rm max}T}{\Gamma}\;.
\end{equation}
Therefore, in the presence of dissipation, the QFI scales linearly with the available resources, and Heisenberg scaling is no longer attainable. It is worth emphasizing that the impossibility of reaching Heisenberg scaling is not a consequence of using critical systems as probes, but rather stems from fundamental quantum mechanical limitations~\cite{Gorecki2025}.
Regarding the optimal measurement strategy in the dynamical protocol, we show in Fig.~\ref{DynamicQFIvsHomodyne}(b) that homodyne detection remains optimal in this case as well. By optimizing the homodyne Fisher information with respect to the quadrature angle \(\varphi\) at each time \(t\), we find that it perfectly saturates the QFI. Finally, although our discussion so far has focused on the zero-temperature case, the main results remain valid at finite temperature. Indeed, even if the analytical expressions differ, the underlying phenomenology and scaling laws are preserved~\cite{alushi2024optimality}.

\begin{figure}[t]
    \centering
    \includegraphics[width=0.49\textwidth]{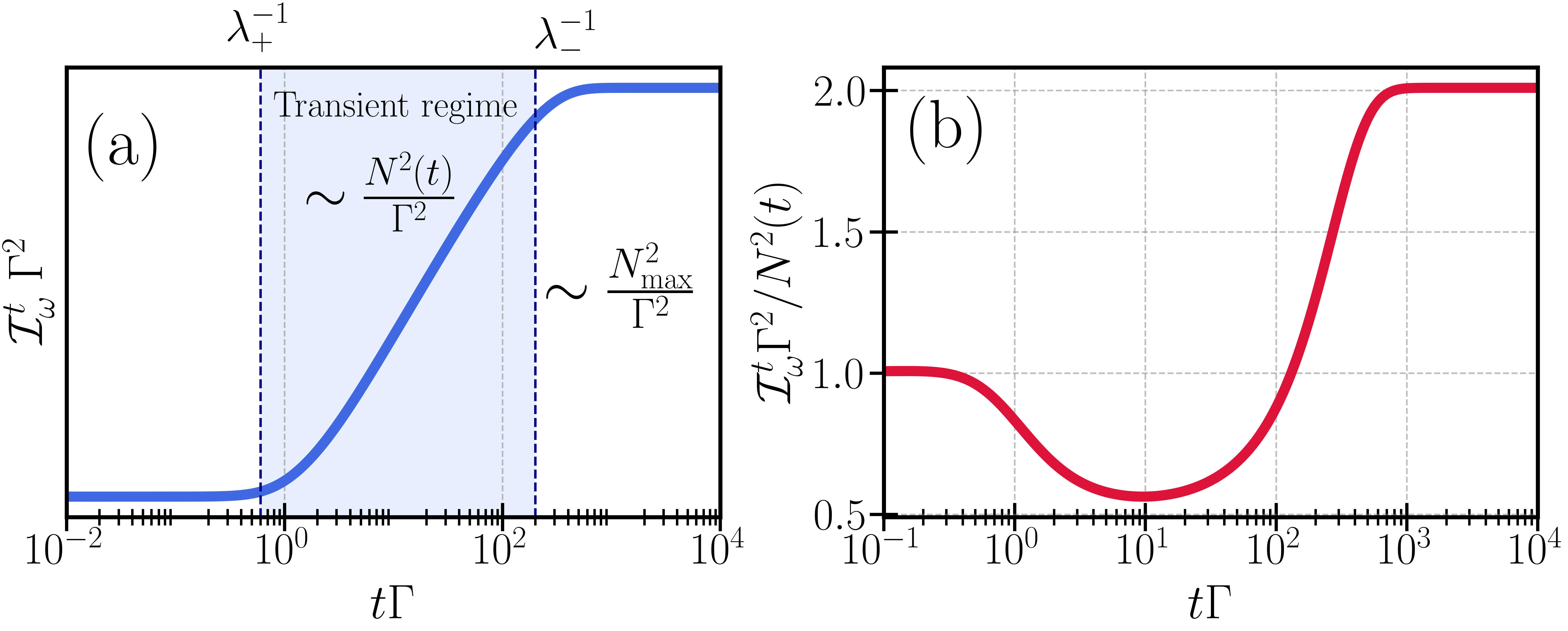}
    \caption{\textbf{Instantaneous QFI of the dynamical protocol.}
    Panel (a) shows the time-dependent QFI of the dynamical protocol, \(\mathcal{I}_\omega^t\), while (b) shows the ratio \(\mathcal{I}_\omega^t/N^2(t)\), as a function of time. In the presence of dissipation, the single-shot QFI scales quadratically with the number of photons---achieving Heisenberg scaling. This is attributed to the emergence of a transient regime as \(\Gamma>0\), during which the QFI can increase. The parameters are chosen such that \(\omega_0=\Gamma, \,N_{\rm max}=100, \epsilon=\epsilon_{\rm opt}=\sqrt{2N_{\rm max}/(1+2N_{\rm max})}\epsilon_c\).}
    \label{DynamicQFI}
\end{figure}

\section{Critical metrology in ultrastrongly coupled light-matter systems}\label{sec:ultrastrong}
In the previous section, we examined a driven-dissipative mechanism in which an external drive generates squeezing in the system's steady state. We now shift focus to a fundamentally different process: interaction-induced frequency renormalization, where the internal interactions among the system’s constituents effectively modify, or \emph{squeeze}, its natural frequency~\cite{gietka2025uscmetrology}. This effect is particularly prominent in cavity quantum electrodynamics, where embedding atoms or other media inside an optical cavity leads to a shift in the cavity’s resonance frequency due to light-matter coupling. 

\subsection{Quantum Rabi model}
To model such light-matter interactions at the most fundamental level, we consider the quantum Rabi model~\cite{Xie2017}, a cornerstone of modern quantum optics and cavity quantum electrodynamics. This model captures the coherent dynamics of a single two-level system---such as an atom, ion, or superconducting qubit---interacting with a single quantized mode of the electromagnetic field, typically confined in an optical or microwave cavity. The simplicity of the setup belies the richness of its physics, which includes nontrivial coupling effects, entanglement generation, and access to regimes far beyond perturbative treatments. The quantum Rabi Hamiltonian is given by
\begin{align}
    \hat H = \omega \hat a^\dagger \hat a + \frac{\Omega}{2}\hat \sigma_z + \frac{g}{2}\left( \hat a + \hat a^\dagger \right)\hat \sigma_x\;,
\end{align}
where \(\omega\) is the frequency of the cavity mode and \(\hat a\), \(\hat a^\dagger\) are the bosonic annihilation and creation operators describing photons in the cavity. The two-level system has an energy splitting \(\Omega\) and is described by the Pauli matrices \(\hat \sigma_x\), \(\hat \sigma_y\), and \(\hat \sigma_z\). The interaction strength between the two-level system and the cavity field is quantified by the coupling constant \(g\). 

The quantum Rabi model naturally divides into three components. The first term describes the energy of the quantized field mode, where each photon contributes energy \(\omega\). The second term accounts for the internal energy of the two-level system, with the ground and excited states separated by an energy \(\Omega\). The third term describes the light-matter interaction, allowing for the exchange of energy between the two subsystems via photon absorption and emission processes. 

Crucially, the interaction term includes both energy-conserving and non-conserving processes, the latter of which are typically neglected in the rotating-wave approximation used in the Jaynes--Cummings model. By retaining these counter-rotating terms, the quantum Rabi model remains valid even in the ultrastrong coupling regime~\cite{Nori2019USCreview,leboiteReview,solano2019rmp,QIN20241}, where the interaction strength becomes comparable to the bare frequencies of the system. This regime, once thought to be of only academic interest, is now experimentally accessible in several platforms, including superconducting circuits, semiconductor microcavities, and trapped ions. In such settings, the quantum Rabi model reveals a host of nonperturbative phenomena, such as significant ground state entanglement, spectral anharmonicity, and enhanced sensitivity to external perturbations~\cite{bastard2005vacuumproperties,Ciuti2006inputoutputUSC,Ciuti2009signaturesUSC,Todorov2010USCpolaritondots,Nori2010qubitoscillatorUSC,blais2011dissipationUSC,savasta2018dissipationUSC,Teufel2019Ulstrastrongmechanicalcavity,uros2023dissipativephasetrans,gietks2023USquezingQRM,stassi2023unvelingveiling,UrosDelic2024OPtomochenicalUSC,chen2024suppressedenergyrelaxationquantum}. 

These features are of particular interest in the context of quantum metrology~\cite{garbe2020criticalmetrology,zhu2023rabi,ying2022critical,plenio2022PRX}, where they can be harnessed to improve parameter estimation beyond what is possible in weakly coupled or classical systems. As we will see, approaching critical regions of the parameter space---where small changes lead to abrupt structural transformations in the system's ground or steady state---opens the door to novel strategies for high-precision sensing based on interaction-induced enhancements.

\subsection{Dispersive limit and the virtual excitations}
To gain analytical insight into the system's behavior, we now focus on the {dispersive limit}, where the energy gap of the two-level system dominates over the cavity frequency, i.e., \(\Omega/\omega \to \infty\). In this regime, the two-level system remains largely in its ground state and can be effectively {eliminated} from the dynamics. This simplification yields an effective Hamiltonian for the cavity field alone~\cite{plenio2015universalityQRM}
\begin{align}
    \hat H = \omega \hat a^\dagger \hat a - \frac{g^2}{4 \Omega}\left(\hat a + \hat a^\dagger\right)^2\;.
\end{align}
The resulting model describes a single harmonic oscillator whose frequency is renormalized---softened---by the light--matter interaction. This type of effective squeezing is a recurring motif throughout this tutorial. However, the situation here is subtly different from the previously discussed driven-dissipative squeezing: the canonical mode operators \(\hat a\) and \(\hat a^\dagger\) do not correspond to the physical excitations of the system. In other words, \(\hat a\) still describes an oscillator of frequency \(\omega\), but the actual physical frequency has shifted due to the interaction. To properly describe the physical observables of the system, we must diagonalize the Hamiltonian via a {Bogoliubov transformation}, introducing a new set of bosonic operators:
\begin{align}
    \hat c = \hat a \cos \xi + \hat a^\dagger \sin \xi\;,
\end{align}
which satisfy the canonical commutation relations and correspond to the new normal mode. The diagonalized Hamiltonian reads
\begin{align}
    \hat H = \omega \sqrt{1 - g^2/g_c^2} \, \hat c^\dagger \hat c\;,
\end{align}
where the critical coupling strength is given by \(g_c = \sqrt{\omega \Omega}\), and the frequency renormalization is encoded in the parameter \(\xi\) through the relation \(\omega \sqrt{1 - g^2/g_c^2} = \omega e^{-2\xi}\). Beyond the critical coupling strength, the effective Hamiltonian describes an inverted harmonic oscillator indicating the breakdown of the approximation in the superradiant phase.

In this basis, the system behaves as a regular harmonic oscillator with no squeezing and no excitations in the ground state, which appears completely classical. However, when expressed in terms of the original cavity mode \(\hat a\), the ground state contains a finite number of excitations:
\begin{align}
    \langle \hat a^\dagger \hat a \rangle = \sinh^2 \xi\;.
\end{align}
These excitations are referred to as {virtual photons}, since they do not correspond to real, measurable photons in the normal mode. Importantly, these virtual photons are not just a theoretical curiosity. They manifest in physical observables and carry quantum correlations that can, in principle, be extracted from the system~\cite{PhysRevA.110.063703}. This has sparked a growing experimental effort aimed at converting virtual excitations into real photons, with potential applications in quantum technologies~\cite{Liberato2007virtualph,Cirio2016vph,Cirio2017vph,Giannelli2024}. In particular, the squeezing embedded in these virtual photons offers a promising resource for {quantum metrology}, especially in regimes close to criticality, where the system becomes extremely sensitive to external perturbations.

\begin{figure}[t]
    \centering
    \includegraphics[width=0.95\linewidth]{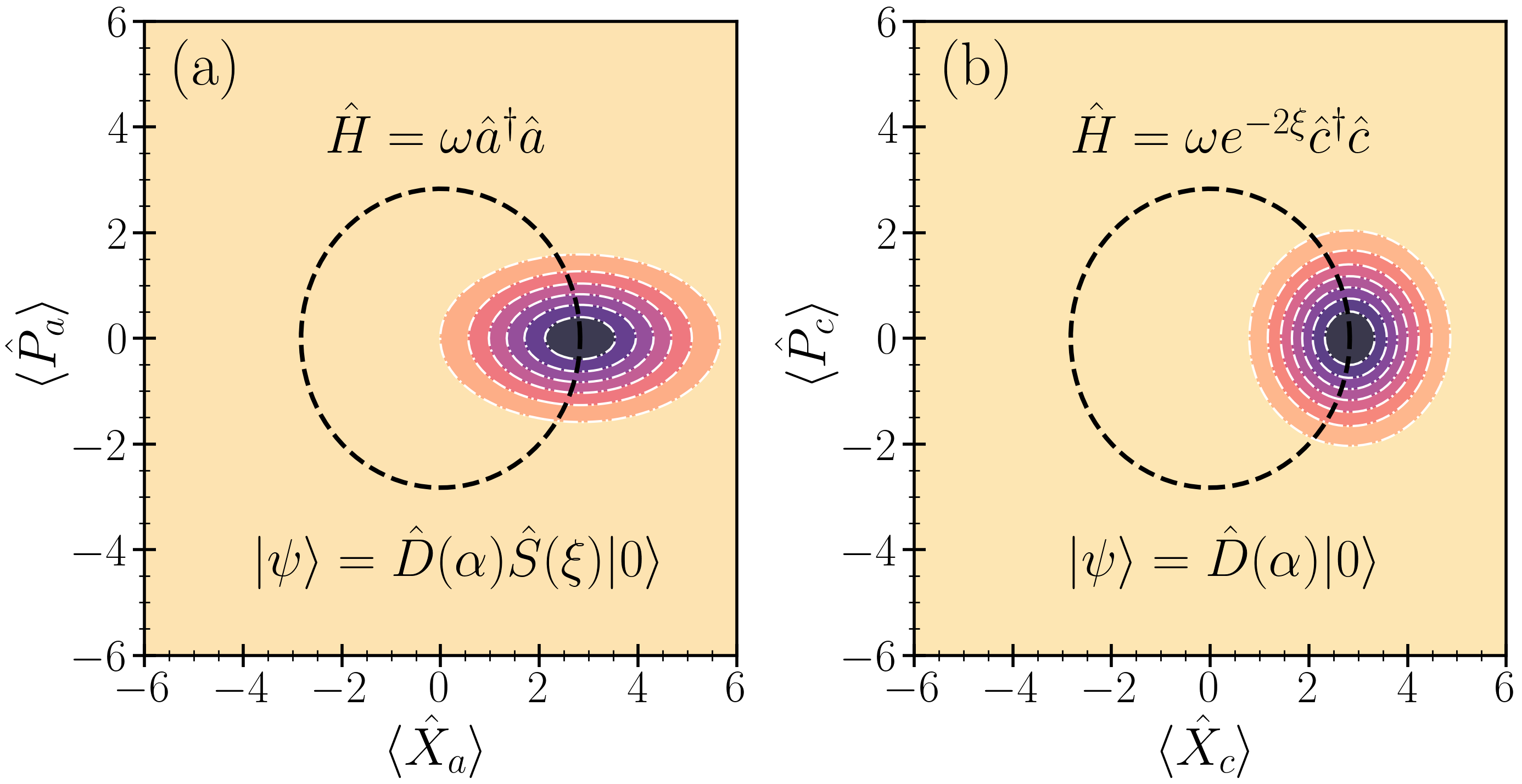}
    \caption{\textbf{Schematic illustration of the concept of frequency renormalization.} (a) In conventional quantum metrology, like Ramsey interferometry, explicit squeezing directly reduces quantum noise along one quadrature. (b) In contrast, virtual squeezing—realized as a modification of the system’s effective frequency—can be exploited. Here, the signal is amplified through the nontrivial dependence of the effective frequency, $\omega e^{-2\xi}$, on the unknown parameter $\omega$, mediated by the squeezing parameter $\xi$. For the same squeezing strength $\xi$ and simple quadrature measurements, virtual squeezing might offer improved sensitivity.}
    \label{fig:virtual}
\end{figure}

\subsection{Estimating the bare frequency from the normal mode}
Interestingly, it is not necessary to measure the system in the original cavity mode \(\hat a\) to take advantage of the squeezing present in the virtual excitations. In fact, all measurements can be performed using the normal mode \(\hat c\), which itself exhibits no squeezing (see Fig.~\ref{fig:virtual}). Nonetheless, this mode retains access to the metrological advantages encoded in the system's virtual photons. To demonstrate this, we now consider a simple yet instructive metrological task: estimating the cavity frequency \(\omega\). Crucially, the system's physical response occurs at the renormalized frequency \(\omega \sqrt{1 - g^2/g_c^2}\), rather than the bare cavity frequency \(\omega\). We assume the system is coherently driven at frequency \(\omega_p\) and coupled to a zero-temperature Markovian environment. The corresponding Hamiltonian reads
\begin{align}
    \hat H = \eta(\hat c + \hat c^\dagger) + \omega \sqrt{1 - g^2/g_c^2} \, \hat c^\dagger \hat c\;,
\end{align}
where \(\eta\) denotes the driving strength. The dynamics of the system, described by the density matrix \(\hat \varrho\), evolve according to the Lindblad master equation~\cite{breuer2002theory},
\begin{align}
    \partial_t \hat \varrho = \mathcal{L}[\hat \varrho] = -i \left[ \hat H, \hat \varrho \right] + \kappa \mathcal D [ \hat \varrho]\;,
\end{align}
where \(\kappa\) is the dissipation rate, and \(\mathcal{L}[\cdot]\) is the Liouvillian superoperator. The dissipator is expressed in terms of the normal mode bosonic ladder operators \(\hat{c}\) as
\begin{align}
    \mathcal D [\hat \varrho] = 2 \hat c \hat \varrho \hat c^\dagger - \hat c^\dagger \hat c \hat \varrho - \hat \varrho \hat c^\dagger \hat c\;,
\end{align}
ensuring the conservation of the total number of measurable rather than virtual excitations~\cite{Ciuti2006inputoutputUSC,blais2011dissipationUSC, Bamba2012,
Bamba2013, DeLiberato2014}.

This driven-dissipative system reaches a steady state described by a coherent state
\begin{align} \label{eq:usc_coh}
    |\alpha\rangle = 2  \sqrt{\frac{\kappa \eta^2}{\kappa^2 + 4 \delta^2}} \, \exp\left[ -i \arctan\left(\frac{\kappa}{2\delta}\right)\right]\;,
\end{align}
where \(\delta = \omega \sqrt{1 - g^2/g_c^2} - \omega_p\) is the detuning between the system’s response frequency and the driving field. The output photon rate is given by
\begin{align}
    \frac{\langle \hat n \rangle}{t} = \frac{4 \kappa \eta^2}{\kappa^2 + 4 \delta^2}\;.
\end{align}
Since the steady state is a coherent state, the QFI for estimating \(\omega\) can be calculated directly:
\begin{align}
    \mathcal{I}_\omega = 4 \left( \partial_\omega \alpha \right)^2 = 4 \left( \partial_\omega A \right)^2 + 4 A^2 \left( \partial_\omega \varphi \right)^2\;,
\end{align}
where \(A\) and \(\varphi\) denote the amplitude and phase of \(\alpha\), respectively. Substituting the expression from Eq.~\eqref{eq:usc_coh}, and focusing on the regime near criticality \(g \sim g_c\), we find that the QFI after time \(t\) becomes
\begin{align}
\begin{split}
    \mathcal{I}_\omega \approx &\,\, 4 |\alpha|^2 \frac{ 4 \delta^2 \Omega  \left(4 \delta^2 + \kappa^2 \right)}{\omega^2 \left(\Omega  \left(4 \delta^2 + \kappa^2 \right) - 4 g^2 \omega \right)^3} \, e^{4\xi} \\
    & + 4 |\alpha|^2 \frac{4 \kappa^2}{\omega^2 \left(\Omega  \left(4 \delta^2 + \kappa^2 \right) - 4 g^2 \omega \right)^2} \, e^{4\xi}\;,
\end{split}
\end{align}
where the first term corresponds to the information extractable from the amplitude, while the second term corresponds to the information extractable from the phase shift (see Fig.~\ref{fig:USC}).  Remarkably, despite operating in a classical coherent state, the QFI retains the imprint of the virtual-photon squeezing via the \(e^{4\xi}\) factor, reflecting the hidden quantum correlations inherited from the original light--matter interaction.

Interestingly, although the drive is classical and the system behaves as a simple harmonic oscillator with a renormalized frequency, the QFI is enhanced by a squeezing-related factor stemming from the redefinition of the resonant frequency \(\omega \rightarrow \omega \exp(-2\xi)\). This enhancement is intimately connected to the presence of virtual photons in the ground state of an ultrastrongly coupled light-matter system,
\begin{align}
    \langle \hat a^\dagger \hat a \rangle \sim \exp(2\xi)\;,
\end{align}
demonstrating that these virtual excitations do not need to be physically extracted to contribute to quantum-enhanced metrology. This observation reveals that even though the normal mode \(\hat{c}\) appears classical and unsqueezed in the dispersive regime, it retains a hidden imprint of the squeezing present in the virtual excitations of the original bosonic mode \(\hat{a}\). Remarkably, standard measurements performed on the driven-dissipative steady state of \(\hat{c}\) are sufficient to access this hidden resource, enabling enhanced sensitivity without direct manipulation or extraction of the virtual photons. The resulting metrological gain, captured by the QFI, reflects the effective frequency renormalization induced by ultrastrong coupling. Thus, virtual photons---despite being inaccessible through conventional means---can still be exploited for high-precision sensing by probing the system's dynamical response. This provides a practical and robust path to leveraging nonclassical resources embedded in the ground states---and the entire energy structure---of ultrastrongly coupled systems. The interplay between criticality and ultrastrong interactions therefore not only challenges conventional notions of metrological resources but also opens new avenues for quantum-enhanced sensing rooted in vacuum fluctuations.

\begin{figure}[t]
    \centering
    \includegraphics[width=0.9\linewidth]{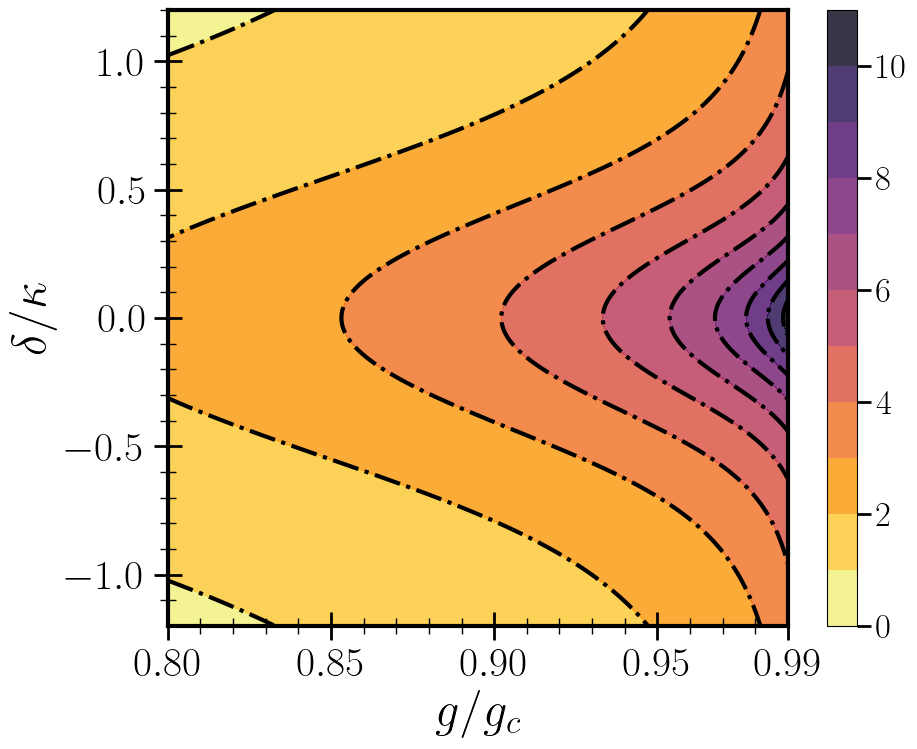}
    \caption{\textbf{Enhancement in the estimation of $\omega$ from the phase measurement as a function of $g/g_c$ and $\delta/\kappa$ with respect to a non-interacting case $g=0$.} The enhancement derives from the fact that the effective frequency of the system $\omega e^{2\xi}$ has a large derivative with respect to the parameter of interest $\omega$ close to the critical point of a phase transition. This enhancement can be traced back to the virtual photons in the ground state.}
    \label{fig:USC}
\end{figure}

\subsection{Finite-size effects}
The calculations above assume the limit \(\omega/\Omega \rightarrow 0\), 
where the system reduces to a harmonic oscillator with a tunable frequency. This captures the essential physics and predicts maximal squeezing, but it is an idealization. For any finite ratio \(\Omega/\omega\), anharmonicities appear, leading to nonlinear effects such as photon blockade~\cite{kimble2005photonblockade}, which limit the achievable coherent field amplitude. Moreover, realistic systems are finite, so the energy gap never fully closes, bounding both the QFI and the accessible squeezing parameter \(\xi\). Thus, the divergence of \(\xi\) predicted in the strict limit cannot be realized in practice. A way around these limitations is to move to the many-body regime described by the Dicke model~\cite{Garraway2011DM}, where a single bosonic mode couples to \(N\) two-level atoms. For large \(N\), the collective spin acts almost classically, restoring effectively harmonic dynamics near resonance without requiring extreme detuning. This collective enhancement provides a practical route to strong squeezing and metrological advantages that are inaccessible in single-atom setups.

Finally, it should be stressed that perfectly harmonic oscillators do not exist: strong driving or ultrastrong coupling always reveals nonlinearities. Managing their interplay with virtual excitations and squeezing is therefore crucial for realistically assessing the metrological potential of ultrastrongly coupled systems.

\graphicspath{{./mlrm_figs/}}

\section{Critical metrology in strongly correlated electron systems}\label{sec_mrlm}

As a final example, we now turn to a fundamentally different platform for critical quantum sensing: fermionic impurity systems hosting boundary quantum critical points. These systems constitute paradigmatic examples of strongly non-Markovian open quantum systems, where extensive multipartite entanglement between the impurity and its reservoirs builds up at low temperatures. As a result, perturbative treatments---such as those applicable to the driven dissipative Kerr resonator and ultrastrong light-matter platforms discussed previously---fail to capture the essential correlations and critical behavior. Remarkably, however, these non-Markovian open-system effects can be treated exactly: either numerically by employing the numerical renormalization group (NRG) originally introduced by Wilson~\cite{wilson1975renormalization}, or analytically at special solvable points of the model~\cite{PhysRevB.46.10812,andrew_entropy_exact}. To illustrate these ideas concretely, we will consider the exactly solvable point of the two-channel Kondo model (2CK), where the strongly correlated electron physics at the critical point can be mapped onto a non-interacting theory of free fermions, reducing a typically difficult problem into a pedagogical solvable model.

\subsection{Two-channel Kondo model}

The Kondo model is one of the paradigmatic models of strongly correlated electron physics~\cite{kondo1964resistance}. Originally introduced to explain the anomalous increase in resistivity observed in metals containing dilute magnetic impurities, it captures how a localized magnetic moment interacts with a sea of conduction electrons via antiferromagnetic exchange. Below an emergent low-energy scale, known as the \emph{Kondo temperature} \( T_K \), the impurity spin becomes dynamically screened by the surrounding electrons, forming a non-trivial many-body singlet state. This correlated screening process accounts for the characteristic rise in resistivity at low temperatures arising from enhanced spin-flip scattering of conduction electrons off the impurity.

From a modern perspective, Kondo physics has found a renewed relevance in nanoscale electronic systems. Semiconductor quantum dots, single-molecule transistors, and hybrid superconductor-semiconductor devices can all realize controllable Kondo effects, where a localized electronic level with strong Coulomb interaction acts as an effective magnetic impurity coupled to metallic leads~\cite{goldhaber1998kondo,potok2007observation,potok2007observation}. In such settings, parameters such as gate voltage, coupling asymmetry, and local magnetic fields can be tuned \emph{in situ}, allowing detailed experimental exploration of strongly correlated phenomena originally discovered in bulk systems. These include the emergence of SU(4) and orbital Kondo effects, spin-orbit-entangled screening, and even Majorana-Kondo hybridization in nanowires~\cite{keller2014emergent,pouse2023quantum,Kurzmann2021,Ricco2021}. Importantly, these effects do not rely on delicate fine-tuning, but rather, emerge inherently from strong electron-electron interactions, reflecting universal many-body phenomena which appear across a wide range of materials and devices. The resulting collective states host robust quantum coherence, entanglement, and enhanced response functions---resources of direct relevance to quantum metrology and quantum information processing, where sensitivity near correlated critical points can be fundamentally amplified. The Kondo model thus continues to serve as both a conceptual test-bed for the study of strong correlations and a practical platform for realizing correlated quantum technologies in mesoscopic devices. 

The two-channel Kondo (2CK) model represents one of the most celebrated examples of non-Fermi liquid physics and boundary quantum criticality in strongly correlated electron systems~\cite{2ck}. Originally introduced as an extension of the conventional Kondo problem, it describes a localized spin-$\tfrac{1}{2}$ impurity symmetrically coupled to two independent conduction electron channels (or ``leads'') whose Hamiltonian reads
\begin{equation}
    \hat{H}_{2CK} = \hat{H}_{\text{leads}} + J_L \hat{\mathbf{S}} \cdot \hat{\mathbf{s}}_{0L} + J_R \hat{\mathbf{S}}\cdot \hat{\mathbf{s}}_{0R}\;,
\end{equation}
with $\hat{\mathbf{S}}$ a spin-\( \tfrac{1}{2} \) operator for a single impurity, exchange coupled to two independent conduction electron channels with antiferromagnetic coupling $J_\alpha > 0$. The impurity couples locally to the spin density of conduction electrons in each channel at the impurity site \( \hat{\mathbf{s}}_{0 \alpha} = \tfrac{1}{2} \sum_{k k' \sigma \sigma'}  \hat{c}^\dagger_{\alpha \sigma k} \boldsymbol{\sigma}_{\sigma \sigma'}^{\phantom{\dagger}} \hat{c}^{\phantom{\dagger}}_{\alpha \sigma' k'}\) where \( \boldsymbol{\sigma}_{\sigma \sigma'}\) is a matrix of Pauli matrices. The leads are given by
\begin{equation}
    \hat{H}_{\text{leads}} = \sum_{\alpha \sigma k} \epsilon_k^{\phantom{\dagger}} \hat{c}^{\dagger}_{\alpha \sigma k } \hat{c}^{\phantom{\dagger}}_{\alpha \sigma k}\;,
\end{equation}
where \( \hat{c}_{\alpha \sigma k}\) ( \( \hat{c}^{\dagger}_{\alpha \sigma k}\)) annihilates (creates) an electron in the single particle momentum state $k$ of lead \( \alpha = L \) or \( R \) with spin \( \sigma = \uparrow \) or \( \downarrow \) and with electronic dispersion $\epsilon_k$. Unlike the single-channel case---where the impurity spin is completely screened at low temperatures---the symmetric \( J_L = J_R \) two-channel model flows to an overscreened non-Fermi liquid fixed point, characterized by a finite residual impurity entropy \( S_{\text{imp}} = \tfrac{1}{2}\ln{(2)}\).

Experimentally, realising the two-channel Kondo effect poses significant challenges. This is a consequence of the difficulty in engineering two independent symmetrically coupled electronic leads that compete to screen a single localized spin. Nonetheless, remarkable progress has been achieved in mesoscopic devices that emulate such boundary criticality. Semiconductor quantum dots coupled to metallic leads have demonstrated signatures consistent with overscreened behaviour: such as non-Fermi liquid scaling of conductance and residual zero-bias anomalies~\cite{Keller2015}. At the same time, the theoretical 2CK model admits an exactly solvable point—corresponding to a special symmetry limit in which the strongly correlated problem can be mapped onto a non-interacting theory of free fermions—providing a rare analytic window into the quantum critical behavior that these experiments seek to emulate~\cite{PhysRevB.46.10812}. We will turn to this solvable limit in the following section to illustrate explicitly how non-Fermi-liquid correlations and boundary criticality manifest in an exactly tractable form.

\subsection{Exactly solvable point of the model}
The intrinsic difficulty of the 2CK problem originates from the presence of two equivalent screening channels competing to screen a single spin-\(\tfrac{1}{2}\) impurity. In the frustrated, symmetric coupling regime \(J_L = J_R\), extensive entanglement between the impurity and a macroscopic number of fermionic degrees of freedom in the leads gives rise to a non-Fermi liquid ground state which cannot be accessed perturbatively in either the weak or strong-coupling limits. Remarkably, however, a special limit of the 2CK model---the so called \emph{Emery-Kivelson point} (EK)---admits an exact analytical solution. This point corresponds to a fine-tuned anisotropic limit of the exchange couplings $J_\alpha$ in which the transverse and longitudinal spin-exchange components satisfy a particular relation allowing an exact mapping of the strongly interacting Hamiltonian onto a quadratic model of fermions. 

The standard approach to this mapping employs bosonization of the conduction-electron fields, followed by a canonical transformation that diagonalizes the longitudinal exchange term. After refermionization of the remaining bosonic modes, the system reduces to an effective Hamiltonian in which only one linear combination of electronic fields couples to the impurity spin. At the EK point, the spin-flip term becomes bilinear in these new fermionic degrees of freedom, and the resulting model can be expressed as a \emph{Majorana resonant-level model} (MRLM) describing a single localized Majorana mode hybridized with a continuum of conduction-electron Majoranas. In its simplest form, the effective low-energy Hamiltonian at the EK point reads~\cite{PhysRevB.46.10812}
\begin{equation}
    \hat{H}_{\text{MRLM}} = \epsilon_d \hat{d}^\dagger \hat{d}^{\phantom{\dagger}}  + \sum_k\epsilon_k \hat{c}_{k}^{\dagger} \hat{c}^{\phantom{\dagger}}_k + V\sum_{k} \big( \hat{c}^\dagger_k + \hat{c}_k^{\phantom{\dagger}} \big) \big(\hat{d}^{\dagger} - \hat{d}^{\phantom{\dagger}}\big)\;,
\end{equation}
where \( \hat{c}_k\) (\( \hat{c}^{\dagger}_k\) ) annihilates (creates) a conduction-electron with energy \( \epsilon_k \), \( \hat{d} \) describes a localized impurity fermion of energy \( \epsilon_d\) and \( V \) is the hybridization. The key feature of this Hamiltonian is the non-standard hybridization which couples specific Majorana components of the conduction and impurity fermions, rather than their full complex forms. By defining the Majorana operators,
\begin{equation}
    \hat{\eta}_k = \frac{1}{\sqrt{2}} \big( \hat{c}_k^{\phantom{\dagger}} + \hat{c}_k^{\dagger} \big)\;, \qquad \hat{\xi}_k = \frac{i}{\sqrt{2}}\big( \hat{c}_k^{\dagger} - \hat{c}_k \big)\;,
\end{equation}
and similarly,
\begin{equation}
    \hat{\eta}_d = \frac{1}{\sqrt{2}} \big( \hat{d}^{\phantom{\dagger}} + \hat{d}^{\dagger} \big)\;, \qquad \hat{\xi}_d = \frac{i}{\sqrt{2}}\big( \hat{d}^{\dagger} - \hat{d} \big)\;,
\end{equation}
one can obtain Majorana operators for the impurity and conduction electron degrees of freedom:
\begin{equation}
    \hat{d} = \frac{1}{\sqrt{2}}\big( \hat{\eta}_d + i \hat{\xi}_d \big)\;, \qquad \hat{c}_k = \frac{1}{\sqrt{2}}\big( \hat{\eta}_k + i \hat{\xi}_k \big)\;,
\end{equation}
where all \( \hat{\eta} \) and \( \hat{\xi }\) operators are Hermitian Majoranas satisfying \( \{ \hat{\eta}_a , \hat{\eta}_b \} = \{ \hat{\xi}_a , \hat{\xi}_b\} = \delta_{ab} \) and \( \{\hat{\eta}_a , \hat{\xi}_b \} = 0\). Substituting these definitions into \( \hat{H}_{\text{MRLM}}\) yields the full Majorana representation:
\begin{equation}
    \hat{H}_{\text{MRLM}} = \frac{i\epsilon_d}{2} \hat{\eta}_d \hat{\xi}_d + \frac{i}{2} \sum_k \epsilon_k \hat{\eta}_k \hat{\xi}_k +  i V \sum_k \hat{\eta}_k\hat{\xi}_d\;.
\end{equation}
In this basis, the hybridization term couples only one of the impurity Majoranas \( \hat{\xi}_d \) to the conduction Majoranas \( \hat{\eta}_c \). This selective coupling captures the physical essence of the 2CK fixed point. Only half of the impurity degree of freedom hybridizes with the conduction bath, while the other half remains free. The MRLM admits a quantum critical point at particle-hole symmetry, \( \epsilon_d = 0\), where the impurity Majorana \( \hat{\eta}_d\) becomes completely decoupled from the Hamiltonian.

This decoupled zero-mode yields a two-fold degenerate ground state and a residual impurity entropy \( S_{\text{imp}} = \tfrac{1}{2} \ln{(2)}\), identifying the overscreened non-Fermi liquid fixed point of the 2CK model. Detuning away from criticality, e.g. by introducing a finite impurity level energy \( \epsilon_d \neq 0\), breaks the particle-hole symmetry, driving the system toward a conventional Fermi-liquid ground state. Physically, this occurs because the term \( \big(i \epsilon_d / 2 \big) \hat{\eta}_d \hat{\xi}_d\) couples the two impurity Majoranas, lifting the degeneracy and generating a second emergent low-energy scale
\begin{equation}
    T^* \sim \frac{\epsilon_d^2}{\Gamma}\;,
\end{equation}
where \( \Gamma = \pi \rho_0 V^2 \) is the hybridization width and \( \rho_0\) the conduction density of states at the Fermi level. This emergent low-energy scale \( T^*\) controls the crossover from non-Fermi liquid to Fermi-liquid behavior. For temperatures \( T \gg T^* \), the system remains in the quantum critical regime characterized by 2CK scaling, whereas for \( T \ll T^*\) the formerly decoupled Majorana becomes weakly hybridized, restoring full screening and conventional Fermi-liquid behavior.  

The energy scale \( \Gamma \) plays a role analogous to the Kondo temperature \( T_K\) setting the energy below which the system flows toward the critical non-Fermi liquid regime. In the low temperature limit \( T \ll \Gamma \), all thermodynamic and dynamical quantities exhibit universal scaling governed by this fixed point. Physical observables---such as the impurity spectral function, conductance, or spin susceptibility---show fractional power laws and logarithmic corrections, reflecting the anomalous dimension of the decoupled Majorana and the boundary quantum criticality of the 2CK problem~\cite{han2022fractional}.

\subsection{Estimating the on-site potential at the boundary driven critical point }

In strongly correlated electronic platforms such as quantum dots, nanowires, and molecular junctions, the natural observables are inherently local and transport based. Nevertheless, the inherent non-Markovian nature of the strong correlations accrued in such systems permits global many-body coherence and entanglement to manifest through such local quantities. In other words, although experimental access is typically restricted to local observables---such as the impurity occupation, current, or conductance---their expectation values and fluctuations retain signatures of the entire correlated environment. In context of critical quantum metrology, we therefore seek to exploit the divergent response of \emph{local} impurity observables, which faithfully encode the boundary quantum criticality, as opposed to relying explicitly on \emph{collective} measurements of many-particle states~\cite{Mihailescu_Thermometry,mihailescu2023multiparameter,mihailescu2024quantumsensingnanoelectronicsfisher}.

\begin{figure}[t]
    \centering \includegraphics[width=\linewidth]{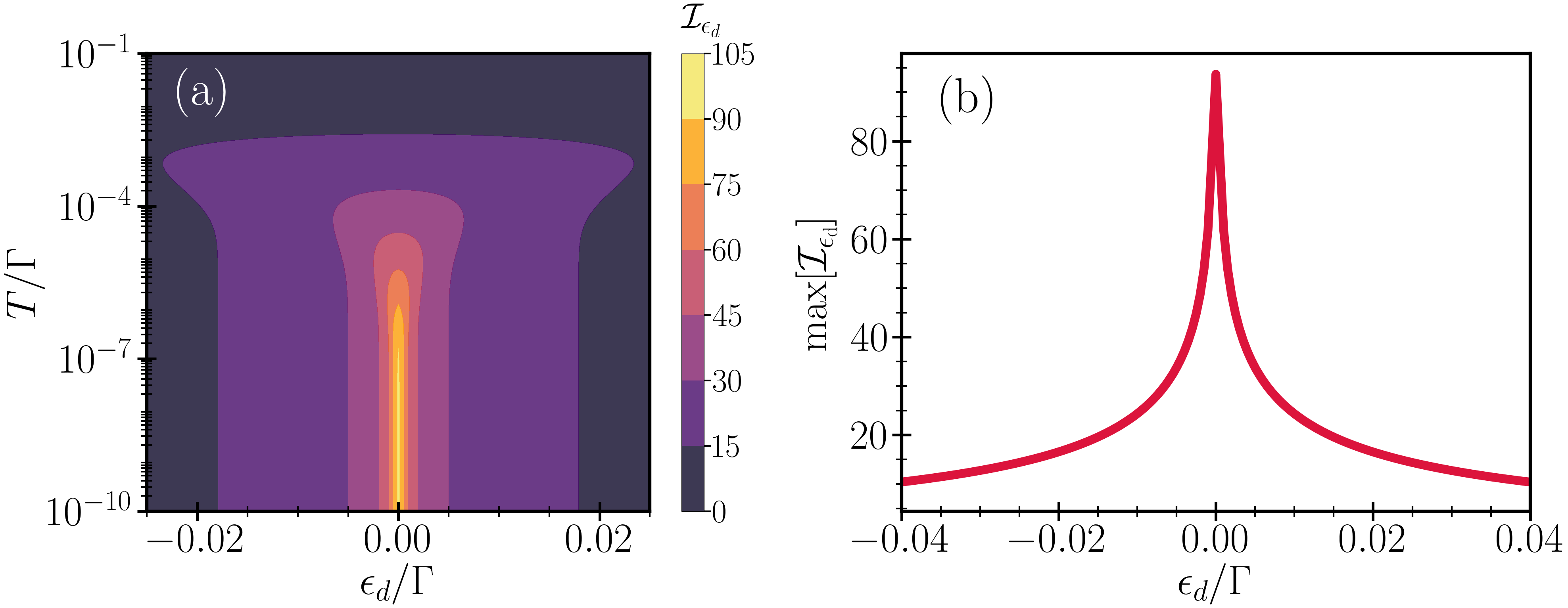}
    \caption{\textbf{Estimation of the on-site potential \( \epsilon_d\) in the 2CK model.} (a) QFI \( \mathcal{I}_{\epsilon_d}\) as a function of on-site potential \( \epsilon_d\) and temperature \(T \) (log-scale). (b) The maximum QFI \( \mathcal{I}_{\epsilon_d}\) sampled over the temperature window \( T \in \{10^{-10},10^{-1} \}\) for a given \(\epsilon_d\). The critical point occurs at \( \epsilon_d = 0\) and the energy scale is set \( \Gamma = 1\).}
    \label{fig:mrlm}
\end{figure}

Therefore, we will utilize the impurity quantum dot as an \emph{in situ} probe to estimate target quantities of interest. In particular, we will consider estimation of the on-site potential \( \epsilon_d \). Even though the impurity probe constitutes a single local degree of freedom, the reduced density matrix of the dot, denoted as \( \hat{\varrho}_{d}\), is strongly influenced by the many-body environment through the hybridization and entanglement with the conduction channels. Crucially, the reduced state of the impurity in the presence of the strongly interacting environment is characterized by a single observable: the average dot occupation \( \langle \hat{n}_d \rangle = \langle \hat{d}^\dagger \hat{d} \rangle \) which determines all elements of \( \hat{\varrho}_d\) in the basis of impurity states \( \{|0\rangle , |1\rangle  \}\) as
\begin{equation}
    \hat{\varrho}_d = \big( 1 - \hat{n}_d \big) |0\rangle \langle 0 | + \hat{n}_d |1 \rangle \langle 1 |\;.
\end{equation}
As the system is quadratic, it can be solved exactly by standard Green's function techniques and this can be exploited to provide a closed form solution of the dot occupation~\cite{vn83mt2v}
\begin{equation}
    \label{eq:dot_occup}
    \langle \hat{n}_d \rangle = \frac{1}{2} - \text{Im}\left\{ \frac{\epsilon_d}{\pi \Delta}\left[ \psi\big( \tfrac{1}{2} + \tfrac{\Gamma + i \Delta}{4 \pi T}\big) - \psi\big( \tfrac{1}{2} + \tfrac{\Gamma - i \Delta}{4 \pi T}\big)\right]\right\}\;,
\end{equation}
where \( \Delta = \sqrt{4 \epsilon_d^2 - \Gamma^2}\) and \( \psi(\cdot) \) the digamma function. Despite involving only a local expectation value, Eq.~\eqref{eq:dot_occup} encodes the full many-body correlations of the impurity with its reservoirs in the vicinity of the critical point. Moreover, since the impurity is a two-level system whose populations are described completely by the occupation \( \langle \hat{n}_d \rangle \), the QFI may be expressed as~\cite{Mihailescu_Thermometry,mihailescu2023multiparameter}
\begin{equation}
    \mathcal{I}_{\epsilon_d} = \frac{[\partial_{\epsilon_d} \langle \hat{n}_d \rangle]^2}{\langle \hat{n}_d \rangle (1 - \langle \hat{n}_d \rangle)}\;.
\end{equation}
We plot this quantity in Fig.~\ref{fig:mrlm}, fixing the energy scale \( \Gamma = 1\). In panel (a) the metrological sensitivity \( \mathcal{I}_{\epsilon_d}\) in estimating the on-site potential \( \epsilon_d \) is shown in the colorscale as a function of on-site potential and temperature. In particular, we note that the QFI diverges at the critical point \( \epsilon_d = 0\) where a metrological enhancement is offered due to boundary criticality, despite making \emph{local} measurements on the impurity probe \( \hat{\varrho}_d \). This represents a clear example of critical metrological amplification achieved without collective measurements, arising solely from the divergent response of local observables near criticality---a consequence of diverging correlation length scales which emerge at critical points. For finite \( \epsilon_d\) the sensitivity becomes reduced. This is because changing the on-site potential induces a change in the emergent energy scale \( T^*  \sim \epsilon_d^2/\Gamma\)---pushing the system out of the critical window.

Panel (b) shows the maximum QFI for each detuning, obtained by sampling over the temperature range in panel (a). This representation highlights the characteristic \emph{quantum critical fan}---a region of enhanced sensitivity that extends to finite temperature and detuning, bounded by the crossover scale \( T^*\). Within this fan, the system retains universal non-Fermi liquid correlations and exhibits critical metrological response, while outside this window the behaviour crosses over to conventional Fermi-liquid scaling. 

\section{Conclusions} \label{sec:conc}

Critical quantum metrology sits at a frontier where several of the most vibrant areas of modern quantum physics converge. It draws simultaneously on parameter estimation theory, information geometry, and the physics of critical phenomena, providing a common language in which concepts from quantum sensing and many-body theory meet. Within this unifying framework, the universality that emerges at critical points---driven by long-range correlations and diverging susceptibilities----naturally manifests as an amplification of parameter sensitivity, making criticality an intrinsic resource for metrology. We begin by introducing the fundamentals of parameter estimation theory, and use Ramsey interferometry as a benchmark against which to contrast critical metrology. From there, we build up from simple toy systems such as the harmonic oscillator and Landau–Zener model to paradigmatic many-body systems including the transverse-field Ising and Lipkin–Meshkov–Glick models: illustrating how the quantum state geometry, universal scaling laws, finite-size effects, and dynamical protocols shape the attainable precision. Along the way, we explore the subtleties of multiparameter estimation in critical regimes, as well as the role of finite temperature, before finally connecting these insights to realistic platforms such as driven–dissipative resonators, ultrastrong light–matter coupling regime, and strongly correlated electron systems, where critical metrology can be tested under experimentally relevant conditions.

A recurring theme has been the dual nature of criticality. On the one hand, diverging susceptibilities and enhanced correlations open the door to precision scaling beyond what is achievable with conventional probes. On the other, critical slowing down and the complexity of non-Gaussian states introduce significant experimental challenges. In particular, special care is required when interpreting scaling laws. The relevant preparation and evolution times typically grow with system size, and once these time scales are taken into account, apparent metrological enhancements may be substantially reduced or even vanish. Several strategies have been discussed to address these limitations. Operating at finite temperature can sometimes enhance sensitivity by populating low-lying excited states. Similarly, explicitly exploiting excited-state quantum phase transitions provides access to metrologically useful correlations beyond the ground state. Dynamical schemes such as sudden quenches or ramps offer another route, turning nonequilibrium critical dynamics into a resource for precision sensing. 

Together, these approaches demonstrate that critical metrology is not limited to ground-state equilibrium physics but can be extended into richer regimes where scaling advantages survive under realistic conditions. Realistic settings such as driven–dissipative resonators and ultrastrong light–matter systems further show that these trade-offs can be mitigated. The tutorial also highlights the importance of accessible measurements. While the quantum Fisher information sets fundamental precision bounds, practical protocols must contend with limitations on which observables can be measured. This makes the design of measurement schemes—ranging from collective spin observables to higher-order correlations and time-dependent signals—an essential ingredient of critical metrology.

\section{Outlook}\label{sec:outlook}
The main challenge in the experimental implementation of critical quantum sensing protocols is to operate systems in proximity of critical points, where the high susceptibility requires high stability of control parameters. Depending on the platform considered, the long timescales can also represent a technical challenge, even if formally the critical slowing down does not prevent achieving the optimal scaling also with respect to time. First experimental implementations already prove the practical feasibility of this approach. For example, many-body critical enhanced metrology for the sensing of external microwave electric fields has been demonstrated in a non-equilibrium Rydberg atomic gas~\cite{ding2022enhanced}. 
An adiabatic scheme on a perturbed Ising spin model with a first-order QPT has been implemented with nuclear magnetic resonance techniques~\cite{Liu2021}. A critical protocol for \emph{global} quantum sensing has been implemented with a two-qubit superconducting device~\cite{yu2025}. A 9-qubit superconducting device has been used to test a critical sensing protocol on a Stark-Wannier localization transition~\cite{li2025nonequilibrium}. A recent experiment demonstrated that quantum criticality in a non-Hermitian topological quantum walk can boost parameter estimation sensitivity, even during the transient dynamics~\cite{xiao2025}. Finally, a quadratic scaling of the estimation precision with respect to the effective system size has been demonstrated exploiting a driven-dissipative phase transition in a parametric superconducting resonator~\cite{beaulieu2025Criticality}. On a more general basis, a recent review collates the various methodologies by which many-body systems can be levered for enhanced parameter sensitivity more generally~\cite{MONTENEGRO20251}

An emerging theme is that critical metrology is not confined to equilibrium physics. Phenomena such as \emph{discrete time crystals}~\cite{montenegro2023quantum,yousefjani2025discretetimecrystalperiodicfield,PhysRevB.111.165117,PhysRevB.111.125159} illustrate how dynamical phases of matter can also be exploited as metrological resources. This perspective extends the scope of critical metrology well beyond traditional settings.  In summary, critical metrology challenges us to move beyond simple paradigms: its success will depend on protocols that combine theoretical elegance with experimental practicality, translating abstract ideas about scaling into real, resource-aware advances in sensing.  Looking ahead, several promising research avenues may define the next phase of critical metrology:

\begin{itemize}
    \item \textbf{Ultrastrong coupling and open systems.}  
    In regimes of strong and ultrastrong light--matter interactions~\cite{Nori2019USCreview,solano2019rmp}, dissipation can carry valuable information because jump operators themselves become parameter dependent~\cite{blais2011dissipationUSC}. Since the eigenstates of the system are drastically modified, open-system dynamics may encode parameter information in entirely new ways. This could lead to the derivation of novel metrological bounds that are specific to strongly and ultrastrongly interacting systems.  

    \item \textbf{Engineering gapped critical-like states.}  
    A long-standing challenge in critical metrology is the closing of the energy gap, which slows down adiabatic state preparation. An exciting direction is to engineer nonclassical, entangled eigenstates that mimic the advantageous sensitivity of states near criticality, but retain or even enlarge the energy gap~\cite{PhysRevLett.133.120601,gietka2022speedup}. Such states would combine enhanced precision with fast preparation times, potentially making critical (or perhaps anti-critical) metrology experimentally much more viable.  

    \item \textbf{Exploring non-equilibrium phases of matter.}  
    Beyond equilibrium quantum phase transitions, time crystals~\cite{montenegro2023quantum,10.21468/SciPostPhys.18.3.100} and other dynamical phases offer fresh metrological resources. Their intrinsic periodicity, rigidity, and noise resilience could inspire new sensing protocols that merge critical amplification with robustness.  

    \item \textbf{Quantum transport.} Experiments with nanoscale devices are capable of probing  charge, spin, and heat flow with high precision~\cite{goldhaber1998kondo,van2000kondo}---where observables such as conductance and current may be highly sensitive to control fields and temperature~\cite{perrin2015single,nowack2007coherent,pustilnik2004kondo,van2002electron,ihn2009semiconductor}. These platforms, long used to study many-body phenomena such as Coulomb blockade~\cite{meirav1990single,park2002coulomb}, Kondo physics~\cite{goldhaber1998kondo,liang2002kondo,keller2014emergent,piquard2023observation,anderson1961localized,PhysRevLett.18.1049,jeong2001kondo,kondo1964resistance,derry2015quasiparticle,sen2023many,mitchell2017kondo}, and quantum Hall effects~\cite{PhysRevLett.45.494,PhysRevB.41.7653,PhysRevLett.48.1559,DEPICCIOTTO1998395,PhysRevLett.50.1395}, may act as natural resources for metrology. Furthermore, quantum criticality has already been observed in such systems~\cite{potok2007observation,roch2008quantum,iftikhar2018tunable,pouse2023quantum,karki2023z,PhysRevB.89.045143}. Unlike optical schemes~\cite{Mabuchi_1996,Gammelmark2014,gammelmark2013bayesian,Landi_Col,kiilerich2010bayesian,plenio2022PRX}, transport devices rely on current-based measurements typically operating in strongly interacting, non-Markovian regimes. This necessitates novel approaches to bound ultimate precision via the quantum Fisher information and to quantify precision of experimentally accessible measurements. Recent progress has laid this foundation~\cite{mihailescu2024quantumsensingnanoelectronicsfisher,khandelwal2025currentbasedmetrologytwoterminalmesoscopic}, positioning transport as a natural arena for quantum metrology—experimentally ubiquitous and supported by mature measurement infrastructure. An exciting avenue for exploration lies in the emerging connection between quantum thermodynamics~\cite{binder2018thermodynamics} and metrology, where recent investigations have begun to reveal deep conceptual and practical links~\cite{campbell2025roadmapquantumthermodynamics,Mehboudi_Thermo_Review,PRXQuantum.5.020201,PhysRevE.111.064107,vn83mt2v}.


\item \textbf{Continuous Measurements.} A recurring message from recent work on continuously monitored quantum systems is that the information contained in the full measurement record is substantially richer than what can be extracted from the time-integrated or time-local signal alone. For both homodyne and photodetection schemes, Refs.~\cite{PhysRevA.94.032103,radaelli2024} show that the information contained in the mean signal captures only part of the available sensitivity, while the full Fisher information arises from temporal correlations in the measurement record. Refs.~\cite{PhysRevX.13.031012,PhysRevLett.132.050801} further demonstrate that standard time-local measurement strategies may fail to extract the full QFI of the emitted field. In such cases, exploiting temporal correlations—either through coherent-absorber schemes or by directly analyzing time correlations in the emission record—can recover the missing information, reveal enhanced scaling, and ultimately saturate the QFI, underscoring the central role of temporal structure in optimal continuous sensing. Finally, Ref.~\cite{plenio2022PRX} analyzes the optimality of continuous measurements in critical quantum systems, identifying the conditions under which continuous monitoring can saturate the QFI and realize the enhanced precision scaling associated with criticality. These results point toward a promising direction for future quantum sensing architectures, where exploiting the full temporal structure of continuous measurement records could unlock the ultimate precision limits of critical quantum systems.

\item \textbf{Global Sensing.} Recent works~\cite{abol2021criticalityglobalsensing, salvia2023CQMrealtimefeedback, PhysRevLett.133.120601} highlight that in critical quantum sensing, the achievable quantum enhancement depends sensitively on the available prior knowledge: any nonadaptive strategy will fail to exploit the critical advantage if the prior is insufficient, and the amount of prior knowledge required scales with the number of photons or particles. Adaptive measurement strategies with real-time feedback can overcome this limitation, enabling sub-shot-noise precision even with a limited number of measurements and substantial prior uncertainty. This underscores that practical implementations of global sensing in critical systems must carefully account for both the system size and the available prior information to fully harness the metrological potential offered by critical quantum sensing.
  
\end{itemize}

\begin{acknowledgments}
S.F. acknowledges financial support from the foundation Compagnia di San Paolo, grant vEIcolo no. 121319, and from
PNRR MUR project PE0000023-NQSTI financed by the European Union – Next Generation EU.
U.A. and R.D. acknowledge financial support from the Academy of Finland, grants no. 353832 and 349199. G.M. acknowledges fruitful exchanges with Scott and is grateful to Andrew Mitchell for suggesting the fermionic example in Sec~\ref{sec_mrlm}. K.G. would like to acknowledge discussions with Linda and Lidia.

\end{acknowledgments}



\bibliography{master_bib}

\end{document}